\documentclass[structabstract]{aa}  
\usepackage[toc,page]{appendix}
\usepackage{graphicx}
\usepackage{microtype}
\usepackage{hyperref}
\usepackage{graphicx}
\usepackage{multirow}
\usepackage{float}
\usepackage{etoolbox}



 \let\oldcitet=\citet
 \let\oldcitep=\citep
 \let\oldcitealp=\citealp

\renewcommand{\citet}[1]{\textcolor[rgb]{0,0,1}{\oldcitet{#1}}}
\renewcommand{\citep}[1]{\textcolor[rgb]{0,0,1}{\oldcitep{#1}}}
\renewcommand{\citealp}[1]{\textcolor[rgb]{0,0,1}{\oldcitealp{#1}}}

\bibpunct{(}{)}{;}{a}{,}{,}
\usepackage{txfonts}
\usepackage[dvipsnames]{xcolor}
\usepackage{orcidlink}

\usepackage{multicol}
\usepackage{longtable}
\usepackage{supertabular}
\usepackage{pdflscape} 
\usepackage{subcaption}
\usepackage{tablefootnote}

\begin{document} 

    \title{PRODIGE - Envelope to Disk with NOEMA\thanks{Based on observations carried out under project number L19MB with the IRAM NOEMA Interferometer. IRAM is supported by INSU/CNRS (France), MPG (Germany) and IGN (Spain)}}
    \subtitle{VI: The missing Sulfur Problem}
    
    \author{ J.~J.~Miranzo-Pastor\inst{1,2}\orcidlink{0009-0005-6689-7038}
    \and
    A.~Fuente \inst{1}
    \and
    D.~Navarro-Almaida \inst{1}\orcidlink{0000-0002-8499-7447}
    \and
    J.~E.~Pineda \inst{3}\orcidlink{0000-0002-3972-1978}
    \and
    D.~M.~Segura-Cox \inst{4,3}\orcidlink{0000-0003-3172-6763}
    \and 
    P.~Caselli \inst{3}
    \and
    R.~Martin-Domenech \inst{1}
    \and
    M.~T.~Valdivia-Mena \inst{3,5}
    \and
    T.~Henning \inst{6}
    \and
    T.-H.~Hsieh \inst{3}
    \and
    L.~A.~Busch \inst{3}
    \and
    C.~Gieser \inst{6}
    \and
    Y.-R.~Chou \inst{3}
    \and
    B.~Commer\c{c}on \inst{7}
    \and
    R.~Neri \inst{8}
    \and
    D.~Semenov \inst{6}
    \and
    A.~Lopez-Sepulcre \inst{9,8}
    \and
    N.~Cunningham \inst{10}
    \and
    L.~Bouscasse \inst{8}
    \and
    M.~Maureira \inst{3}
    }              
    \institute{Centro de Astrobiolog\'{\i}a (CAB), CSIC-INTA, Ctra. de Torrej\'on a Ajalvir km 4, 28850, Torrej\'on de Ardoz, Spain 
    \and
    Departamento de Física de la Tierra y Astrofísica, Facultad de Ciencias Físicas, Univ. Complutense de Madrid, 28040, Madrid, Spain
    \and
    Max-Planck-Institut f\"{u}r extraterrestrische Physik, Giessenbachstrasse 1, D-85748 Garching, Germany
    \and 
    {Department of Physics and Astronomy, University of Rochester, Rochester, NY 14627-0171, USA}
    \and
    {European Southern Observatory, Karl-Schwarzschild-Strasse 2, 85748 Garching, Germany}
    \and
    {Max-Planck-Institut f\"{u}r Astronomie, K\"{o}nigstuhl 17, D-69117 Heidelberg, Germany}
    \and
    {Centre de Recherche Astrophysique de Lyon/ENS Lyon}
    \and
    {Institut de Radioastronomie Millim\'{e}trique (IRAM), 300 rue de la Piscine, F-38406, Saint-Martin d'H\`{e}res, France}
    \and
    {Univ. Grenoble Alpes, CNRS, IPAG, 38000 Grenoble, France}
    \and
    {SKA Observatory, Jodrell Bank, Lower Withington, Macclesfield SK11 9FT, United Kingdom}
    }

    \abstract 
    {Determining the amount of sulfur in volatiles and refractories in the interstellar medium remains one of the main problems in astrochemistry. The detection of H$_2$S ices, which are thought to be one of the main sulfur reservoirs, is still a great challenge and has not been achieved yet, and the only sulfur-bearing species detected in the ices to date is OCS. PRODIGE (PROtostars and DIsks: Global Evolution) large survey observations with the NOrthern Extended Millimeter Array (NOEMA) of several Class 0/I protostars in the Perseus Molecular Cloud provide a perfect opportunity to study the H$_2$S and OCS composition of the ices through the volatiles sublimated in the warm inner core (T>100K, $n$$\sim$10$^6$cm$^{-3}$) of these protostars.}
    {Our aim is to determine the H$_2$S/OCS ratio in the warm inner core of the protostars of our sample in order to study how it is affected by different factors during its evolution.}
    {We used the NOEMA millimeter observations from the PRODIGE program of H$_2$S, H$_2^{33}$S, OCS, OC$^{33}$S and OC$^{34}$S to estimate the H$_2$S and OCS column densities in the warm inner core of 24 protostars of Perseus. In addition, we used SO and SO$_2$ data from the Atacama Large Millimeter/submillimeter Array (ALMA) archive to complete the sulfur budget and give a rough estimate of the total sulfur abundance in each of the sources. We explore the chemistry of H$_2$S and OCS in the warm cores using chemical and dynamical simulations of the collapse of a dense core to form a protostar.}
    {The compound H$_2$S is detected in 21 protostars and OCS in 17 protostars of our sample. The estimated H$_2$S/OCS ratio reveals a segregation of the sources into ``OCS-poor'' and ``OCS-rich'' protostars, where the OCS-poor protostars present higher H$_2$S/OCS ratios than the OCS-rich ones. Total sulfur abundance, which is always dominated by either H$_2$S or OCS, grows with evolution during the Class 0 phase, reaching a minimum depletion of a factor $<$8 in the Class 0/I objects, and decreasing again in the Class I. Simulations show that temperature changes in the pre-stellar phase and during the collapse can produce substantial differences in the H$_2$S and OCS (ice and gas-phase) abundances and in the H$_2$S/OCS ratio.}
    {Our analysis shows that the H$_2$S/OCS ratio is strongly influenced by the environment and the initial conditions of the cloud.}
  
    \maketitle

\section{Introduction}
\label{sec:intro}
Sulfur is the tenth most abundant element in the Universe and is known to play a significant role in biological systems \citep{Leman2004, Chen2019}.  It is found in a wide variety of biomolecules, such as amino acids, nucleic acids, sugars, and vitamins. In fact, along with hydrogen, carbon, oxygen, nitrogen, and phosphorus, it is considered to be one of the six elements fundamental to life. Moreover, some sulfur compounds, such as hydrogen sulfide (H$_2$S) have been proposed as necessary catalyst to form amino acids in the interstellar medium (ISM) \citep{Olson2016}. 
Nowadays, we are aware of the existence of more than 300 molecules in interstellar regions (\citealp{Muller2005,McGuire2022}, CDMS\footnote{\url{https://cdms.astro.uni-koeln.de/}}). Out of the currently detected interstellar molecules, only 33 contain sulfur atoms. The apparent  paucity of sulfur-bearing molecules detected in the ISM is somewhat reflective of a great problem in astrochemistry: while the observed gaseous sulfur seems to account for its total cosmic abundance (S/H$\sim$1.5 $\times$ 10$^{-5}$, \citealp{Asplund2009, Daflon2009}) in diffuse clouds and photodissociation regions (PDRs) \citep{Neufeld2015, Goicoechea2021, Fuente2024}, the sum of the abundances of the gas-phase sulfur bearing molecules detected constitute  $<$1\% of the expected amount \citep{Ruffle1999, Vastel2018, Rodriguez-Baras2021}. In protoplanetary disks, observations also reveal $<$1\% of the expected amount of sulfur \citep{Semenov2018, LeGal2019, Riviere-Marichalar2020}. One could think that in these cold and dense  regions, most of the sulfur is locked in the icy mantles that cover the dust grain surfaces. However, the detection of sulfur-bearing species in interstellar ices remains elusive. Nowadays, s-OCS (‘s-’ means that the molecule is in the icy grain mantles) is the only compound firmly detected in interstellar ices \citep{Palumbo1995,  Boogert2022}.  Different authors have published tentative detections of  the 7.5 $\mu$m band of s-SO$_2$  \citep{Rocha2024}, but this band is overlapped with intense bands of abundant complex organic molecules (COMs, C-bearing molecules containing six atoms or more) and CH$_4$, which hinders its confirmation (Taillard et al., in prep).  Although s-H$_2$S is the most abundant sulfur species in comets \citep{Calmonte2016}, and is predicted to be the main sulfur reservoir in interstellar ices, its has not been detected in the ISM, yet. One main problem is that its band at 3.9 $\mu$m is overlapped with an intense methanol band (Taillard et al., in prep). By now,  only upper limits to its abundance relative to water, which are within the range of  $\sim$0.1\% -- 1\%, have been determined \citep{Jimenez-Escobar2011, Rocha2024}. These upper limits are roughly consistent with some chemical models (see \citealp{Taillard2025}) and suggest that H$_2$S might not be as abundant as predicted, at least in the observed environments. It has been proposed that H$_2$S is destroyed within the ice to eventually form other sulfur compounds such as OCS \citep{elakel2022}. \citet{Laas2019} modeled sulfur chemistry assuming enhanced accretion of cations (in particular, S$^+$) on negatively charged grains and found that in timescales similar to the free-fall time, some species such as s-SO, s-OCS or s-HSO could be more abundant than s-H$_2$S. 

All in all, the abundances of all the species detected in gas and ice contribute to less than 5\% of the total sulfur, leaving around 95\% unaccounted for. Several theories have been proposed to explain this $"$missing$"$ sulfur. One possibility is that sulfur exists as neutral atomic sulfur, which is undetectable under the conditions typical of molecular clouds. Indeed, high abundances of atomic sulfur have been detected towards the Orion Bar by \citet{Goicoechea2021, Fuente2024}. Also, \citet{Hily-Blant2022} showed, indirectly with the NS/N$_2$H$^+$ ratio, that atomic sulfur could be the main carrier in dense cores.
Another compelling hypothesis is that sulfur is sequestered in sulfur allotropes and hydrogen sulfides  (S$_{\rm x}$, H$_2$S$_{\rm x}$), which could serve as semi-refractory 
sulfur reservoirs \citep{Jimenez-Escobar2012, Shingledecker2020, Fuente2023, Cazaux2022, Carrascosa2024}.  This idea is supported by the detection of these compounds 
 in comets \citep{Calmonte2016} and meteorites \citep{Aponte2023}, as well as the detection of S$_2$H towards the Horsehead nebula \citep{Fuente2017}.
Other proposed reservoirs include ammonium hydrosulfide \citep{Altwegg2022} and iron sulfides that have been detected in protoplanetary disks and meteorites \citep{Keller2002} . By now, the primary sulfur reservoir in cores ---where the depletion timescale is shorter than the dynamical timescale--- remains unidentified,  which highlights a critical gap in our understanding of sulfur chemistry.

Millimeter observations of gas-phase molecules can help to disentangle which is the most abundant sulfur compound in ices. We know that ices are formed in the cold pre-stellar phase where grain temperature decreases to $<$10 K and most molecules become frozen on their surfaces \citep{Walmsley2004, Pineda2022, Caselli2022}.  As the collapse proceeds,  a young protostellar object is formed in the interior of the dense core and heats the surrounding material. Dust and gas temperatures of hundreds of K can be reached in the inner $\sim$100 au around the central protostar \citep{MartinDomenech2021, Bianchi2022}. As the temperatures increase above 100 K, the water-rich ice mantles sublimate, injecting molecules into the gas phase (see, e.g. \citealp{Collings2003}). In some cases these warm regions present a rich chemistry with the detection of several COMs. In these cases,  we refer to this compact warm region as the hot corino \citep{Ceccarelli2007, Jorgensen2020}. The chemical composition of these warm regions should reflect to some extent the composition of the ices formed during the collapse, modulated by the effect of the different desorption mechanisms at play in the region. Binding energies determine the sublimation temperature of the different species, and one would expect a layered structure where the most volatile compounds appear at larger radii while the less volatile ones are only detected in the hot interior \citep{Ruaud2016}. Grain surface molecules can also be released to the gas-phase due to the accretion shocks formed when the envelope material falls onto the disk \citep{Villarmois2022}, and/or to the energetic shocks produced by the bipolar outflow \citet{Caselli1997, Schilke1997, Jimenez-Serra2008, Holdship2016}. Interferometric observations are needed to spatially resolve the interior of protostars and to determine the main desorption mechanism at work, a mandatory step to infer the connection between the gas and ice chemistry.
 
In Sections \ref{sec:obs} and \ref{sec:results}, we present high spatial resolution observations ($\sim$, corresponding to 1$\arcsec$ $\sim$ 300 au) of H$_2$S and OCS of 24 protostars located in Perseus. These compounds are expected to be the most abundant sulfur species in ices and present similar binding energies (E$_{bin}$ = 2400 K for OCS and 2700 K for H$_2$S, \citealp{Wakelam2017}), which ensures a fair comparison.
In Section \ref{sec:coldens-abund} we estimate the column density and abundances of H$_2$S, OCS and their isotopologues in all the sources of the sample, and in Section \ref{sec:ratios}, we determine the H$_2$S/OCS to explore possible variation of the ice composition along the sample. We combine our observations with previous SO and SO$_2$ observations to determine the sulfur budget in Section \ref{sec:sulfur_budget}. In Section \ref{sec:model}, we run a set of theoretical simulations to explore the parameters that could affect the different H$_2$S/OCS ratios. Finally, in Section \ref{sec:discussion}, we discuss the astrochemical implications of our results.

\begin{figure*}
    \centering
    \includegraphics[width=\linewidth]{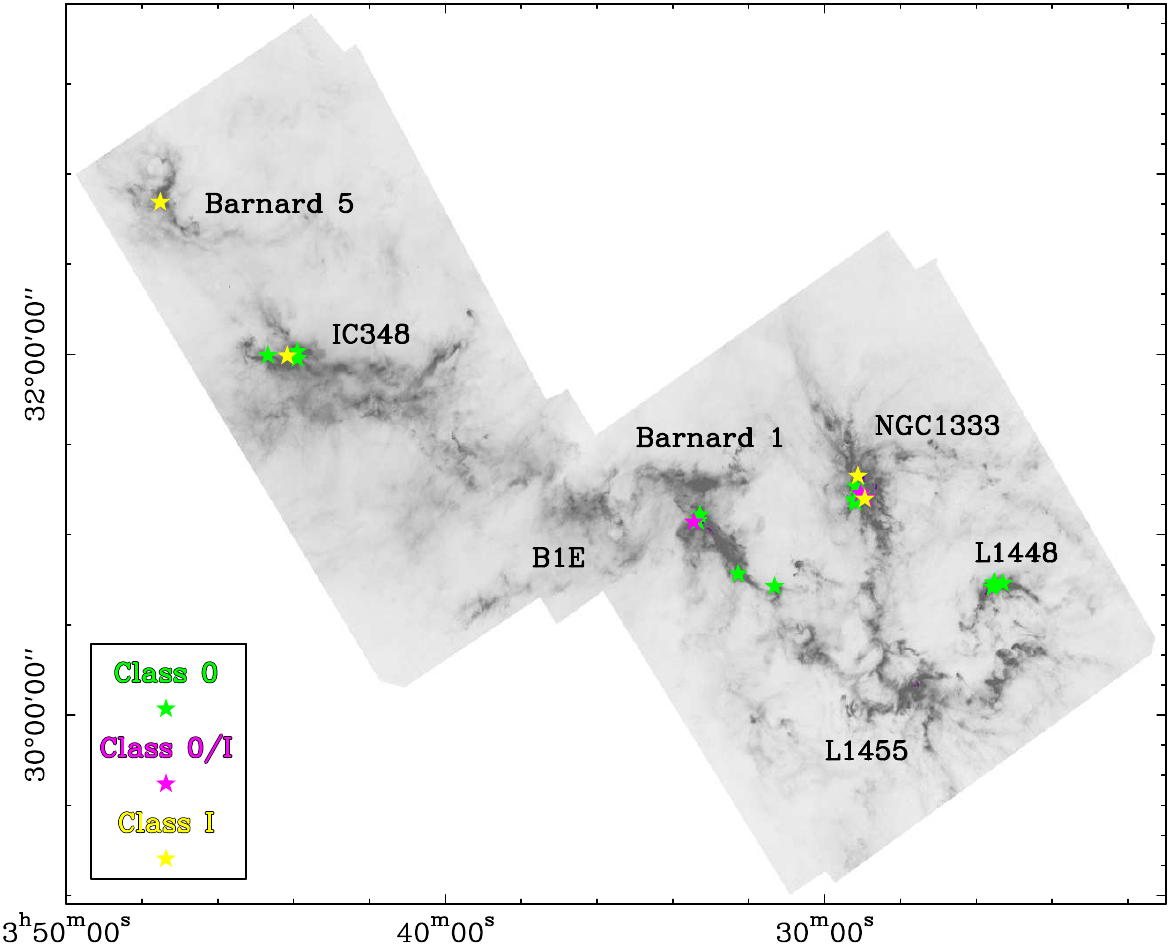}
    \caption{Map of the Perseus Molecular Cloud and its several subregions. Represented with stars, the positions of the 24 protostars observed in this work. Their color shows the evolutionary stage of the source. Gray-scale background shows far-infrared and submillimeter emission from the Herschel Gould Belt Survey \citep{Andre2010,HGBS2020}}
    \label{fig:perseus-map}
\end{figure*}

\section{Observations}
\label{sec:obs}

This paper is based on observations of the MPG-IRAM Observing Program PRODIGE (PROtostars and DIsks: Global Evolution, Project ID: L19MB)\footnote{\url{https://iram-institute.org/science-portal/proposals/lp/miop/miop-002/}} PIs: Paola Caselli, Thomas Henning. Within this program, a series of 32 Class 0/I protostars have been observed with the NOrthern Extended Millimeter Array (NOEMA). In this program, the Band 3 receiver and the PolyFix correlator were used. PolyFix provides $\sim$15.5 GHz of bandwidth (divided into two 7.744 GHz wide sidebands separated by 15.488 GHz). The whole band was observed with a channel width of 2 MHz. 
Moreover,  several windows with high spectral resolution,  62.5 kHz channel width, were placed to observe selected lines.

We processed the NOEMA observations from uncalibrated data using the standard observatory pipeline in the Grenoble Image and Line Data Analysis Software (GILDAS) package Continuum and Line Interferometer Interferometer Calibration (CLIC). We used 3C84 and 3C454.3 as bandpass calibrators. 0333+321 and 0333+322 were used for phase and amplitude calibration, with observations for these calibrators taken every 20 min. Flux calibration sources were LKHA101 and MWC349. The uncertainty in flux density was $10\%$. The continuum was bright enough to allow for phase self-calibration. More details about calibration and imaging can be found in Section 2.1 of \citet{Gieser2024}.

\setlength{\tabcolsep}{3pt}
\renewcommand{\arraystretch}{1.5}
\begin{table}[h]
\caption[List of studied NOEMA lines in order of increasing wavelength with atomic data retrieved from Splatalogue and CDMS.]{List of studied NOEMA lines in order of increasing wavelength with atomic data retrieved from Splatalogue\tablefootnote{\url{https://splatalogue.online}} and CDMS\tablefootnote{\url{https://cdms.astro.uni-koeln.de/}}.}
\label{tab:lines}
\centerline{
\begin{tabular}{ccccccc}
\hline\hline
$\nu$ &  Species & \multicolumn{1}{c}{Transition} &    A$_{ij}$               &  E$_u$  &  $\Delta\nu$ & RMS \\
\scriptsize (GHz)  &  & \multicolumn{1}{c}{\scriptsize Upper - Lower} & \scriptsize log(s$^{-1}$) & \scriptsize (K) & \scriptsize (kHz) & \scriptsize (mJy/beam) \\
\hline
215.5029  &  H$_2^{33}$S &  2$_{2,0,4}$--2$_{1,1,4}$ &  -4.374  &  81.6  &  2000  &  [2.0-2.9] \\
216.7104  &  H$_2$S      &  2$_{2,0}$--2$_{1,1}$   &  -4.312  &  84.0  &  2000  &  [2.0-2.9] \\  
216.1474  &  OC$^{33}$S  &  J = 18 $-$ 17     &  -4.420  &  98.6  &  62.5  &  [15-60] \\
218.9034  &  OCS         &  J = 18 $-$ 17     &  -4.517  &  99.8  &  62.5  &  [10-50] \\
231.0615  &  OCS         &  J = 19 $-$ 18     &  -4.446  &  111   &  62.5  &  [20-80] \\
237.2736  &  OC$^{34}$S  &  J = 20 $-$ 19     &  -4.411  &  120   &  2000  &  [2.5-3.5] \\
\hline 
\end{tabular}
}
\begin{flushleft}
    \tiny\textbf{Notes:} For H$_2^{33}$S and OC$^{33}$S, the A$_{ij}$ were calculated from the available data in CDMS. $\Delta$$\nu$ is the channel width with which the transition was observed. In the last column (RMS) we give a range for the typical noise in the datacubes of each transition.
\end{flushleft}
\end{table}

This work relies in a subset of 24 Class 0/I protostars from the previously mentioned observations (see Table~\ref{tab:sources}, Fig.~\ref{fig:perseus-map}), which were the ones observed and reduced between 2019 and 2023. This includes all Class 0 (the younger) sources and also some of the Class 0/I and I protostars.

\section{Results}
\label{sec:results}

In this section, we show the moment 0 maps of the 24 sources in our sample, the spectra of H$_2$S, OCS(18-17) and OC$^{34}$S (see Table~\ref{tab:lines}) and their Gaussian fits, and we list the detection rates of each species.

In Fig.~\ref{fig:moment-0-maps} we present the moment-0 maps of the 24 protostars, showing the emission of the main species H$_2$S, OCS(18-17) and OCS(19-18) lines. We created the moment-0 maps by integrating the data cubes over specific velocity ranges for each map. Integration was done over a threshold of 3$\sigma$, for each channel individually. For a more accurate description of the maps, the different velocity ranges and $\sigma$ values are detailed in the corresponding table of the repository referenced in Appendix \ref{sec:appendix_tables}. The emission is presented in the form of a contour map, within a window of 4$\arcsec$$\times$4$\arcsec$, centered in the peak of the continuum (see Table~\ref{tab:sources}). The contour lines delineate regions were emission is within 10\%, 30\%, 50\%, 70\% and 90\% of the peak emission. In addition, to complement this representation, we also give the maximum value of the emission and the size of the beam, represented by a purple ellipse in the bottom-left corner. We have also marked the position of the peak of the continuum, which is represented with a purple star in the center of each image.

In the maps, we see that the H$_2$S line usually appears centered in the peak of the continuum, showing an approximately symmetric distribution. In the sources where OCS was not detected, H$_2$S emission may have a more irregular shape. We also find that OCS is only observed in sources where H$_2$S was detected, and that OCS(18-17) and OCS(19-18) are always detected together, although sometimes they may have rather different distributions.

Regarding the isotopologues, we find that many of the OCS detections are also detected in OC$^{34}$S. This is a clear indicator of optically thick OCS lines, as we will see later in Sec.~\ref{sec:coldens-abund}. We have assumed an $^{32}$S/$^{34}$S isotopic ratio of 22.5 \citep{AndersGravesse1989}. $^{33}$S is even scarcer than its 34 isotope counterpart, but the value of the $^{34}$S/$^{33}$S ratio has not been uniformly established, as it may change depending on the region. Standard values are 5.61 (solar, \citealp{AndersGravesse1989}) and 6.27$\pm$1.01 \citep{Chin1996}.

\hfill\newline
In this work, we are interested in the H$_2$S/OCS ratios in the warm inner core (T>100K, $n$$\sim$10$^6$cm$^{-3}$) of 24 Class 0/I protostars in Perseus. To carry out this analysis, we needed to integrate the spectra of only the most interior region of the protostar, where the core is found. By comparing with the typical extension of the H$_2$S (2$_{2,0}$--2$_{1,1}$) emission, we decided to extract the spectra of a circular region of 1.5$\arcsec$ in diameter around the peak of the continuum for every source. After extracting all the spectra, we fitted a Gaussian profile to the detected lines using the {\tt CLASS} software from GILDAS\footnote{\url{https://www.iram.fr/IRAMFR/GILDAS}}.

In Fig~\ref{fig:specs}, we present the integrated spectra of the H$_2$S 2$_{2,0}$--2$_{1,1}$, the OCS (18--17) and the OC$^{34}$S (20--19) lines in the warm inner core (1.5$\arcsec$$\times$1.5$\arcsec$) of the 24 protostars from our sample. We find clear detections (>5$\sigma$) of both H$_2$S and OCS in most of the sources while OC$^{34}$S detections are usually more faint (>3$\sigma$) or tentative (<3$\sigma$). Note that, after integrating the spectra over the 1.5$\arcsec$ region, measured emission can be over 3-5$\sigma$ due to the increased signal to noise ratio after the integration, even if emission inside one beam was below 3$\sigma$ for some channels.

We have also extracted the spectra of the OCS (19--18), the OC$^{33}$S (18--17) and the H$_2^{33}$S 2$_{2,0}$--2$_{1,1}$ lines. H$_2^{33}$S has hyperfine structure and its emission is divided in 10 different lines. We show an example of the H$_2^{33}$S 2$_{2,0}$--2$_{1,1}$ spectral distribution in Fig.~\ref{fig:H233S}. OC$^{33}$S also has hyperfine structure; however, for high $J$ transitions, the line splitting cannot be resolved, and we do not account for it in our fits and calculations.

Table~\ref{tab:detections} summarizes the detected and undetected lines in the 24 protostars of Perseus observed in this work. We can see that in all the sources where OCS was detected, H$_2$S emission could also be observed, but not vice versa (see B5-IRS1, L1448NW, Per-emb-08). Additionally, all OCS detections were observed in both J=18$-$17 and J=19$-$18 transitions.

Regarding the 17 Class 0 sources, we observed the H$_2$S 2$_{2,0}$--2$_{1,1}$ line in 16 sources ---although B1bN, the only Class 0 source not observed in H$_2$S, was actually detected in its isotopologue, H$_2^{33}$S--- and we detected both OCS lines in 14 of them. That is a $\sim$94$\%$ detection rate of H$_2$S and an $\sim$82$\%$ detection rate of OCS in Class 0 protostars (or $\sim$88$\%$ detection rate of OCS in Class 0 protostars with H$_2$S). On the other hand, for the seven Class 0/I and Class I protostars, we detected H$_2$S in four of the sources and we observed OCS emission in only three of them. That corresponds to a $\sim$57$\%$ detection rate of H$_2$S and a $\sim$43$\%$ for Class 0/I and Class I protostars (or $\sim$75$\%$ detection rate of OCS in sources with H$_2$S emission). Globally, we have a $\sim$83$\%$ detection rate of H$_2$S and a $\sim$71$\%$ detection rate of OCS (or, equivalently, a $\sim$85$\%$ detection rate of OCS in protostars with H$_2$S).

When it comes to the isotopologues, in the Class 0 protostars, we found OC$^{34}$S emission towards ten of the 14 sources with OCS emission, and we detected OC$^{33}$S in seven of them. This is a $\sim$71$\%$ detection rate of OC$^{34}$S and a $\sim$50$\%$ detection rate of OC$^{33}$S in Class 0 protostars with OCS (or $\sim$70$\%$ detection rate of OC$^{33}$S where detected in OC$^{34}$S). For the Class 0/I and Class I objects, we found OC$^{34}$S in only three of the seven protostars, and OC$^{33}$S in just two of them. This makes a $\sim$38$\%$ detection rate of OC$^{34}$S and a $\sim$25$\%$ detection rate of OC$^{33}$S in Class 0/I and Class I sources where OCS was observed (or $\sim$67$\%$ detection rate of OC$^{33}$S in the sources with OC$^{34}$S emission). In general, we find that OC$^{34}$S has a detection rate of $\sim$76$\%$ and that OC$^{33}$S has a $\sim$53$\%$ detection rate in protostars with detected OCS (which is a $\sim$69$\%$ detection rate of OC$^{33}$S in sources with OC$^{34}$S).
Additionally, we found that six sources presented H$_2^{33}$S emission in the warm inner core, which is a 30$\%$ detection rate of H$_2^{33}$S in the 20 sources where H$_2$S was detected. Also, one more H$_2^{33}$S detection corresponding to the <3$\sigma$ H$_2$S emission of B1bN was found. This makes a total of seven detections and a detection rate of $\sim$29$\%$ in the full sample. Of these seven detections, five of them were in Class 0 protostars and two in Class 0/I sources, making it a $\sim$29$\%$ detection rate in both Class 0 and Class 0/I + I sources.

\begin{figure}[h]
    \centering
    \includegraphics[width=\linewidth]{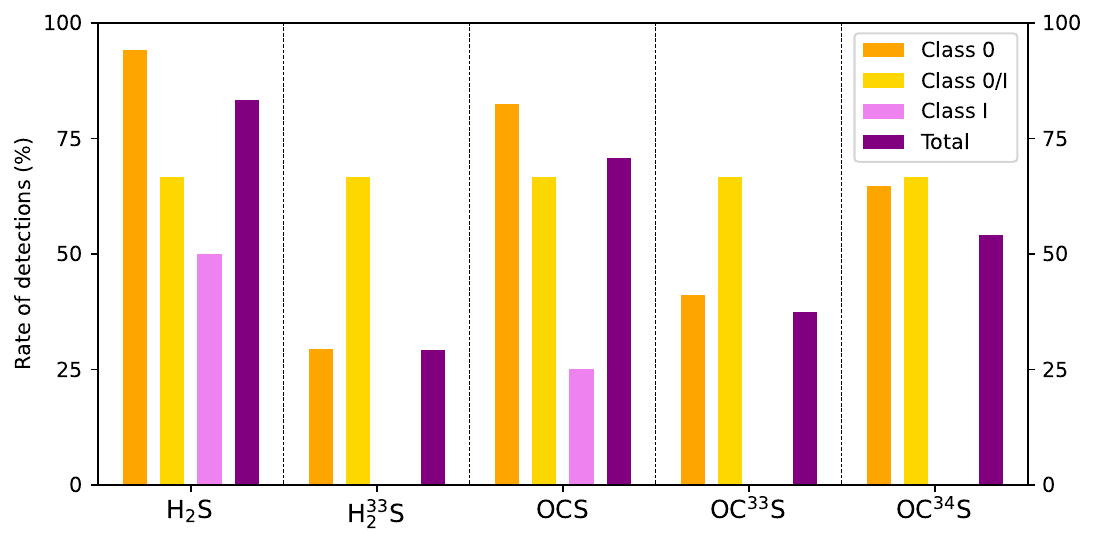}
    \caption{Detection rate of the different molecules in the warm inner core of the 24 sources of our sample. Of the total 24 protostars, 17 of them are Class 0 sources, three of them are classified as Class 0/I and the last four sources are Class I.}
    \label{fig:detection-rate}
\end{figure}

\begin{figure*}
    \centering
    \begin{subfigure}{0.47\textwidth}
        \includegraphics[width=\linewidth]{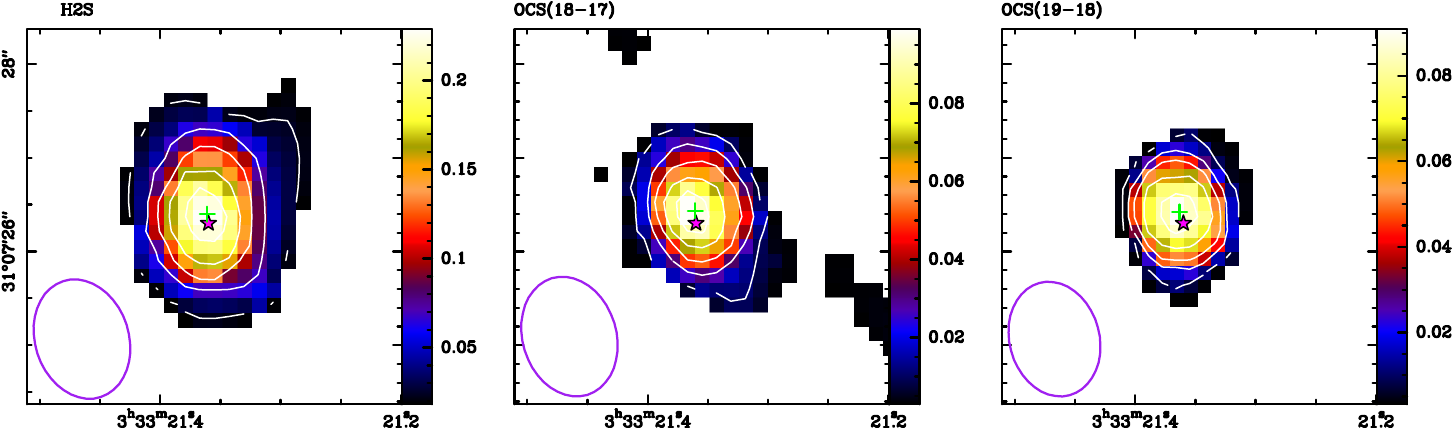}
        \caption{B1bS}
        \label{fig:map_B1bS}
    \end{subfigure}
    \hspace{0.9cm}
    \begin{subfigure}{0.47\textwidth}
        \centering
        \includegraphics[width=\linewidth]{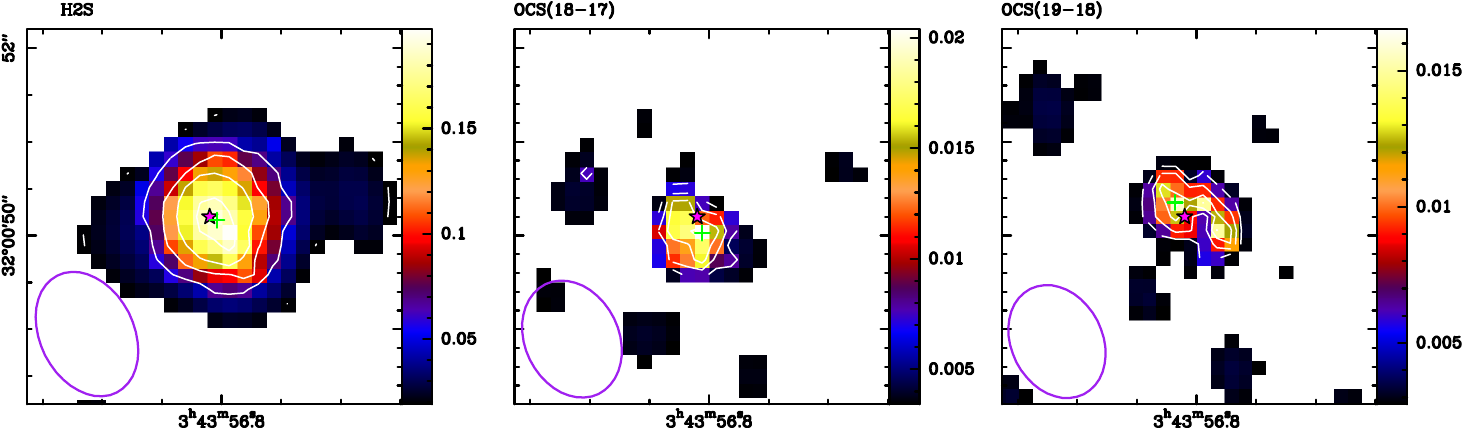}
        \caption{HH211MMS}
        \label{fig:map_HH211}
    \end{subfigure}

    \vspace{0.5cm}
    \begin{subfigure}{0.47\textwidth}
        \centering
        \includegraphics[width=\linewidth]{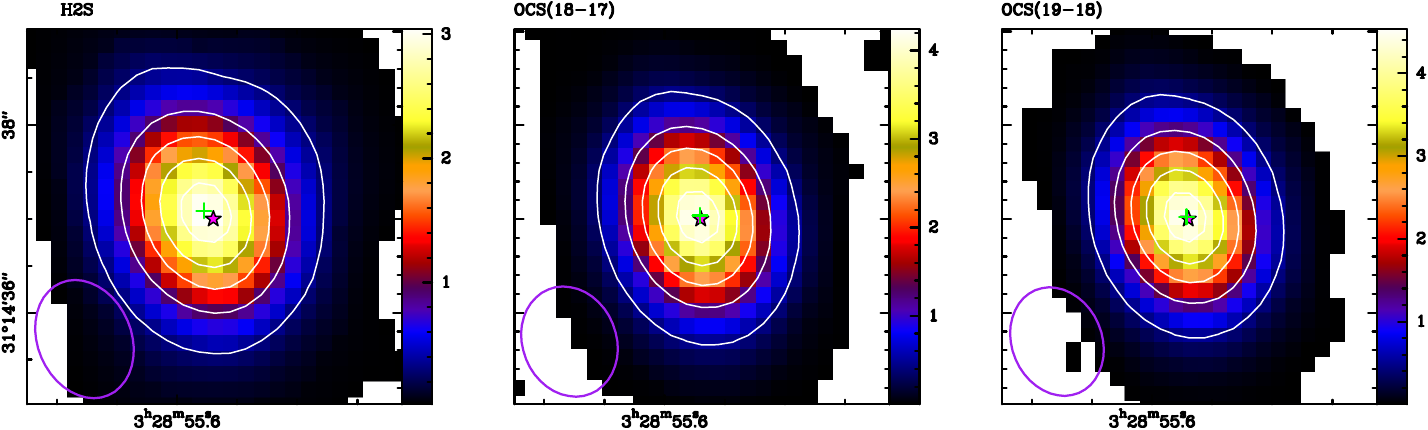}
        \caption{IRAS2A}
        \label{fig:map_IRAS2A}
    \end{subfigure}
    \hspace{0.9cm}
    \begin{subfigure}{0.47\textwidth}
        \centering
        \includegraphics[width=\linewidth]{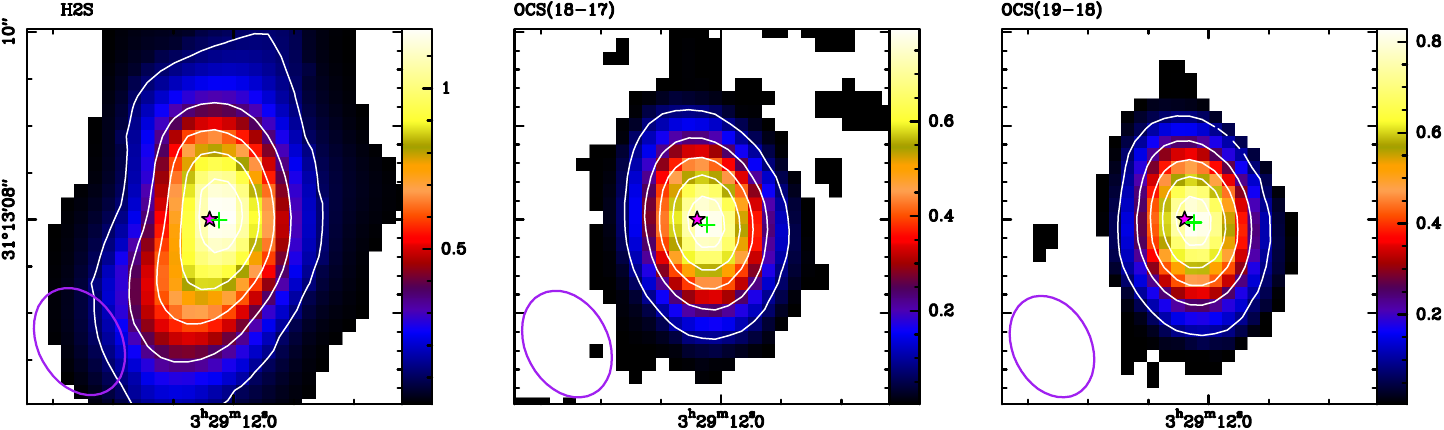}
        \caption{IRAS4B}
        \label{fig:map_IRAS4B}
    \end{subfigure}

    \caption{Emission of the main species (H$_2$S, and both OCS lines) in the inner core of protostars B1bS, HH211MMS, IRAS2A, IRAS4B. The colormap represents the >3$\sigma$ emission integrated images in a 4$\arcsec$$\times$4$\arcsec$ square region. The color scale, shown at the right of each map, is the brightness temperature in K. The white contours represent 10\%, 30\%, 50\%, 70\% and 90\% of the peak temperature. The pink star shows the position of the protostar, determined by the position of the maximum emission in the continuum \citep{Tobin2016}. The green cross marks the point with maximum emission of the line.}
    \label{fig:moment-0-maps}
\end{figure*}

\begin{figure*}
    \centering
    
    \begin{subfigure}{0.47\textwidth}
        \includegraphics[width=\linewidth]{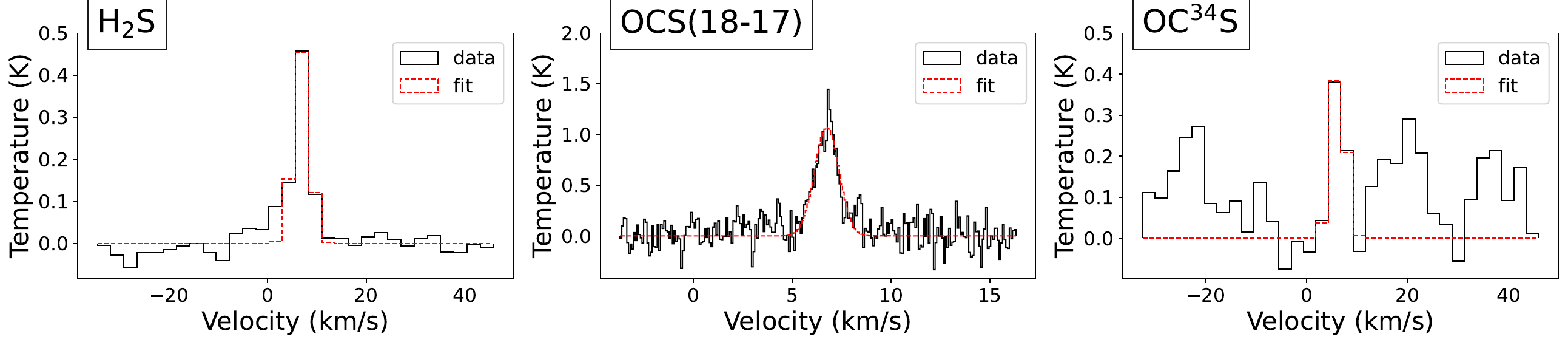}
        \caption{B1bS}
        \label{fig:spec_B1bS}
    \end{subfigure}
    \hspace{0.9cm}
    \begin{subfigure}{0.47\textwidth}
        \centering
        \includegraphics[width=\linewidth]{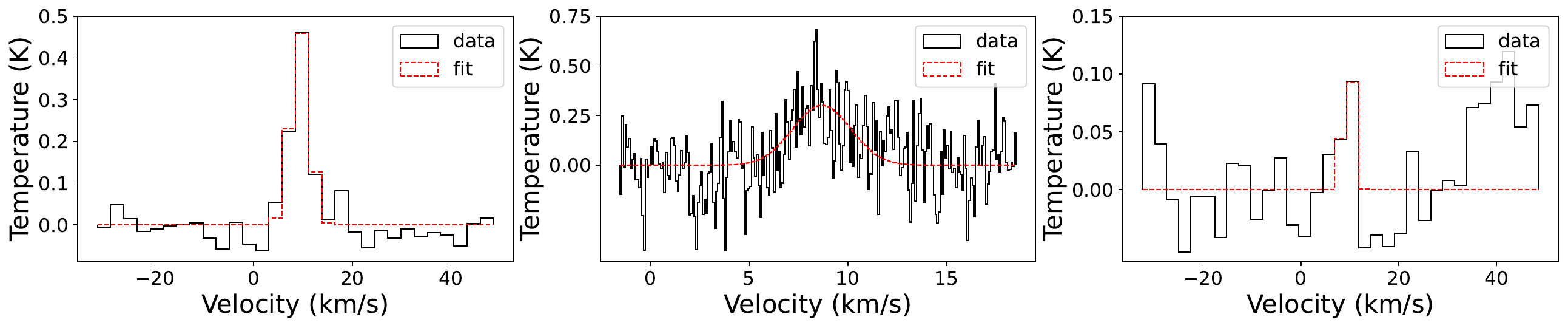}
        \caption{HH211MMS}
        \label{fig:spec_HH211}
    \end{subfigure}

    \begin{subfigure}{0.47\textwidth}
        \centering
        \includegraphics[width=\linewidth]{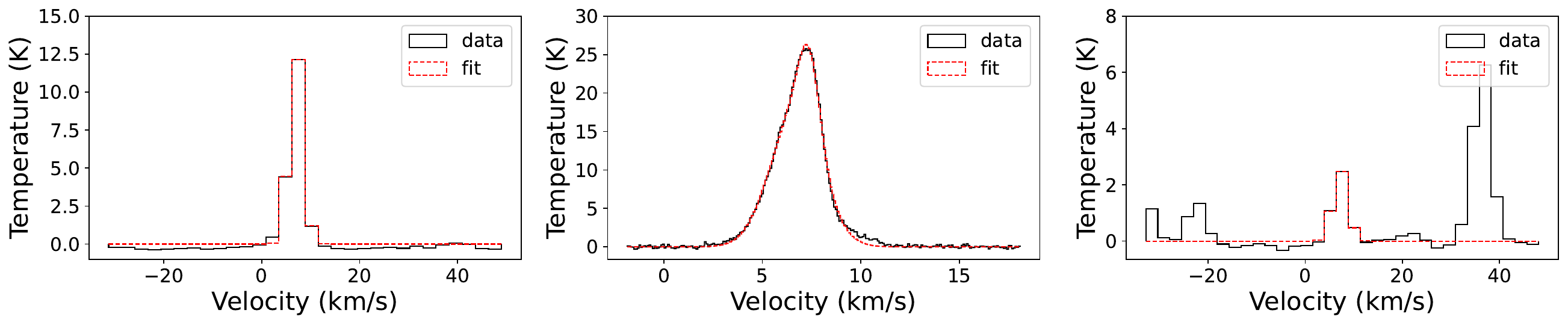}
        \caption{IRAS2A}
        \label{fig:spec_IRAS2A}
    \end{subfigure}
    \hspace{0.9cm}
    \begin{subfigure}{0.47\textwidth}
        \centering
        \includegraphics[width=\linewidth]{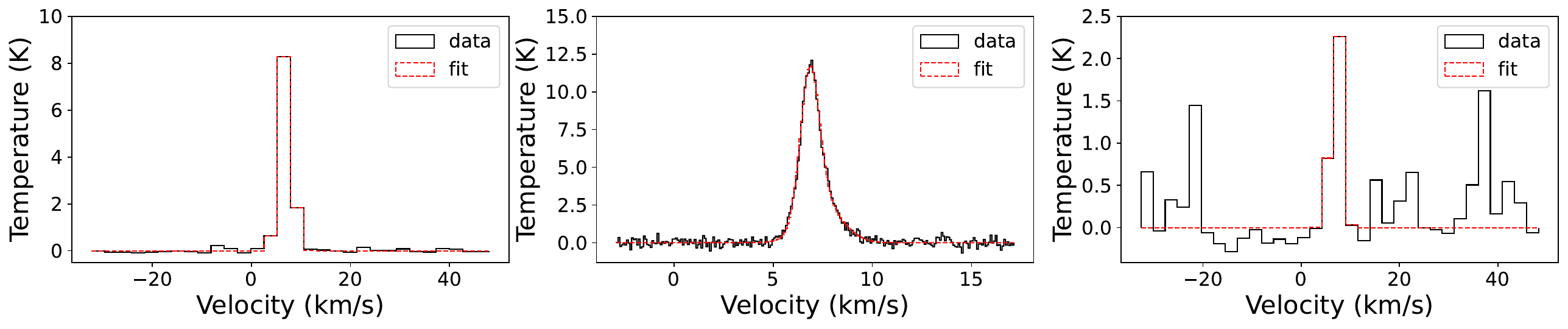}
        \caption{IRAS4B}
        \label{fig:spec_IRAS4B}
    \end{subfigure}

    \caption{Spectra of the H$_2$S 
    (2$_{2,0}$--2$_{1,1}$), OCS (18--17) and OC$^{34}$S (20--19) lines in protostars B1bS, HH211MMS, IRAS2A, IRAS4B. The red dotted lines represent the values of a Gaussian fit to each of the detected lines. In some cases, the combination of two Gaussian fits was necessary for the fit. The OCS (18-17) line was observed with a spectral resolution of 62.5 kHz while the H$_2$S and OC$^{34}$S lines were observed with a spectral resolution of 2 MHz. Gaussian fits have been plotted for the >3$\sigma$ detections, except for the OC$^{34}$S were we also show the <3$\sigma$ fits when the OC$^{33}$S counterpart was detected with >3$\sigma$.}
    \label{fig:specs}
\end{figure*}

Now, we will make a brief analysis of the sources in our sample. All 24 protostars are located in the Perseus Molecular Cloud (Fig.~\ref{fig:perseus-map}, and segregated in several smaller denser regions: Barnard 1, IC348, NGC1333, L1448 and Barnard 5.

\subsection{B1bN}

B1bN is the northern member of the multiple system embedded in B1b \citep{Hirano1999}, inside the Barnard 1 region (comprised by objects B1bN, B1bS and Per-emb-41). The low velocity of the outflow and the high degree of turbulence categorize the object as a first hydrostatic core (FHSC) \citep{Gerin2015}, even younger than its companion B1bS, as B1bN almost free-of-emission spectra suggests \citep{Marcelino2018}. There is a tentative detection of H$_2$S below 3$\sigma$, but the detection of the isotopologue H$_2^{33}$S (with $\sim$3$\sigma$) suggests a high opacity for the main species (see Fig.~\ref{fig:spec_B1bN}). OCS or its isotopologues are not detected in the protostar.

\subsection{B1bS}

B1bS is the southern member of B1b \citep{Hirano1999}, and is separated from B1bN by 17.4$\arcsec$ ($\sim$5218 au; \citealp{Tobin2018}). Interferometric NOEMA and ALMA observations identified B1bS as a very early Class 0 object due its young and slow outflow \citep{Hirano2014, Gerin2015} and small and compact structures, including an incipient disk \citep{Gerin2017}. Additionally, \citet{Fuente2017} provided evidence of pseudo-disk rotation by imaging this source in NH$_{2}$D emission using the NOEMA interferometer. Unlike its northern counterpart, B1bS is considered a hot corino, showing a high chemical richness in COMs \citep{Marcelino2018}. This richness is also present in our analysis. In Fig.~\ref{fig:map_B1bS}, we find the observed S-bearing molecules emission centered in the peak of the continuum. All species, both H$_2$S and OCS and all three isotopologues, are detected. Similarly to B1bN, the lines of the main species are optically thick (Table~\ref{tab:col_dens_and_ratios}). The OCS line shown in Fig.~\ref{fig:spec_B1bS} seems to present very faint wings, which can be evidence of the molecular outflow.

\subsection{B5-IRS1}

B5-IRS1 is a protostellar object in the Barnard 5 star-forming core. It is identified as a Class I protostar from its spectral energy distribution \citep{Enoch2009}. The protostar presents a wide-angle bipolar outflow \citep{Zapata2014}, and a streamer injecting new material from the filaments into B5-IRS1, characterized by \citet{Valdivia-Mena2023}. They also found processed and unprocessed material from the core infalling toward the filaments. Fig.~\ref{fig:map_B5-IRS1} shows an asymmetric distribution of the H$_2$S emission, perpendicular to the outflow direction, which could be explained by the material flow from the streamer (see Fig.~14 from \citealp{Valdivia-Mena2023}). H$_2$S spectrum presents a rather irregular profile (see Fig.\ref{fig:spec_B5-IRS1}). This, again, may be produced by the streamer or by the outflow present in B5-IRS1 ---or both phenomena combined. We did not find emission of OCS or from the H$_2^{33}$S isotopologue coming from the warm inner core.
 
\subsection{HH211MMS}

HH211MMS, located in the IC348 complex, is one of the most studied and well-characterized outflows, featuring a highly collimated jet and a molecular outflow \citep{Gueth1999, Lee2009, Ray2023, CarattioGaratti2024}. It shows a wide variety of atomic and ionic line emission, as well as molecular and PAH features \citep{Dionatos2010, Dionatos2018}. The central source has been classified as a low-mass, low-luminosity Class 0 protostar \citep{Froebrich2005}. In Fig.~\ref{fig:map_HH211} we see the warm inner core of HH211MMS, where we detect irregular distributions of H$_2$S and OCS. H$_2$S emission is extended in the outflow direction, while OCS spatial distributions are much more compact. We also find a tentative detection (<3$\sigma$) of the OC$^{34}$S isotopologue in the warm inner core (Fig.~\ref{fig:spec_HH211}). 

\subsection{IC348MMS}

IC348MMS is another Class 0 object in the IC348 region. It has a close companion, JVLA3a \citep{Rodriguez2014}, 2.95$\arcsec$ southwest, but we have not been able to observe it in any of the lines. \citet{Palau2014} suggests that JVL3a may be coincident with the origin of the outflow. Previous observations have not found any substructure in IC348MMS \citep{Tobin2015-2}. We detect emission of all main species and isotopologues. The spatial distribution of all the species is centered in the peak of emission, as shown in Fig.~\ref{fig:map_IC348MMS}. The detection of high column densities of OCS and H$_2$S isotopologues suggests a high opacity in the lines of the main species (Table~\ref{tab:col_dens_and_ratios}).

\subsection{IRAS2A}

IRAS2A (see, e.g., \citealp{Jennings1987}) is a Class 0/I protostellar system \citep{Brinch2009} in the NGC1333 region that hosts a protobinary as revealed by the high-resolution observations of the VANDAM survey \citep{Tobin2015}, separated by only 0.62$\arcsec$ ($\sim$186 au; \citealp{Tobin2018}). These two sources produce the two outflows, orthogonal to one another, known to originate from this system (see, e.g. \citealp{Jorgensen2004, Tobin2015}). Molecular emission in COMS has been observed toward this system \citep{Jorgensen2005, Maury2014}. IRAS2A also forms a wide binary with IRAS2B. The moment 0 maps in Fig.~\ref{fig:map_IRAS2A} show largely extended emission for all the lines, especially in the south-north outflow direction (see also Fig.\ref{fig:map_IRAS2A_BIG}), and centered at the emission peak. The spectra of the OCS lines, Fig.~\ref{fig:spec_IRAS2A}, clearly reveal the presence of a bipolar outflow. The H$_2$S line may present this same characteristic, but the lower spectral resolution complicates the interpretation. We also detect emission in the inner core from all three isotopologues.

\subsection{IRAS2B}

IRAS2B is a Class I protostar in the NGC1333 region, with the wide companion IRAS2A separated by 31.4$\arcsec$ ($\sim$9426 au; \citealp{Tobin2018}). IRAS2B also has a close companion, separated by 0.31$\arcsec$ ($\sim$93 au; \citealp{Tobin2018}), that we cannot resolve in our observations. In the observations, we can only detect the main species, H$_2$S and the two OCS lines. The H$_2$S moment 0 map in Fig.~\ref{fig:map_IRAS2B} shows a barely resolved emission along the beam's major axis direction and an asymetric spatial distribution along the beam's minor axis direction. The OCS emission in this source is faint, but above the 5$\sigma$ threshold (Fig.~\ref{fig:spec_IRAS2B}). The spatial distributions of these species are also very different which suggests that the OCS and the H$_2$S may trace different regions of the protostar (e.g. disk and core). 

\subsection{IRAS4B}

IRAS4B is a Class 0 young stellar object \citep{Sandell1991, Hirota2008, Enoch2009} characterized by a deeply embedded 0.24 $M_{\odot}$ disk \citep{Jorgensen2009} inside a 2.9 $M_{\odot}$ envelope \citep{Jorgensen2002}. It is located in the NGC1333 region and there are several protostars in its near vicinity: IRAS4A (not studied in this work), IRAS4B and IRAS4C, and their close companions (IRAS4A' and IRAS4B'). IRAS4B is located 29.74$\arcsec$ away from IRAS4A ($\sim$8922 au; \citealp{Tobin2018}) and $\sim$54$\arcsec$ away from IRAS4C. IRAS4B close companion, IRAS4B', is separated by 10.65$\arcsec$ (3196 au; \citealp{Tobin2018}, \citealp{Looney2000}) east of the main source. This embedded system exhibits hot corino chemistry, with emission of COMS as shown in, for instance, the Perseus ALMA Chemistry Survey (PEACHES) \citep{Yang2021}. Water emission has also been identified in its outflow \citep{Herczeg2012}. IRAS4B has very spread emission of the two main species. The moment 0 maps in Fig.~\ref{fig:map_IRAS4B} show extended emission of H$_2$S, OCS(18-17) and OCS(19-18) in the core but also in two lobes in the outflow direction (see also Fig.\ref{fig:map_IRAS4B_BIG}): both OCS maps show a main emission centered in the peak of the continuum and second emission clearly visible in the northern part of the map; in the case of the OCS lines, also a third emission region appears in the south. These additional outer detections represent material that belongs to the outflow. On the other hand, the isotopologues are only detected in the warm inner core. High column densities have been estimated for all isotopologues, but specially H$_2^{33}$S, revealing high opacity particularly in the H$_2$S (Table~\ref{tab:col_dens_and_ratios}). The irregular but slim shape of the spectra in Fig.\ref{fig:spec_IRAS4B} suggests that the velocity of the outflow is low. These narrow lines could also appear if the outflow is almost on the plane of the sky or if the emission is coming from the inner cavity walls.

\subsection{IRAS4C}

IRAS4C is a Class 0 protostar in the NGC1333 nebula \citep{Looney2000}. Although it is in the IRAS4 subregion, it is quite far from the nearest protostar, IRAS4B, which is separated by 54$\arcsec$. The observations reveal similar charateristics to those of IRAS2B (Fig.~\ref{fig:map_IRAS4C}), and the OCS emission is faint, but above the 5$\sigma$ threshold too (Fig.~\ref{fig:spec_IRAS4C}). None of the isotopologues are observed in IRAS4C.

\subsection{L1448-IRS3A}

L1448-IRS3A is a Class I protostar (see e.g. \citealp{Reynolds2021}) in the L1448 star-forming region. It forms a multiple system with other 5 sources: L1448-IRS3B's three components and L1448NW binary; it is located 7.32$\arcsec$ ($\sim$2195 au; \citealp{Tobin2018}) away from L1448-IRS3B. Recently, \citet{Gieser2024} found an elongated gas bridge with bright emission that connects the
IRS3A and IRS3B protostellar systems. They also revealed infalling material toward IRS3A from both the red- and blueshifted sides. In Fig.~\ref{fig:map_L1448-IRS3A} we present the moment 0 maps of L1448-IRS3A, where we observe the object in H$_2$S and OCS, with the OCS emission localized in a compact region around the peak of the continuum. The H$_2$S emission is also compact but more stretched, in the direction of the outflow, and displaced from the peak of the continuum. In Fig.~\ref{fig:spec_L1448-IRS3A}, we see a tentative detection of OC$^{34}$S but we do not observe either OC$^{33}$S nor H$_2^{33}$S. The 4 lines observed in this source present a double peak (tentatively for the OC$^{34}$S line), which could be the product of a bipolar outflow or could be tracing the infalling material revealed by \citet{Gieser2024}. The double peaked emission could also appear due to the disk/envelope rotation.

\subsection{L1448-IRS3B}

L1448-IRS3B is a Class 0 triple protostar in L1448 region \citep{Tobin2016}, with the separation of the two companions from the main component being $\sim$79 and $\sim$238 au \citep{Tobin2018}. The formation of the triple system has been associated with the fragmentation of a gravitationally unstable disk, supported by the detection of prominent spiral arms surrounding the three protostars \citep{Tobin2016-L1448IRS3B}. The protostellar system is embedded, together with L1448-IRS3A, in extended molecular gas that extends well beyond 6600 au \citep{Gieser2024}. The moment-0 maps of L1448-IRS3B, Fig.~\ref{fig:map_L1448-IRS3B}, show extended emission of H$_2$S in the outflow direction, but a compact spatial distribution of OCS. In Fig.~\ref{fig:spec_L1448-IRS3B}, we present the spectra of the six observed lines, with the detection of the two main species, H$_2$S and OCS, and also a tentative detection (<3$\sigma$) of the OC$^{34}$S isotopologue.

\subsection{L1448C}

L1448C is a Class 0 object in the L1448 region \citep{Tobin2007} and a well known example of an outflow-driving source with extremely-high velocity (EHV) jets. These jets have been detected in CO \citep{Bachiller1990} and in several transitions of SiO \citep{Bachiller1991, Dutrey1997}. High angular resolution observations \citep{Hirano2010} revealed that L1448C consists of two sources, L1448C(N) and L1448C(S), with L1448C(N) being responsible for the EHV jets. In Fig.~\ref{fig:map_L1448C} we show the 0-moment maps of the H$_{2}$S and the two OCS lines observed toward this source. The peak of emission of the three lines is centered in the peak of the continuum. In the spectra, shown in Fig.~\ref{fig:spec_L1448C}, wings are present in the OCS 18--17 line and linewidths are much larger than most objects in the sample. This suggests that the emission of these molecules may be perturbed by the extremely high-velocity jets coming from the source and the close binary.

\subsection{L1448NW}

L1448NW is a binary Class 0 protostar in the L1448 region \citep{Barsony1998}. Its two components are separated by 0.25$\arcsec$ ($\sim$58 au, \citealp{Tobin2016}) and the binary is 21.5$\arcsec$ ($\sim$6450 au) away from L1448-IRS3B \citep{Tobin2018}. The source presents a bipolar outflow, were the blueshifted emission is much less visible possibly due to an asymmetric distribution of the surrounding gas \citep{Lee2015}. In the observations of the source (Fig.~\ref{fig:map_L1448NW}, we have only detected H$_2$S emission in a streched and compact distribution perpendicular to the outflow direction. We also present the spectra of the lines in Fig.~\ref{fig:spec_L1448NW}. 

\subsection{Per-emb-02}

Per-emb-02 is a binary protostar in the Barnard 1 region and has its two components separated by only 0.08$\arcsec$ ($\sim$24.0 au; \citealp{Tobin2018}). It is a chemically poor Class 0 YSO \citep{Yang2021}, as no COMs have been detected besides CH$_{3}$OH. A streamer feeding this source with chemically fresh gas has been detected in carbon chain emission \citep{Taniguchi2024, Pineda2020}, and outflows in the northwest-southeast direction were detected \citet{Stephens2019}. This protostar is located in the Barnard 1 region. In Fig.~\ref{fig:map_Per-emb-2}, we present the moment-0 maps of Per-emb-02, where we find extended emission of H$_2$S. In Fig.~\ref{fig:spec_Per-emb-2}, we also show the detection of the two OCS lines in the warm inner core of the source. There are no signs of emission of any of the isotopologues.

\subsection{Per-emb-05}

Per-emb-05 is also a Class 0 binary protostar, with both components separated by 0.10$\arcsec$ ($\sim$29.1 au; \citealp{Tobin2018}). The evidence of a companion embedded within the extended dust continuum structure is interpreted as evidence of ongoing disk fragmentation by \citet{Tobin2016}, though the surrounding structure is not confirmed to be rotationally supported. We present the H$_2$S and OCS moment-0 maps in Fig.~\ref{fig:map_Per-emb-5}. OCS emission is localized in a compact region around the peak of the continuum while H$_2$S emission is slightly more extended. Part of this OCS emission is stretched perpendicular to the south-eastern outflow, and is specially visible for the J=19$-$18 transition. OC$^{34}$S is clearly detected in the spectrum, displayed in Fig.~\ref{fig:spec_Per-emb-5}. We think the emission corresponds to OC$^{34}$S because it is centered at the same velocity as the other lines; however, it is not clear if the larger width of the line is only produced by the isotopologue or also by other nearby transitions that may have contaminated the spectrum. OC$^{33}$S is tentatively detected within $\sim$3$\sigma$.

\subsection{Per-emb-08}

Per-emb-08 is a Class 0 protostar in the IC348 region that forms a multiple system with binary Class I object Per-emb-55, separated by 9.3$\arcsec$ ($\sim$2870 au; \citealp{Tobin2018}), which is not studied in this work. It is associated with a north-south bipolar outflow and is surrounded by a warped protostellar envelope, suggesting a highly turbulent environment or strong interactions between this source and its companions in Per-emb-55 \citep{Lin2024}. It is chemically poor, as COMs have not been detected in this source \citep{vanGelder2022}. The moment-0 maps, Fig.~\ref{fig:map_Per-emb-8}, reveal H$_2$S emission in the central region of the protostar. The spatial distribution is very asymmetric, with two peaks of emission, west and east of the source's position, and a third weaker peak north-east of the source. The peculiar distribution might be related with the important disk present in this source, which is very differentiated from the warm inner core \citep{Tobin2018}. This could also explain the potentially double-peaked emission in the H$_2$S spectrum, displayed in Fig.~\ref{fig:spec_Per-emb-8} ---note that it is very uncommon to find a double peaked emission for H$_2$S in this work due to the lower spectral resolution for this line---. A large velocity gradient in Per-emb-08 (>10 km/s) is expected.

\subsection{Per-emb-18}

Per-emb-18 is a Class 0 binary protostar in the NGC1333 region, with both components separated by only 0.085$\arcsec$ ($\sim$25.6 au; \citealp{Tobin2018}). It forms a wider multiple system with Per-emb-21, separated by 13.2$\arcsec$ ($\sim$3976 au), and Per-emb-49, which is 27.5$\arcsec$ ($\sim$8240 au) away from Per-emb-18 \citep{Tobin2018} ---the two additional sources are not studied in this work---. It was previously observed in CS, SO and SO$_2$ in \citet{Zhang2023}, were SO was found to trace asymmetric accretion shocks near the
edge of the circumbinary disk. Fig.~\ref{fig:map_Per-emb-18} displays the moment-0 maps of the source, where we can find similar emission of OCS and H$_2$S, with the peak of emission somewhat displaced to the south-west from the peak of the continuum. In the spectra, Fig.~\ref{fig:spec_Per-emb-18}, we find a double-peaked emission in both OCS lines, supported by the broader width of the H$_2$S and OC$^{34}$S spectra. This line profile could be produced by the close binary, by a strong outflow driven by the protostar, or enhanced by the accretion shocks and an alignment with magnetic fields, as suggested in \citet{Zhang2023}.

\subsection{Per-emb-22}

Per-emb-22 is a Class 0 binary protostar in the L1448 region, with both components separated by 0.75$\arcsec$ ($\sim$225 au; \citealp{Tobin2018}), which we cannot spatially resolve. It is a source with observed outflow activity in continuum emission whose shape resembles that of outflow cavities \citep{Yang2021}. Emission of H$_2$S is detected in the inner core of the source, but rather displaced to the west of the peak of the continuum, as shown in Fig.~\ref{fig:map_Per-emb-22}. OCS emissions show a smaller but similar distribution, with the addition of some more extended emission in the NE-SW direction, perpendicular to the outflow. The spectra in Fig.~\ref{fig:spec_Per-emb-22} might suggest the presence of an outflow, but the low level of OCS emission, and the low spectral resolution of the H$_2$S observations make it difficult to determine certainly. None of the isotopologues are observed in this object.

\subsection{Per-emb-29}

Per-emb-29 is a Class 0 object in the Barnard 1 region. It is a hot corino that has been observed to have rich emission lines of gas-phase COMs \citep{vanGelder2020}, as well as a high velocity outflow \citep{Jorgensen2006}. \citet{Chen2024} recently detected several COMs also in the form of ices. We present the moment-0 maps of Per-emb-29 in Fig.~\ref{fig:map_Per-emb-29}, were we find a regular and round distribution around the peak of the continuum for the three main lines. The spectra (Fig.~\ref{fig:spec_Per-emb-29}) show a self absorption in OCS, which is produced by a high opacity. The detection of the OC$^{34}$S isotopologue, which peaks inbetween the two peaks of the OCS line, supports this fact. In this source, we also detected emission of the H$_2^{33}$S isotopologue.

\subsection{Per-emb-30}

Per-emb-30 is a very faint Class 0/I protostar in the Barnard 1 region. We have not detected any of the transitions in this object.

\subsection{Per-emb-50}

Per-emb-50 is a Class I protostar in the NGC1333 region. It is a very bright and massive object, compared with other Class I protostars (around $\times$10, \citealp{Enoch2009},\citealp{Fiorellino2021}), and emission of some sulfurated species has been observed in the warm inner core (e.g. \citealp{Valdivia-Mena2022}). However, emission of H$_2$S and OCS has not been detected in our work (see Figs. \ref{fig:map_Per-emb-50}, \ref{fig:spec_Per-emb-50}). A streamer depositing material close to the edge of the gas disk from roughly 1500 to 3000 au from the protostar was also identified in \citet{Valdivia-Mena2022}.

\subsection{Per-emb-62}

Per-emb-62 is a Class I protostar in the IC348 region. We have not detected any of the species in this source either.

\subsection{SVS13A}

SVS13A is a Class 0/I protostar in the NGC1333 region. It belongs to the SVS13 multiple system, whose components are SVS13A, SVS13B and SVS13C ---this last one not being studied in this work---. SVS13A is a triple system where the two main components are separated by 0.3$\arcsec$ ($\sim90$ au, \citealp{Tobin2018}). The third component, SVS13A2, is located 5.3$\arcsec$ south-west of the main source, SVS13A1; however, we do not detect it in our observations. The three components are surrounded by a molecular envelope \citep{Lefloch1998} and drive a molecular outflow associated with the Herbig-Haro object HH711 \citep{Reipurth1993} as well as other younger flows \citep{Lefevre2017}. The system has been associated with a hot corino using deuterated water observations \citep{Codella2016}, and, more recently, many sulfurated species were observed in the warm inner core \citep{Codella2021}. Several COMs were also detected in this source, likely not only arising from the hot corino, but also from shocked gas at disk scales \citep{Hsieh2024}. A streamer, possibly infalling,
with a length of $\sim$700 au was identified by \citet{Hsieh2023} and can be observed with multiple tracers. SVS13A moment-0 maps, in Fig.~\ref{fig:map_SVS13A}, present extended emission in both H$_2$S and OCS, centered in the peak of the continuum. OCS(18-17) spectrum (Fig.~\ref{fig:spec_SVS13A}) shows a self-absorption feature which, supported by the detection of the OC$^{34}$S isotopologue that peaks inbetween the two peaks of the OCS line. H$_2^{33}$S is also detected in this source. Apart from this, OCS(18-17) slightly asymmetric spectrum could be a sign of the present outflows and flows in the system.

\subsection{SVS13B}

SVS13B is a Class 0 protostar in the NGC1333 region \citep{Grossman1987}\citep{Sadavoy2014}. It is located 14.9$\arcsec$ south-west of SVS13A ($\sim$4479 au, \citealp{Tobin2018}) and we can find both sources in the same field of view. The object presents a collimated outflow that could contribute to other flows around the HH711 \citep{Bachiller1998}. We present the moment-0 maps of the three main transitions for SVS13B in Fig.~\ref{fig:map_SVS13B}, and we can see a compact spatial distribution around the continuum peak for all of them. We only find emission of the two main species and the OC$^{34}$S isotopologue (Fig.~\ref{fig:spec_SVS13B}). Neither OC$^{33}$S nor H$_2^{33}$S were detected in the warm inner core.

\section{Column densities and abundances}
\label{sec:coldens-abund}

We calculated column densities for both OCS transitions, as well as for the H$_2$S transition. In the cases where H$_2$S was detected but OCS emission was not detected, we estimated an upper limit to the column density of OCS. We proceeded similarly with the 3 isotopologues. For the typical conditions of the warm inner cores (T>100K, $n$$\sim$10$^6$cm$^{-3}$) we assumed that all of the transitions are thermalised, and so, we did a local thermodynamic equilibrium (LTE) analysis to estimate the column densities.

Assuming that emission is optically thin and that the emission fills the beam, the column density of the upper level ($N_u$) of a given transition is (see e.g. \citealp{Goldsmith1999}):
\begin{equation}
    \label{eq:Nu}
    N_u = \frac{8\pi k\nu^2 W}{hc^3 A_{ul}},
\end{equation}
where $k$ is Boltzmann's constant, $\nu$ is the frequency of the molecular transition, $W\equiv\int T_a \textrm{d}$v is the integrated line intensity after the spectral fitting (with $T_a$ the brightness temperature and v the velocity), $h$ is Planck's constant, $c$ is the speed of light and $A_{ul}$ is Einstein's coefficient for espontaneous emission. Notice that, since $N_u$ depends on the integrated intensity, the spectral dilution in the lower spectral resolution lines has no effect in the result, as the area of a gaussian remains constant under dilution effects. Then, the total column density using the partition function, $Q(T_{\textrm{ex}})$, is:
\begin{equation}
    \label{eq:Ntot_Q}
    N = \frac{N_u\,Q(T_{\textrm{ex}})\,e^{\,E_u\,/\,kT_{\textrm{ex}}}}{g_u},
\end{equation}

Total column densities have been calculated assuming a $T_\textrm{ex}$ of 100 K, or $T_\textrm{ex}=T_\textrm{bol}$ if $T_\textrm{bol}>100 K$, in the warm inner core, for all species and sources. We calculated two different N(OCS) values, one from the J=18--17 transition and another from the J=19--18 one (which should actually be the same). We decided the total column density of OCS to be the mean of both values, given that both quantities are similar in all cases (less than a factor of 2). We estimated an additional value for the column densities of OCS and H$_2$S, when possible, using their isotopologues OC$^{34}$S and H$_2^{33}$S (see Section~\ref{sec:ratios}). The column densities of H$_2$S and H$_2^{33}$S have been calculated assuming the statistical value of the orto/para ratio of 3. Regarding uncertainties in the temperature, a factor of 2 in the temperature translates in a factor of $\sim$1.5 in OCS, OC$^{34}S$ and H$_2$S column densities, and slightly major changes in the H$_2^{33}$S column density. The effect of the temperature is smaller in the H$_2$S/OCS ratio, as long as the same temperature is assumed for both species. In fact, the maximum change of the H$_2$S/OCS ratio in all the sample is of a factor $\sim$1.5, usually within the uncertainties.

We then proceeded to the calculation of the abundances of each species, with respect to molecular hydrogen (H$_2$) gas-phase abundance. We calculated the column density of H$_2$ using the data from our observations of the continuum, following the process in \citet{Kauffmann2008}. We start from the expression of the intensity emitted by a source of temperature $T$, optical depth $\tau_\nu$ and frequency $\nu$, given by the equation of radiative transfer:
\begin{equation}
    \label{eq:radiative-transfer}
    I_\nu = B_\nu (T) ( 1 - e^{\tau_\nu} ) \,\,,
\end{equation}
where $B_\nu$ is Planck's blackbody function, and the optical depth can be expressed as:
\begin{equation}
    \tau_\nu = \int\kappa_\nu\,\rho\,\mathrm{d}s \,\,,
\end{equation}
with $\kappa_\nu$ being the absorption coefficient and $\rho$ the gas density.

If most of the hydrogen is in its molecular form, H$_2$, column density can be related with optical depth, using the relation:
\begin{equation}
    \label{eq:NH2_prev}
    N_{\textrm{H}_2} = \int n_{\textrm{H}_2}\,\mathrm{d}s = \int\frac{\rho}{\mu_{\textrm{H}_2}\,m_\textrm{H}}\,\mathrm{d}s = \frac{\tau_\nu}{\mu_{\textrm{H}_2}\,m_\textrm{H}\,\kappa_\nu} \,\,,
\end{equation}
where $\mu_{\textrm{H}_2}$=2.8 is the molecular weight per per hydrogen molecule, and $m_\textrm{H}$ is the H-atom mass.

On the other hand, $B_\nu$ cannot be simplified using the Rayleigh-Jeans limit as the necessary $\lambda$ regime is not achieved in this conditions; for $\sim$1.2 mm observations, in most of the cases \citep{Kauffmann2008}:
\begin{equation}
    \lambda\gg1.44\textrm{mm}\left(\frac{T}{10\textrm{K}}\right)^{-1} \,\,.
\end{equation}

Apart from this, we also look for an easy-to-use relation between flux per beam and intensity. In this case, we can start with:
\begin{equation}
    \label{eq:Fbeam}
    F_{\textrm{beam},\nu} = \int I_\nu\,P\,\mathrm{d}\Omega \,\,,
\end{equation}
where $P$ is the normalised power pattern of the telescope, and the solid angle of the antenna is defined as $\Omega_A=\int P\,\mathrm{d}\Omega$. If we assume the beam solid angle of the telescope is similiar to a Gaussian function, we have that:
\begin{equation}
    P = P(\theta) = e^{-\theta^2/2\theta_0^2} \,\,,
\end{equation}
with $\theta_0 = \theta_\textrm{HPBW}/\sqrt{8\ln{2}}$. With this condition, we can then integrate the antenna's solid angle:
\begin{equation}
    \Omega_A = \frac{\pi}{4 \ln{2}}\theta^2_\textrm{HPBW} \,\,.
\end{equation}

Eventually, we can rewrite Eq.~\ref{eq:NH2_prev} using the optically thin approximation in Eq.~\ref{eq:radiative-transfer} and using the average intensity from Eq.~\ref{eq:Fbeam} as $\left<I_\nu\right>=F_{\textrm{beam},\nu}/\Omega_A$ to obtain:
\begin{equation}
    N_{\textrm{H}_2} = \frac{\tau_\nu}{\mu_{\textrm{H}_2}\,m_\textrm{H}\,\kappa_\nu} = \frac{I_\nu}{\mu_{\textrm{H}_2}\,m_\textrm{H}\,\kappa_\nu\,B_\nu(T)} = \frac{F_{\textrm{beam},\nu}}{\mu_{\textrm{H}_2}\,m_\textrm{H}\,\kappa_\nu\,B_\nu(T)\,\Omega_A}
\end{equation}

Using useful units, we can recover Eq.~A.27 from \citet{Kauffmann2008}:
\begin{multline}
    \label{eq:NH2_def}
    N_{\textrm{H}_2} = 2.02\times10^{20}\textrm{cm}^{-2}\times\left(e^{1.439(\lambda/\textrm{mm})^{-1}(T/10\textrm{K})^{-1}}-1\right)\times\left(\frac{\lambda}{\textrm{mm}}\right)^3\times \\
    \times\left(\frac{\kappa_\nu}{0.01\textrm{cm}^2/\textrm{g}}\right)^{-1}\times\left(\frac{F_{\textrm{beam},\nu}}{\textrm{mJy/beam}}\right)\times\left(\frac{\theta_{\textrm{HPBW}}}{10\,\textrm{arcsec}}\right)^{-2}
\end{multline}
which we used to calculate the H$_2$ column density from the flux of our continuum observations. 

The magnitudes represented in Eq.~\ref{eq:NH2_def} are the following: $\lambda$ is the wavelength of the continuum observations, 1.29 mm; the temperature is the respective $T_{\textrm{dust}}$, which we have estimated to be the $T_{\textrm{bol}}$ of each source; $\kappa_\nu$ is the dust opacity, for which we used the value of 0.01056 cm$²$ g$^{-1}$, following \citet{ArturDeLaVillarmois2023} and \citet{OssenkopfHenning1994}, for a gas-to-dust ratio of 100; $\theta_{\textrm{HPBW}}$ is the width of the region were we integrate the flux, which in this case is a circular region of 1.5$\arcsec$ of diameter; $F_{\textrm{beam},v}$ is the integrated flux per beam of the dust continuum map inside the 1.5$\arcsec$ region around the peak, in mJy/beam.

\setlength{\tabcolsep}{12pt}
\begin{table*}[h]
    \centering
    \caption{Column densities and H$_2$S/OCS ratios in Perseus.}
    \label{tab:coldens-ratio}
    \begin{tabular}{lcccccc}
    \hline\hline
    Protostar  &  Class  &  $T_\textrm{bol}$  &  $T_\textrm{kin}$  &  N(OCS)  &  N(H$_2$S)  &  H$_2$S / OCS  \\
     & & (K) & (K) & (cm$^{-2}$) & (cm$^{-2}$) & \\
     \hline
    B1bN & 0 & 14.7 & 100.0 & 0.0 & (1.54$\pm$1.31) $\times10^{15}$ $^{(b)}$ & - \tabularnewline
    B1bS & 0 & 17.7 & 100.0 & (2.88$\pm$1.12) $\times10^{15}$ $^{(a)}$ & (2.65$\pm$0.08) $\times10^{16}$ $^{(b)}$ & 9.21$^{+4.33}_{-2.79}$ \tabularnewline
    B5-IRS1 & I & 287.0 & 287.0 & <7.15 $\times10^{13}$ & (5.04$\pm$1.26) $\times10^{14}$ & >7.04 \tabularnewline
    HH211MMS & 0 & 27.0 & 100.0 & (1.60$\pm$0.87) $\times10^{15}$ $^{(a)}$ & (2.10$\pm$0.17) $\times10^{14}$ & 0.132$^{+0.125}_{-0.053}$ \tabularnewline
    IC348MMS & 0 & 30.0 & 100.0 & (1.97$\pm$0.30) $\times10^{15}$ $^{(a)}$ & (2.94$\pm$0.01) $\times10^{16}$ $^{(b)}$ & 14.9$^{+1.9}_{-2.0}$ \tabularnewline
    IRAS2A & 0/I & 69.0 & 100.0 & (1.85$\pm$1.16) $\times10^{16}$ $^{(a)}$ & >5.69 $\times10^{16}$ $^{(b)(c)}$ & >3.08 \tabularnewline
    IRAS2B & I & 106.0 & 106.0 & (1.06$\pm$0.21) $\times10^{14}$ & (1.03$\pm$0.15) $\times10^{14}$ & 0.973$^{+0.333}_{-0.283}$ \tabularnewline
    IRAS4B & 0 & 28.0 & 100.0 & (1.44$\pm$0.39) $\times10^{16}$ $^{(a)}$& (2.06$\pm$0.01) $\times10^{17}$ $^{(b)}$ & 14.3$^{+3.6}_{-3.1
    }$ \tabularnewline
    IRAS4C & 0 & 31.0 & 100.0 & (3.89$\pm$1.38) $\times10^{13}$ & (1.15$\pm$0.13) $\times10^{14}$ & 2.97$^{+1.53}_{-1.02}$ \tabularnewline
    L1448-IRS3A & I & 47.0 & 100.0 & (2.17$\pm$1.16) $\times10^{15}$ $^{(a)}$ & (2.18$\pm$0.46) $\times10^{14}$ & 0.101$^{+0.115}_{-0.049}$ \tabularnewline
    L1448-IRS3B & 0 & 57.0 & 100.0 & (1.61$\pm$0.76) $\times10^{15}$ $^{(a)}$ & (2.71$\pm$0.12) $\times10^{14}$ & 0.168$^{+0.112}_{-0.059}$ \tabularnewline
    L1448C & 0 & 47.0 & 100.0 & (3.82$\pm$3.69) $\times10^{15}$ $^{(a)}$ & (1.31$\pm$0.03) $\times10^{15}$ & 0.34$^{+6.44}_{-0.17}$ \tabularnewline
    L1448NW & 0 & 22.0 & 100.0 & <3.43 $\times10^{13}$ & (9.61$\pm$2.63) $\times10^{13}$ & >2.81 \tabularnewline
    Per-emb-2 & 0 & 27.0 & 100.0 & (2.79$\pm$1.11) $\times10^{13}$ & (3.00$\pm$0.15) $\times10^{14}$ & 10.8$^{+5.5}_{-3.4}$ \tabularnewline
    Per-emb-5 & 0 & 32.0 & 100.0 & (1.91$\pm$0.58) $\times10^{15}$ $^{(a)}$ & (1.48$\pm$0.11) $\times10^{14}$ & 0.078$^{+0.030}_{-0.022}$ \tabularnewline
    Per-emb-8 & 0 & 43.0 & 100.0 & <2.92 $\times10^{13}$ & (1.16$\pm$0.30) $\times10^{14}$ & >3.98 \tabularnewline
    Per-emb-18 & 0 & 59.0 & 100.0 & (3.65$\pm$0.55) $\times10^{15}$ $^{(a)}$ & (2.77$\pm$0.13) $\times10^{14}$ & 0.076$^{+0.013}_{-0.013}$ \tabularnewline
    Per-emb-22 & 0 & 43.0 & 100.0 & (2.37$\pm$0.15) $\times10^{14}$ & (3.79$\pm$0.13) $\times10^{14}$ & 1.60$^{+0.13}_{-0.15}$ \tabularnewline
    Per-emb-29 & 0 & 48.0 & 100.0 & (4.51$\pm$2.03) $\times10^{15}$ $^{(a)}$ & (5.36$\pm$0.09) $\times10^{16}$ $^{(b)}$ & 11.9$^{+6.8}_{-3.8}$ \tabularnewline
    Per-emb-30 & 0/I & 78.0 & 100.0 & 0.0 & 0.0 & - \tabularnewline
    Per-emb-50 & I & 128.0 & 128.0 & 0.0 & 0.0 & - \tabularnewline
    Per-emb-62 & I & 378.0 & 378.0 & 0.0 & 0.0 & - \tabularnewline
    SVS13A & 0/I & 188.0 & 188.0 & (4.33$\pm$2.70) $\times10^{16}$ $^{(a)}$ & >1.48 $\times10^{17}$ $^{(b)(c)}$ & >3.42 \tabularnewline
    SVS13B & 0 & 20.0 & 100.0 & (1.39$\pm$0.64) $\times10^{15}$ $^{(a)}$ & (2.68$\pm$0.19) $\times10^{14}$ & 0.19$^{+0.13}_{-0.07}$ \tabularnewline
    \hline
    \end{tabular}
    \begin{flushleft}
    \tiny
        \textbf{Notes:} $^{(a)}$ Estimated using the OC$^{34}$S isotopologue column density. $^{(b)}$ Estimated using the H$_2^{33}$S isotopologue column density. $^{(c)}$ The broad CH$_3$CHO line (see Appendix~\ref{sec:apendix_h233s}) complicates the fit of the H$_2^{33}$S lines, but a lower limit could be estimated.
    \end{flushleft}
\end{table*}

\section{H$_2$S/OCS ratio}
\label{sec:ratios}

\begin{figure*}
    \centering
    \includegraphics[width=\textwidth]{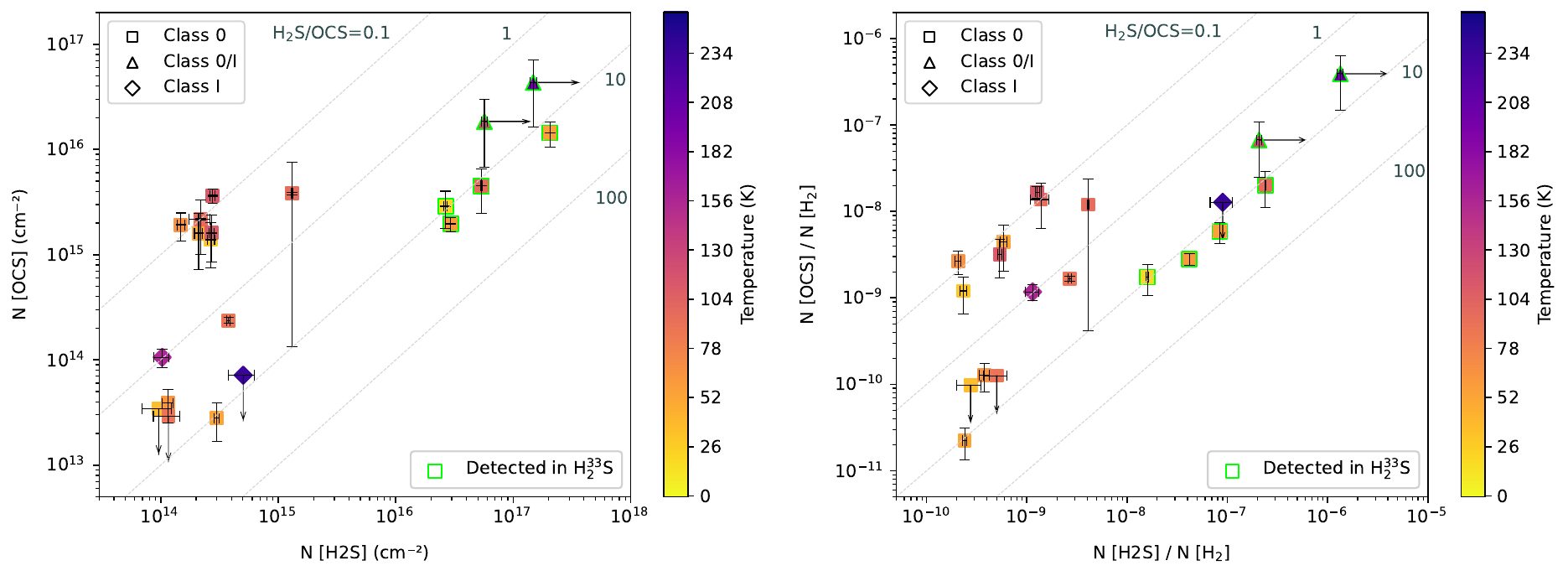}\caption{OCS vs. H$_2$S comparison in the sources where H$_2$S was detected. The color of the markers represents bolometric temperature (side colorbar). \textit{Left:} Column densities. OCS column density increases with increasing H$_2$S column density. We see two different trends in this graph: 1$-$Several sources have H$_2$S/OCS ratios around $\sim$7-10, which we have classified as OCS-poor sources. 2$-$The rest of the sources present H$_2$S/OCS ratios in the $\sim$0.1-1 range, and most of them are arranged in a cluster of several points in the \{N(OCS)$\sim$2$\times$10$^{15}$cm$^{-2}$, N(H$_2$S)$\sim$2$\times$10$^{14}$cm$^{-2}$\} range   . They are classified as OCS-rich sources.
    \textit{Right:} Abundances. We find similarities with the first graph: OCS abundances grow together with H$_2$S abundances, the lower H$_2$S/OCS ratio sources follow a linear (log) trend and the rest are very differentiated from the fitted line. In this case, the fit corresponds to H$_2$S/OCS ratio of 0.147$\pm$0.022.} 
    \label{fig:h2s-ocs-temp}
\end{figure*}

We have calculated the averaged column densities of H$_2$S,  H$_2^{33}$S, OCS, OC$^{34}$S, and  OC$^{33}$S in a 1.5$"$ diameter circle around the continuum peak using
the averaged interferometric spectra and following the procedure described in Sect.~\ref{sec:coldens-abund}. One important problem is the impact that the possible high opacities of the main isotopologue lines can have in our column density estimates, leading to a severe underestimation of the real ones. 
To avoid this problem we use the observations of the less abundant isotopologues to estimate the column density of the most abundant one. We have detected the OC$^{34}$S line in 13 of the total 17 targets detected in OCS. Significant deviations of the $^{32}$S/ $^{34}$S ratios from the solar value are not detected in the interstellar medium and/or comets \citep{Gratier2016, Calmonte2017}. Therefore, we use OC$^{34}$S as a proxy for OCS by calculating N(OCS)=22.5 $\times$N(OC$^{34}$S). These are the values shown in Table~\ref{tab:coldens-ratio} and used for the following discussion. We have also detected the OC$^{33}$S line in 9 protostars.  The values of  N(OC$^{34}$S)/N(OC$^{33}$S) in these targets are $>$ 2.5, proving that the OC$^{34}$S line is not optically thick. In the case of H$_2$S, we have detected the  H$_2^{33}$S 2$_{2,0}$--2$_{1,1}$ in 7 sources. In these sources, we used the  H$_2^{33}$S column density to compute that of  H$_2$S assuming $^{32}$S/ $^{33}$S = 126 \citep{AndersGravesse1989}. The column densities thus calculated are significantly higher than those estimated from the main isotopologue line. The most extreme case is B1bS, in which the column density estimated from  H$_2^{33}$S  is $\sim$144 times higher than that obtained from the main isotopologue observations. However, the difference is lower, $\sim$15 to $\sim$85, for the other protostars. We think that the extreme case of B1bS is caused by a very high opacity in the source, due to an extense and thick envelope surrounding the very young protostar, similarly to B1bN. This is supported by the very low $^{34}$S/$^{33}$S ratio (see Table~\ref{tab:col_dens_and_ratios}) which suggests that even the OC$^{34}$Sisotopologue is optically thick in this source.

In Fig.~\ref{fig:h2s-ocs-temp}, we plot N(OCS) vs N(H$_2$S) for the targets where H$_2$S has been detected. There is a large scatter between the two magnitudes, specially at low values of  N(H$_2$S), reflecting significant changes in the N(H$_2$S)/N(OCS) ratio (hereafter, H$_2$S/OCS). The values of H$_2$S/OCS range from $\sim$0.1 to $\sim$10, thus  expanding over 2 orders of magnitude (see Table~\ref{tab:coldens-ratio} and Fig.~\ref{fig:ratios}). In fact, we can differentiate two groups. The first one, characterized by H$_2$S/OCS $\sim$7 - 10, is formed by the 7 sources detected in H$_2^{33}$S and one additional source with low H$_2$S column density and we will refer to them as $``$OCS-poor$"$ targets. 
The second one is composed of 8 sources with N(H$_2$S) < 2$\times$10$^{15}$ cm$^{-2}$ and  presents H$_2$S/OCS $\sim$0.1 - 1. We will refer to these targets as $``$OCS-rich$"$ protostars. In 4 sources, we have not detected OCS and only upper limits to the  H$_2$S/OCS ratio can be derived. These limits imply H$_2$S/OCS $>$5 and consequently, they can be considered as  $``$OCS-poor$"$ protostars. One could think that this difference can be related to a different evolutionary stage but, as shown in 
Fig.~\ref{fig:h2s-ocs-temp}, there is no correlations between H$_2$S/OCS and the bolometric temperature. We consider that this differentiated chemistry is more likely related with a different composition of the sublimated ice. In Sect.~\ref{sec:model}, we discuss the origin of this chemical  diversity.

In the following, we discuss some observational uncertainties that can influence the derived  values of  H$_2$S/OCS and therefore, our classification in 
$``$OCS-poor$"$ and $``$OCS-rich$"$ protostars. One could think that the limited angular resolution of our observations can bias the estimated values. The unknown size of the emitting area would definitely have a significant impact on the estimated column densities but we do not expect a big impact on the calculated H$_2$S/OCS since the transitions observed present similar excitations conditions (E$_u$ $\sim$ 80 - 100 K) and more likely come from the same region.  A second problem would be the possible impact of the opacity in our column density estimates. As explained above, we have avoided this problem through the observation of the less abundant isotopologues. In particular, we have observed transitions of OCS, OC$^{34}$S, and  OC$^{33}$S, which allow us to derive accurate OCS column density estimates. In the case of H$_2$S, there is a significant fraction of sources in which we have not detected the  H$_2^{33}$S 2$_{2,0}$--2$_{1,1}$ line and our estimates rely on the observations of the main isotopologue. Looking at the ratios between the H$_2$S column densities derived from the $^{33}$S isotopologue over those using $^{32}$S (see Table~\ref{tab:col_dens_and_ratios}), we expect that the opacity of the main isotopologue line towards the targets without H$_2^{33}$S detection is less than $\sim$15. This would mean that our estimates are reliable within an order of magnitude in these sources. This is also corroborated by the lower column densities and opacities of the OCS lines, which suggest that these protostars host less massive warm regions. Even considering the worst case scenario where all OCS-rich protostars with OC$^{34}$S detection would scale up their H$_2$S by a factor 37 ---which is the geometrical mean of the scalling factors for the sources detected in H$_2^{33}$S (without B1bS; notice opacity scales exponentially, so the appropiate mean is the geometrical and not the arithmetic one)---, we would still find the two-family differentiation between OCS-rich and OCS-poor sources. In this scenario, which we show in Figure~\ref{fig:ratio_comparison}, L1448C would be the only case where we have classified a source as OCS-rich that would prove to be an OCS-poor one, but the high uncertainty in the H$_2$S/OCS ratio makes this comparison a rather bad example.

\section{Completing the Sulfur budget}
\label{sec:sulfur_budget}

\begin{figure*}
    \centering
    \includegraphics[width=\textwidth]{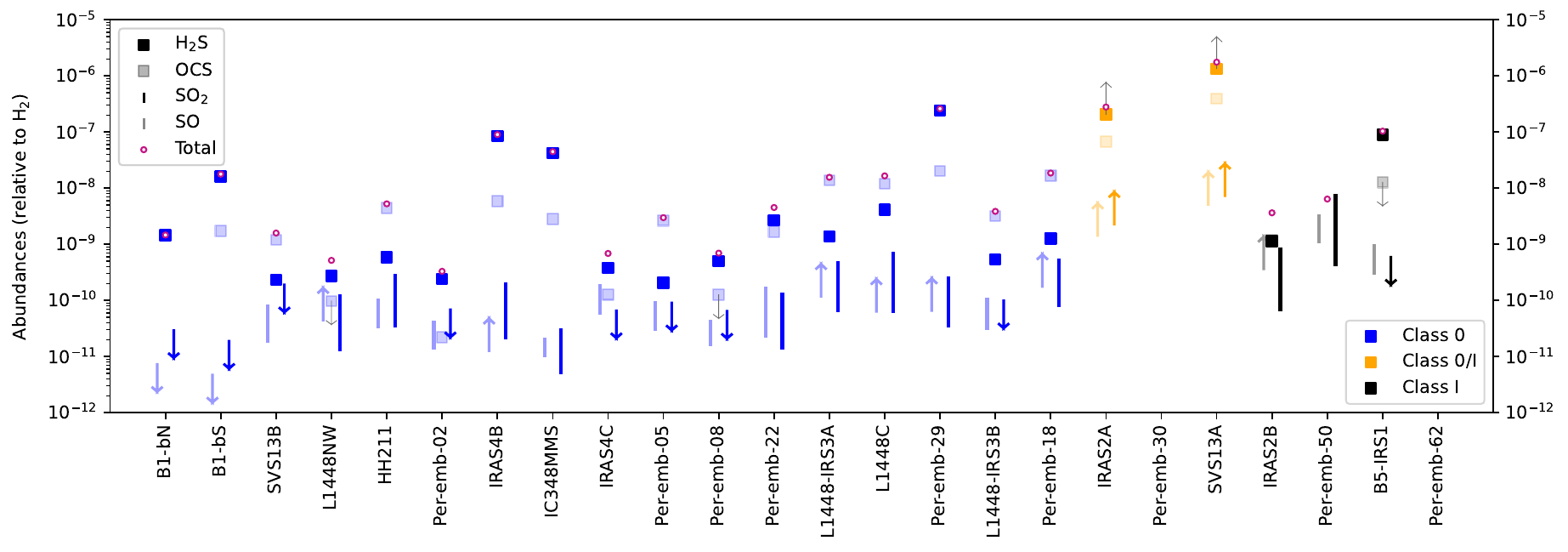}\caption{Abundances of the main sulfuretted species (H$_2$S, OCS, SO, SO$_2$) in each of the protostars in our sample. The purple marks represent the total sulfur abundance produced by this four sulfuretted molecules. Sources are sorted first by their Class and then by $T_{\textrm{bol}}$. Total abundances seem to present a slight growing trend with $T_{\textrm{bol}}$ and Class. We find that, with the data available, SO and SO$_2$ are tipically 1-2 orders of magnitude below H$_2$S and OCS, but optically thick lines (specially the H$_2$S lines not detected in H$_2^{33}$S) may play a role in refuting or supporting this findings. Notice that H$_2$S and OCS abundances are very similar in IRAS2B and hardly distinguishable in the figure.} 
    \label{fig:abundances}
\end{figure*}

To determine sulfur depletion is an important open question in astrochemistry.  In the dense and cold regions of molecular clouds, most of the sulfur is thought to be  locked in the icy mantles that cover the dust grain surfaces. Chemical models \citep{NavarroAlmaida2020} and comet observations \citep{Calmonte2016}  suggest that H$_2$S is the most abundant sulfur compound in the ice. However, its detection in interstellar ices remains elusive even with the high sensitivity provided by the JWST. One factor contributing to this is its relatively low binding energy as calculated by \citet{Bariosco2024}. Nowadays, OCS is the only compound firmly detected in interstellar ices \citep{Palumbo1995,  Boogert2022}.  Different authors have published tentative detections of  the 7.5 $\mu$m band of SO$_2$  \citep{Rocha2024}, but this band is overlapped with intense bands of abundant complex organic molecules (COMs) and CH$_4$, which hinder its confirmation (Taillard et al., in prep).  
Ices are sublimated in the warm innermost regions of protostars where the dust temperature is $>$ 100 K. The chemical composition of this warm region can therefore inform on the 
ice composition as long as the gas-phase chemistry has not significantly altered it. Even if it were the case, these warm regions, where all the volatiles are in gas phase, provide an excellent opportunity to determine the total sulfur budget in volatiles, i.e., sulfur depletion.  
In Sect.~\ref{sec:ratios}, we have calculated the abundances of H$_2$S and OCS, that are expected to be the most abundant sulfur-bearing species in ices. Here, we use the SO and SO$_2$ column densities  reported  from the data of the PEACHES survey \citep{Yang2021} (ALMA project codes: 2016.1.01501.S and 2017.1.01462; PI: N. Sakai) by \citet{ArturDeLaVillarmois2023}, obtained to complete the sulfur budget in most of our targets. 
Unfortunately, observations of Per-emb-30 and Per-emb-62 in SO and SO$_2$ are not included in the sample studied by \citet{ArturDeLaVillarmois2023}.

The synthesized beam of PEACHES observations, $\theta_{\textrm{HPBW}}$$\sim$0.5$\arcsec$, is three times smaller than the region size ($\theta_{\textrm{HPBW}}$=1.5$\arcsec$) used for our calculations, and it is also beyond the maximum resolution of our continuum images ($\theta_{\textrm{HPBW}}\sim$1.3$\arcsec$). Therefore, the comparison of the column densities derived from PEACHES observations with those listed in Table~\ref{tab:coldens-ratio} is not straightforward, since it depends on the unresolved chemical and physical structure towards each target. 
To account for that, we have considered 2 limiting cases to estimate the SO and SO$_2$ abundances in our sample. In the first case, we assume that the emission of SO and SO$_2$ is extended and the continuum emission uniformly fills the 1.5$\arcsec$ integration region. In this case, we calculate the SO and SO$_2$ abundances by adopting the column densities reported by
\citet{ArturDeLaVillarmois2023} and the values of $N$(H$_2$) calculated as described in Sect.~\ref{sec:coldens-abund}. In the second case, we assume that the continuum emission is unresolved, and that all of it comes from the innermost $\sim$0.5$\arcsec$ region, similar to the the beam used to calculate the SO and SO$_2$ column densities; then, we calculate the H$_2$ column density using the peak flux and using equation~\ref{eq:NH2_def} with  $\theta$=0.5$\arcsec$. By considering these two scenarios, we define an upper and lower limit to the SO and SO$_2$ abundances. Every other possibility should fall within this range of values. This uncertainty in the spatial distribution of the different emission lines and dust emission would introduce a large uncertainty, up to a factor of $\sim$9, in the derived abundances. However, as commented below, it does not have a big impact on the total sulfur budget.

In Fig.~\ref{fig:abundances}, we present the abundances of the main sulfuretted species ---H$_2$S, OCS, SO, SO$_2$--- in the 24 protostars of our sample, as well as the total sulfur abundance calculated as the sum of the abundances of these species. There are large differences between sources in the abundances of all S-bearing molecules: SO and SO$_2$ abundances spread over almost three orders of magnitude, while OCS and H$_2$S cover ranges of more than four orders of magnitude. H$_2$S appears to be the most abundant species in most of the sources, with a maximum >1.3$\times$10$^{-6}$ in SVS13A, which is in accordance with the value measured in Orion KL by \citet{Crockett2014} of (3.1$\pm^{1.1}_{1.9}\,)\times$10$^{-6}$. The second most abundant sulfuretted molecule is OCS, which in some cases is greater than the H$_2$S abundance (OCS-rich sources). However, these OCS-rich sources have less H$_2$S in the gas phase than the OCS-poor ones, and the sum of H$_2$S and OCS abundances is less than $\sim$10$^{-7}$ (see Fig.~\ref{fig:abundances}). When it comes to SO and SO$_2$, we have given a range of values for their abundances, taking into account the range of column densities calculated as explained above. In all the sources, the abundances of these two species are far less abundant than H$_2$S and OCS, $<$10$^{-8}$. 
If the gas phase abundances we are measuring reflects the ice composition in the protostellar phase, the main sulfur reservoir in the Perseus Cloud would be H$_2$S ice in $\sim$ 60\%, of the sources and OCS in the rest. In addition, we find that the Class I sources gas-phase sulfur total abundance declines. The SO and SO$_2$ remain with higher abundances in the Class I, starting to gain more weight in the total sulfur abundance.

According with the data shown in Fig.~\ref{fig:abundances}, the amount of sulfur in gas phase increases during the Class 0 stage with the maximum in the Class 0/I sources (IRAS2A, SVS13A). One could think that this is the consequence of missing some important sulfur species, mainly CS, CCS, C$_3$S, and H$_2$CS.
A more likely explanation is that the warm region, i.e., the region in which ices mantles are evaporated, is smaller in these young objects. Higher angular resolution observations and the observation of a wide variety of sulfur species are needed to probe these hot environments. Also interesting, the amount of sulfur seems to decrease again in the Class I stage. It is true, however, that our sample of Class I objects is too small to draw firm conclusions.

\section{Chemical modeling}\label{sec:model}

    In this section we describe a chemical model for evolved Class 0 and Class I warm inner cores aimed at understanding the chemical variety found in our sample and, in particular, the factors that could decide whether a source is H$_{2}$S or OCS-rich. Class 0 and Class I stages are the bridge between the pre-stellar phase and protoplanetary disks. While the degree of chemical processing along these stages is still unclear, chemical surveys (see, e.g., \citealp{Bianchi2019, Mercimek2022}), cometary abundances \citep{LeRoy2015, Drozdovskaya2018, Altwegg2022}, and chemo-dynamical models \citep{NavarroAlmaida2024} suggest that there is a non-negligible fraction of the chemical composition that survives throughout this evolutionary process. Following this nature vs. nurture idea, in this section we investigate whether the chemical diversity of our sample is a product of properties such as the observed bolometric temperature, whether it is inherited from the pre-stellar core from which they were born, or both.
    
    Our chemical model for a Class 0 warm core consists of two steps. First, we set a pre-stellar phase model based on the Barnard-1b model presented in \citet{NavarroAlmaida2020} by taking its physical structure. With it, we then ran the 3-phase gas-grain chemical model \texttt{NAUTILUS} \citep{Ruaud2016} in its latest version \citep{Wakelam2024} for $10^6$ yr. Note that, in this chemical model, the gas-grain exchange processes considered are: thermal desorption, UV photodesorption, cosmic-ray induced photodesorption, cosmic-ray induced heating, chemical (reactive) desorption, cosmic-ray sputtering and diffusive chemistry. We then extracted the chemical abundances at the innermost point in the model as the representative chemical composition of the pre-stellar core and as the initial chemical abundances for the protostellar collapse. In a second step, the gravitational collapse of the pre-stellar core into the young Class 0 object was described by the density and temperature evolution of Lagrangian tracer particles present in the MHD simulation reported in \citet{Hennebelle2016} and \citet{Gerin2017}. This simulation has also been used to investigate the impact of grain-growth on the chemical composition of the hot corino in \citet{NavarroAlmaida2024} and describes the collapse of 1 $M_\odot$ of gas reaching up to the first hydrostatic core (FHSC) phase $(t\sim 10^4\ {\rm yr})$.

    \subsection{The role of collapse in the H$_{2}$S/OCS ratio}\label{sect:mhdH2SOCS}

        After obtaining the pre-stellar chemical composition of both gas and ice phases, we ran a core collapse model taking the density and temperature evolution of a tracer particle that belong to the warm core (T$_{\rm gas}$$\sim$ 500K) at the end of the MHD simulation. As discussed in \citet{NavarroAlmaida2024}, the resulting chemical abundances do not change significantly among tracer particles that belong to the warm core by the end of the simulation and so we can use a random representative one to investigate the effect of the dynamical history on the H$_2$S/OCS ratio.

        \begin{figure*}
			\centering
			\includegraphics[width=\textwidth]{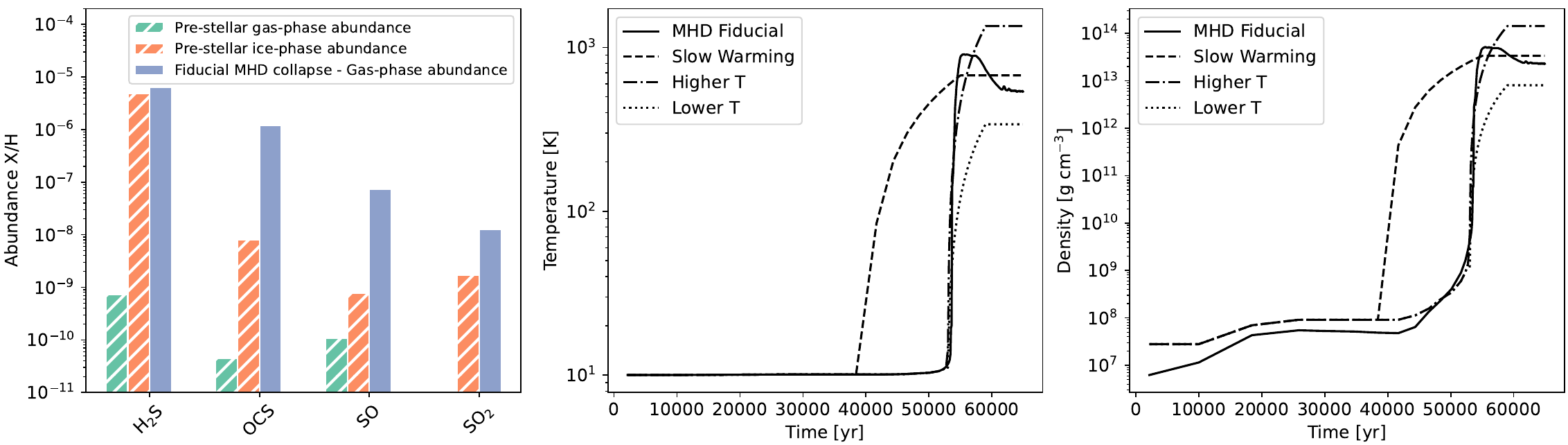}
			\caption{Evolution of the models with different warming curves. \textit{Left:} Initial ice and gas-phase components of the H$_2$S, OCS, SO and SO$_2$ species in the inner core before the collapse. Comparison with the final abundance of these species in the gas-phase after the Fiducial MHD collapse. \textit{Center:} Comparison of the Fiducial MHD warming curve with the rest of warming curves explored in the section. \textit{Right:} Comparison of the Fiducial MHD density curve with the rest of warming curves explored in the section. The different density curves are obtained using the barotropic equation of state from \citet{Machida2006}.
            }
			\label{fig:histoCompPreMHD}
		\end{figure*}

        \begin{figure*}
			\centering
			\includegraphics[width=\textwidth]{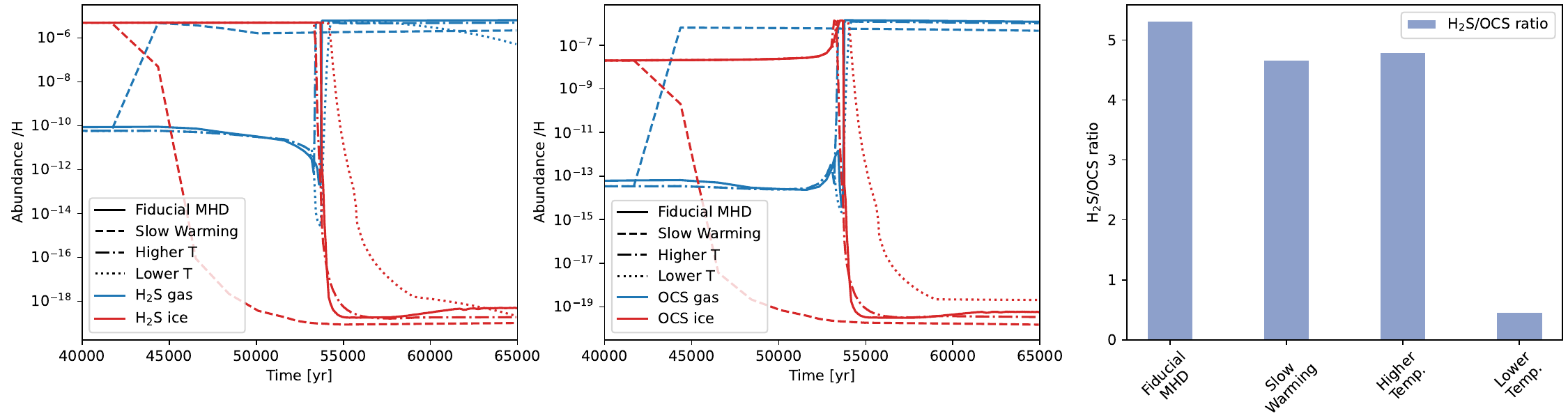}
			\caption{Comparison of the H$_2$S and OCS chemical evolution during the final stages of the collapse, with different warming curves. \textit{Left:} H$_2$S ice and gas-phase abundances during the collapse, with the different warming curves. The starting time of the figure has been set to 4$\times$10$^4$ years to show a closer detail of the major changes. The products after 5.5$\times$10$^4$ are very similar in all the models but, in the later stages, the model with a lower final temperature,``Lower T'', shows a drop of the gas-phase H$_2$S abundance. \textit{Center:} OCS ice and gas-phase abundances during the collapse, with the different warming curves. The starting time of the figure has been set to 4$\times$10$^4$ as in the left panel. The products after 5.5$\times$10$^4$ are very similar in all the models. \textit{Right:} H$_2$S/OCS ratio after the collapse with the different warming curves. All models show a ratio close to 5, except for the model with a lower final temperature.
            }
			\label{fig:warmingAlternatives}
		\end{figure*}
        
        The gas and ice-phase abundances of H$_{2}$S, OCS, SO, and SO$_{2}$ in the pre-stellar phase (first step) are shown in the left panel of Fig. \ref{fig:histoCompPreMHD}. As expected from cold core chemistry, most of molecular abundances are found in ices, with H$_{2}$S abundance being more than two orders of magnitude higher than that of OCS. SO$_{2}$ ice is also more abundant than that of SO. We then ran the MHD model of core collapse using these abundances as initial input. The evolution of density and temperature in this collapse is shown in the central and right panels of Fig. \ref{fig:histoCompPreMHD} as the fiducial MHD model. After the collapse, the resulting gas-phase abundances are shown in the left panel of Fig. \ref{fig:histoCompPreMHD}. Comparing the ice-phase abundances of the pre-stellar phase and the gas-phase abundances at the end of the collapse, it is apparent that the H$_{2}$S in the warm core is mainly coming from the thermal desorption of the ice, with no significant chemical processing. This is not the case for OCS, with an enhancement of two orders of magnitude with respect to the total ice abundance. OCS ice abundance is increased when temperatures reach $\sim 20$ K, before being too high and thus thermally desorbing it. This is due to the enhanced diffusion of the reactants to form icy OCS. When temperatures reach several hundreds K, OCS ice is thermally desorbed and it is mainly found in the gas-phase, with an abundance that matches that of the icy OCS right before its desorption. The degree of chemical processing of OCS is therefore higher, in line with the results of \citet{NavarroAlmaida2024}. Chemical processing is also found in SO and SO$_{2}$, molecules whose abundances are highly dependent on hot gas chemistry (see, e.g., \citealp{Wakelam2011, Esplugues2013}). 
        In this setting, the H$_{2}$S/OCS ratio yields H$_{2}$S/OCS $\sim 5.32$. Compared to the H$_{2}$S/OCS ratio measured toward the members of our sample, this ratio is high, only compatible with the OCS-poor sources, and close to what it is observed toward IRAS4C.
        
Column densities of  OCS and SO$_2$ have been derived  towards a sample of 26 massive hot cores by \citet{Santos2024}. Moreover, they compared their abundances with that of methanol and concluded that the OCS/CH$_3$OH ratio is quite uniform in the sample, and similar to the values measured in the ices towards low-mass and massive star forming regions. They interpreted this similarity as an evidence that the abundances of these two species are inherited from the pre-stellar phase, which is in disagreement with our model results (see \citealp{NavarroAlmaida2020}). As they commented, this conclusion can be biased by the fact that OCS ice has only been detected towards two low-mass starless regions. Further observations are needed to understand the variety of ice composition in pre-stellar cores and test our model predictions.

\begin{figure*}
			\centering
			\includegraphics[width=\textwidth]{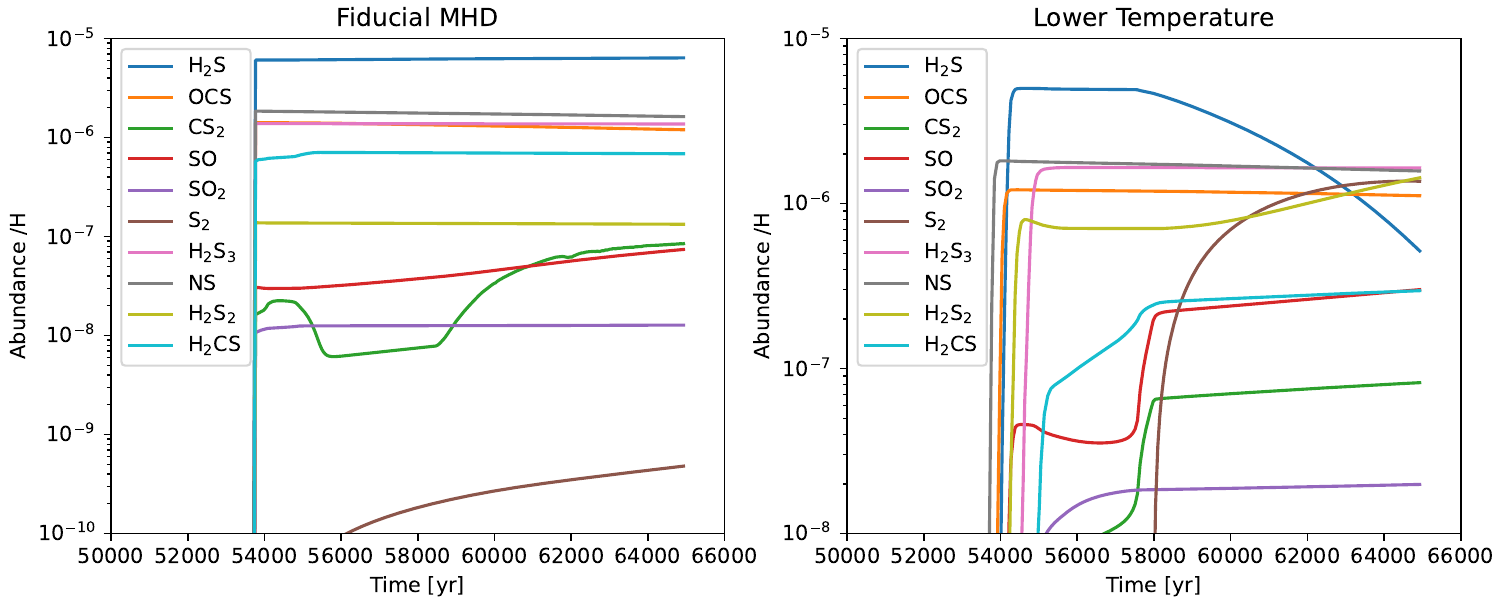}
			\caption{Species' abundance comparison between the Fiducial MHD model and the Lower Temperature model. The starting time of the figure has been set to 5$\times$10$^4$ years to show a closer detail of the major changes. \textit{Left:} In the Fiducial case, we see that the chemistry is mostly inherited from the ices; once the molecules sublimate (at $\sim$5.4$\times$10$^4$ years) their abundances remain almost constant, except for CS$_2$ and S$_2$. \textit{Right:} In the Lower Temperature model, different molecules sublimate at different temperatures, and some need a few thousand years to stabilize their gas-phase abundance (e.g. CS$_2$, SO, SO$_2$ and H$_2$CS). After $\sim$5.8$\times$10$^4$ yr, H$_2$S abundance starts to fall, while H$_2$S$_2$ and S$_2$ start to form. S$_2$ in particular increases its abundance abruptly by several orders of magnitude, reaching an abundance of $\sim$10$^{-6}$, which is 10$^4$ times more than in the Fiducial case.}
			\label{fig:SBearingMolsComp}
		\end{figure*}

        Aiming at investigating why the H$_{2}$S/OCS ratio varies among the members of our sample and the physical or chemical factors behind these changes, we ran the chemical code Nautilus in several alternative collapse scenarios. These are shown in the middle and right panels of Fig. \ref{fig:histoCompPreMHD}. Modifications were introduced in the final temperature of the warm core and the warming rate to make it slower. The corresponding changes in density were computed using the piece-wise barotropic equation of state presented in \citet{Machida2006}:
        \begin{equation}
            T=T_0\sqrt{1+\left(\frac{n}{n_1}\right)^{2g_1}}\left(1+\left(\frac{n}{n_2}\right)\right)^{g_2}\left(1+\left(\frac{n}{n_3}\right)\right)^{g_3}\,,
        \end{equation}
        with $n$ the total density, $T_0$=10K and
        \begin{align}
            & n_1=10^{11}\,\textrm{cm}^{-3}; && n_2=10^{16}\,\textrm{cm}^{-3}; & n_3=10^{21}\,\textrm{cm}^{-3};\\
            & g_1=0.4; && g_2=-0.3; & g_3=0.56667.
        \end{align}
        The evolution of gas and ice-phase abundances of H$_{2}$S and OCS for the different collapse scenarios is shown in the left and central panels of Fig. \ref{fig:warmingAlternatives}, and the H$_{2}$S/OCS ratio is in the right panel of Fig. \ref{fig:warmingAlternatives}. While higher temperatures or a slower warming rate of the warm core do not produce significant changes in the H$_{2}$S/OCS abundance ratio, it decreases when the final temperature is lower than the $\sim 500$ K of the fiducial model. Therefore, gas temperature seems to be an important factor governing the H$_{2}$S/OCS abundance ratio. Only in the the colder scenario, the gas-phase abundance of H$_{2}$S drops from the value set in the pre-stellar phase, while OCS is mainly unaffected (see Figs. \ref{fig:warmingAlternatives} and \ref{fig:SBearingMolsComp}). We recall that the angular resolution of our observations does not allow to resolve the protoplanetary disk. One would expect to have regions with different temperature within the beam, and the exact distribution of these regions could have a strong impact on the average value of the H$_{2}$S/OCS ratio.

        To investigate why the gas-phase H$_{2}$S abundance presents different behaviors, we show the evolution of the main sulfur carriers in Fig. \ref{fig:SBearingMolsComp}. We noted the drastically different evolution of molecular sulfur S$_{2}$ between the fiducial MHD model and the model with a lower final temperature. It is the main culprit in the decline of gas-phase H$_{2}$S abundance, as it is the only sulfur bearing molecule whose abundance is highly enough enhanced by the time H$_{2}$S abundance starts to drop. Comparing the fiducial model with the low temperature model (Fig. \ref{fig:SBearingMolsComp}), the abundance of S$_{2}$ in the latter is up to four orders of magnitude higher than in the former scenario. On the one hand, according to the chemical network in the lower temperature case, H$_{2}$S is mainly destroyed by its neutral-neutral reaction with H such that H + H$_{2}$S $\rightarrow$ H$_{2}$ + HS, with a reaction rate of $k=3\times 10^{-4}$ cm$^3$ s$^{-1}$. H$_{2}$S formation is less efficient, produced on the surface of grains by the radical-radical reaction between sulfanyles HS + HS $\rightarrow$ S + H$_{2}$S, with a reaction rate of $k=1\times 10^{-4}$ cm$^3$ s$^{-1}$. On the other hand, S$_{2}$ is being formed by the reaction of HS with atomic sulfur, such that S + HS $\rightarrow$ H + S$_{2}$ $(k=1\times 10^{-4}$ cm$^3$ s$^{-1})$. Therefore, the sequestration of HS in the formation of S$_{2}$ at a similar rate as the formation of H$_{2}$S and the more efficient destruction of H$_{2}$S are the factors responsible for the decline of gas-phase abundance of H$_{2}$S and the enhancement of S$_{2}$ abundance. In the fiducial MHD scenario, H$_{2}$S production and destruction are again mediated by HS in the neutral-radical reaction of HS with molecular hydrogen H$_{2}$ + HS $\rightleftharpoons$ H$_{2}$S, with its creation (forward) being more efficient than its destruction (backward). Furthermore, in this scenario, S$_{2}$ is no longer formed by reactions with HS, but in the neutral-ion reaction NO + S${_{2}}^{+}$ $\rightarrow$ S$_{2}$ + NO$^{+}$ with the rate $k = 3.73\times 10^{-8}$ cm$^{-3}$ s$^{-1}$. This leads to the less efficient formation of S$_{2}$ and the stable gas-phase abundance of H$_{2}$S we see in Fig. \ref{fig:SBearingMolsComp}.

        Polysulfides H$_{2}$S$_{x}$ have recently been proposed as compounds that could be involved in the synthesis of sulfur chains. Large sulfur chains are refractory, remaining locked on grain surfaces even at high temperatures. These compounds are candidates for sulfur reservoirs and might be where the missing sulfur in molecular clouds is \citep{Cazaux2022, Carrascosa2024}. We noted the high abundance of the polysulfides H$_{2}$S$_{2}$ and H$_{2}$S$_{3}$ in the different models of collapse. Of particular importance is the lower temperature scenario, in which H$_{2}$S$_{2}$ abundance is enhanced one order of magnitude with respect to the fiducial case. In both cases however, H$_{2}$S$_{3}$ abundance is similar.

    \subsection{H$_{2}$S/OCS ratio and the initial conditions}
            
        After analyzing the effect of different warming rates and final temperatures in the final chemical composition surrounding the protostar, we explored the role that initial conditions play in the chemical abundances of the molecules studied thus far. This is motivated by how H$_{2}$S is inherited from the pre-stellar phase since the main contribution to the total H$_2$S in gas phase after the collapse comes from the sublimation of H$_{2}$S ices, as shown in the left panel of Fig.~\ref{fig:histoCompPreMHD}.

        \renewcommand{\arraystretch}{1}
        \setlength{\tabcolsep}{3.5pt}
        \begin{table}[]
            \centering
            \caption{Comparison of physical properties between models.}
            \begin{tabular}{l|c|c|c|c|c}
            \hline\hline
                \multirow{2}{*}{Model} & $n_{\rm H}$ & $T_0$ & $A_{\rm v}$ & \multirow{2}{*}{$\chi_\textrm{UV}$} & $\zeta_{\rm H_{2}}$ \\
                    & \tiny(cm$^{-3}$) & \tiny(K) & (mag) & & \tiny(s$^{-1}$) \\
            \hline
            & & & & & \\
                Model 7.1 & 2.0$\times$10$^{4}$ & 10 & 15 & 25 & 4.0$\times$10$^{-17}$ \\
                    & & & & & \\
                Model 7.2 & 2.0$\times$10$^{4}$ & 15-25 & 15 & 100 & 5.0$\times$10$^{-17}$\\
                    & & & & & \\
            \hline
            \end{tabular}
            \label{tab:models}
            \begin{flushleft}
            \tiny
                    \textbf{Notes:} The five magnitudes compared in the table are, in order from left to right: Initial Gas Density, Initial Gas Temperature, Initial Visual Extinction, UV Flux and Cosmic-rays Ionization Rate.
            \end{flushleft}
        \end{table}
            
\begin{figure*}
            \begin{subfigure}{0.47\textwidth}
                \centering
                \includegraphics[width=\linewidth]{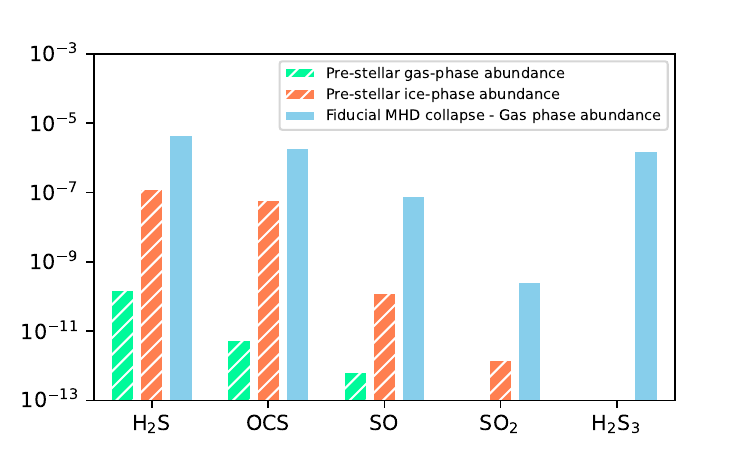}
                \caption{Initial Temperature before the pre-phase: 15 K}
                \label{fig:comparison_15K}
            \end{subfigure}
            \hspace{0.9cm}
            \begin{subfigure}{0.47\textwidth}
                \centering
                \includegraphics[width=\linewidth]{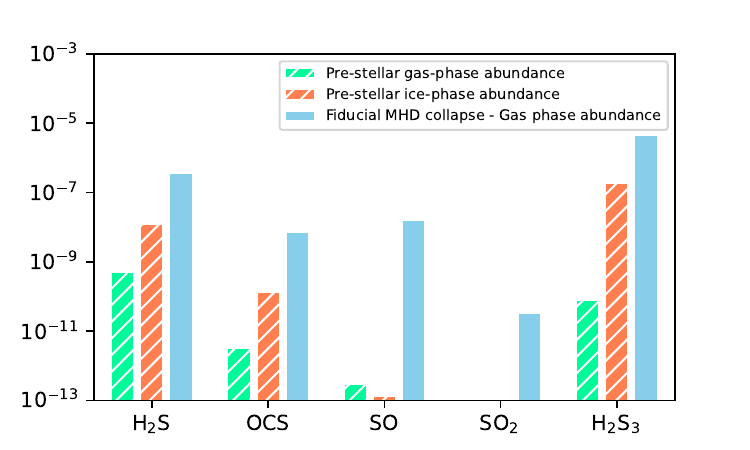}
                \caption{Initial Temperature before de pre-phase: 25 K}
                \label{fig:comparison_25K}
            \end{subfigure}
            \caption{Comparison between the 15 K and 25 K models. Initial ice and gas-phase components of the H$_2$S, OCS, SO, SO$_2$ and H$_2$S$_3$ species in the inner core before the collapse. Comparison with the final abundance of these species in the gas-phase after the collapse. There is a significant fall of H$_2$S, OCS, SO and SO$_2$ in the ices after the 25 K pre-phase, compared with the 15 K pre-stellar phase. On the contrary, H$_2$S$_3$ is formed in the ices during the warmer pre-phase, something that did not happen in the cooler one. After the collapse of the 25 K model, the final H$_2$S, SO and SO$_2$ drops in $\sim$1 order of magnitude with respect to the 15 K simulation. This is not true for OCS, whose abundance falls more than 2 orders of magnitude. H$_2$S$_3$ final gas-phase abundance is greater in the warmer model, reaching an abundance of $\sim$5$\times$10$^{-6}$.}
            \label{fig:model_comparison}
            \end{figure*}

        To do so, we ran a set of zero-dimensional (0-D) pre-stellar models with \texttt{Nautilus} for $10^{6}$ years. These models are warmer alternatives to the pre-phase considered in Sect. \ref{sect:mhdH2SOCS} (see Table~\ref{tab:models}), that is, the innermost point of the Barnard 1b model presented in \citet{NavarroAlmaida2020}, but covering a range of dust and gas temperatures between 10 K and 25 K, spaced by 5 K. These models would be appropriate to describe, for instance, intermediate-mass star forming regions illuminated by a stronger interstellar FUV field or clusters subjected to intense stellar feedback. After $10^6$ years, the chemical abundances become the input of the corresponding core collapse. We modified the density and temperature evolution of the collapsing core in such a way that they take into account the rising temperature of the pre-phase: if at any given time step the temperature is lower than the temperature set at the pre-phase, it gets updated to the pre-phase value. The density is then updated accordingly following the barotropic equation of state. The results of the models with $T=15$ K and $T=25$ K are shown in Fig. \ref{fig:model_comparison}. On one hand, we found that, even at 15 K, a temperature slightly higher than in the fiducial MHD model, icy H$_{2}$S at the end of the pre-phase is less abundant than in the fiducial case. The final gas-phase abundance of H$_{2}$S is not directly inherited as the thermal desorption of the ice content, but it is increased in the warm up of the collapsing core, reaching a comparable value with respect to the fiducial model. On the other hand, OCS ice-abundance is higher than its predicted value by the fiducial model and comparable to that of H$_{2}$S. This also results in a higher gas-phase abundance of this molecule at the end of the collapse, yielding a H$_{2}$S/OCS ratio of $\sim 2.4$. This is no longer the case as we increase the temperature of the pre-phase. In the right panel of Fig. \ref{fig:model_comparison} we show the results of the warmest model among the set of 0-D models. The ice-phase abundance of H$_{2}$S drops by one order of magnitude with respect to the $15$ K case, while icy OCS abundance declines significantly, by three orders of magnitude. The corresponding gas-phase abundances at the end of the collapse are lower compared to the fiducial or the $15$ K case, resulting in a H$_{2}$S/OCS ratio of $\sim 53$.

            The general decreasing H$_{2}$S abundances in warmer scenarios is accompanied by the abundant formation of H$_2$S$_3$ ices during the pre-stellar phase (right panel of Fig. \ref{fig:model_comparison}) that were not present at lower temperatures (left panel of Fig. \ref{fig:model_comparison}). These high abundances then translate into higher abundances in the gas-phase after the MHD collapse. H$_{2}$S$_{3}$ abundances can be even higher than those of H$_{2}$S (right panel of Fig. \ref{fig:model_comparison}). With a final gas-phase abundance of $4.5\times10^{-6}$ and given that this molecule contains three sulfur atoms, it accounts for a great fraction of the cosmic sulfur abundance. According to the chemical network, icy H$_{2}$S$_{3}$ is formed by the diffusion and reaction of HS with H$_{2}$S$_{2}$ in the ice matrix. This diffusion process is only efficient when dust temperature is higher than $\sim$22 K and, consequently, only appears in the ice of the warmest models considered here. The warm-up during the collapse present in each model also triggers this diffusion of the reactants to form H$_{2}$S$_{3}$. This is why the gas-phase H$_{2}$S$_{3}$ abundance is high at the end of the collapse in colder models even when its was not present in the pre-phase. Nevertheless, this result should be considered with caution since the production and destruction of sulfur allotropes are not yet well understood and the family of sulfides H$_2$S$_x$ is included in the chemical network only for $x\leq 3$. However, \citet{Shingledecker2020} included the formation of allotropes in their code, finding that S allotropes can become a major sink of sulfur, which agrees with our results for the warmer models.

            In conclusion, our simulations show that both the absolute abundances of H$_2$S and OCS as well as the H$_2$S/OCS are very sensitive to the gas and dust temperature in the pre-stellar phase and also the final temperature in the inner core. High temperatures in the pre-stellar phase drive to higher H$_2$S/OCS ratio but lower absolute abundances of these molecules, since S is sequestered in hydrogen sulfides. The final temperature of the warm inner core has also a strong impact on the chemical composition of the warm gas. Low values of the H$_2$S/OCS ratio are consistent with regions with $T\sim100$ K. Summarizing, the H$_2$S and OCS are very sensitive to the thermal history of the gas and dust, producing a high range of these molecules abundances and changing the H$_2$S/OCS ratio.

\section{Discussion: Where is sulfur?}
\label{sec:discussion}

This section is dedicated to the discussion of the gas-phase depletion of sulfur-bearing species in the Class 0/I protostars of Perseus Molecular Cloud and how it compares to other environments.
Sulfur depletion is defined as the ratio of the cosmic sulfur abundance over the amount of sulfur in volatiles:
\begin{equation}
    D_S = \frac{[\textrm{S/H}]_{cosmic}}{[\textrm{S/H}]_{volatiles}}\,,
\end{equation}
Since some of the most abundant sulfur bearing species cannot be easily observed, the value of  $\textrm{S/H}_{volatiles}$ needs to be indirectly estimated through chemical modeling (see, e.g., \citealp{Laas2019, NavarroAlmaida2020, Fuente2023}). Warm cores in the nuclei of young protostars are specially favorable environments to estimate $D_S$ because we can assume that the ices have sublimated and all the volatile sulfur is in gas phase. Thus, we can estimate the amount of sulfur in volatiles by adding the abundances of the most abundant sulfur-bearing molecules in this kind of environment. We have used the H$_2$S and OCS abundances derived with PRODIGE data and the SO and SO$_2$ abundances estimated from PEACHES observations, to have an estimate of sulfur depletion in our sample (see Table \ref{tab:abundances}). These species are expected to be the most abundant sulfur compounds in warm inner cores and lock essentially all the sulfur atoms (see, e.g., \citealp{Vidal2018}), their sum providing a good estimate of  D$_{\rm S}$. Note, however, that this is true only if a negligible fraction of S is in allotropes \citep{Shingledecker2020}. Thus calculated, we found that the highest sulfur abundance in our sample is measured toward SVS13A, of around $\sim$1.7$\times$10$^{-6}$, which corresponds to a sulfur depletion of $D_S$$\sim$8. This value is comparable to the depletion observed in sources in Orion A \citep{Fuente2023, Fuente2024}. However, the rest of the sources present higher depletion of sulfuretted species, with most of the values between $D_S$$\sim$500--5000, and the least abundant ones reaching values up to $D_S$$\sim$5$\times$10$^5$ (e.g. Per-emb-02, L1448NW). These values are higher than those found by \citet{Fuente2023} in Orion, Taurus and Perseus molecular clouds, suggesting that sulfur depletion increases during the collapse of a starless core. Yet, there are some observational and theoretical factors that can contribute to these large values of $D_S$.

The first one implies that the sulfur-bearing species may be still in the ices in some regions within our beam.  We remind that the beam of our observations, $\sim$1.3$\arcsec$ ($\sim$390 au) may not be enough to resolve the smaller disks \citep{Tobin2024}. Even in the case of Class I larger protoplanetary disks, it may not be enough to resolve the warmest areas around the protostar. The majority of the sources in our sample are very young (Class 0), and the temperature in the warm inner core might not be as high as in the more evolved ones. This would have a strong effect on the position of the snow line, therefore changing the amount of ices ---and, with them, the amount of sulfuretted molecules--- sublimated into the gas phase, which is the only component we observed in this work. This hypothesis is supported by the results shown in Fig.\ref{fig:abundances}, where the protostars' total abundance of sulfur is growing with evolution (given by $T_\textrm{bol}$) until they reach Class 0/I.  
A lower bolometric temperature in the core entails a closer snow line, as we explained previously, which would produce a concentration of gas-phase species in the most inner part that would remain unresolved by our observations. Possible examples could be L1448-IRS3A or L1448NW, among others. As the protostar evolves, the snow lines would move farther from the protostar. In the opposite side of the evolutionary track, in a evolved disk (Class I), where the protoplanetary disk is thinner than in the Class 0 sources and the hot material is concentrated in the innermost region  (R $<$ 50 au) of the disk and its surfaces, while the outer disk is mostly cold. Again, our observations will not resolve the inner region and we would be essentially detecteting the colder outer disk. This phenomena could give an explanation to the later fall of sulfur abundance in the Class I objects (e.g. B5-IRS1) and the lack of detections of the sulfuretted species (e.g. Per-emb-50, Per-emb-62). Our limited spatial resolution would affect the absolute value of the H$_2$S and OCS estimated abundances but not the H$_2$S/OCS ratio since all the observed transitions are expected to come from the warm region. The large scattering in the values of H$_2$S/OCS demonstrates important chemical differences within our sample.

It has been suggested that sulfur depletion depends on the environment \citep{Fuente2016, NavarroAlmaida2020, Fuente2023} in molecular clouds and, therefore, the estimated values of D$_S$ could depend on the location of the protostars of our sample. In a more crowded region, protostars are exposed to greater fluxes of UV rays, which can lead to higher levels of photo-desorption, and shocks associated to bipolar outflows, which can release sulfur-bearing species to the gas-phase. Although all the protostars in our sample are located in Perseus, there is a patent diversity of conditions in which these objects are present. The most clustered zone of Perseus is SVS13, where, apart from the cluster of protostars, there is a Herbig Haro object (HH 711) in its vicinity \citep{Hatchell2008, Hatchell2009}. IRAS2A, IRAS4B, Per-emb-18, and SVS13A are located in this active star forming region \citep{Hatchell2013, Codella2021}. Another interesting region is  IC 348, where the UV flux is expected to be higher because its proximity to a cluster of massive stars and a second Herbig Haro object is located, HH 211. In these regions, we find values of D$_S$ $\sim$10 to 300, with variations of more than one order of magnitude. Although environment is surely playing a role, we do not find a clear correlation between the protostar's location and D$_S$ because other factors, such as the evolutionary stage, have also an important impact.

The question here is whether the total sulfur abundance ---in gas phase and in the ices--- is the same, or similar, in all these sources in spite of the large dispersion of $D_S$ values. Chemical models predict the presence of large sulfur compounds that possess sublimation temperatures higher than that of water and thus are not expected to be sublimated in warm inner envelope of young protostars, where the dust temperature is, at most, $\sim$500 K. In particular, both laboratory experiments and theoretical work show that sulfur allotropes, such as S$_8$ and large hydrogen sulfides (H$_2$S$_\textrm{x}$), could be important sulfur reservoirs \citep{Jimenez-Escobar2012, Shingledecker2020, Cazaux2022, Fuente2023, Carrascosa2024}.The sublimation temperature of these compounds when x $>$ 3 is $\gg$100~K and can be considered as semi-refractory material \citep{Perrero2024, Carrascosa2024}. Our simulations show a very efficient formation of H$_2$S$_3$ under some physical conditions, locking most of the sulfur atoms (see Section~\ref{sec:model}) while the abundances of H$_2$S and OCS remain lower than 10$^{-8}$. The formation and destruction processes of allotropes are not well understood yet but our predictions demonstrate the formation of these compounds in molecular clouds. Several laboratory experiments confirm that large hydrogen sulfides are formed during the irradiation of  H$_2$S and H$_2$S:H$_2$O ices \citep{Cazaux2022, Carrascosa2024}. More recently, experiments reported by \citet{Martin2024} suggest that sulfur chains (S$_\textrm{X}$) can be formed in CO:CS$_2$ and CO$_2$:CS$_2$. Morevoer, sulfur chains have been detected in comets \citep{Calmonte2016} and the compound S$_2$H was detected in the Horsehead nebula by \citet{Fuente2017-s2h}. The existence of these molecules in the interstellar medium is proved and the relative importance of these compounds as sulfur reservoirs in different environments deserves further study.

Other compounds such as NH$_4$SH have also been suggested as important sulfur reservoirs in dense regions \citep{Vitorino2024}. This compound sublimates together with water, and is released to the gas phase increasing the abundances of H$_2$S and NH$_3$. The formation of this molecule cannot be therefore responsible for the low abundance of H$_2$S in our sample of warm cores. Finally, some molecules not considered in this work such as CS, CCS, and H$_2$CS could lock a significant fraction of sulfur atoms. Recent observations of these compounds towards SVS 13A unveiled that CCS, and H$_2$CS column densities are one order of magnitude lower than those of the molecules in this work. Regarding CS, the column density is very uncertain and could be of the same order, even higher, than that of H$_2$S and OCS \citep{Codella2021}. High angular observations of these molecules could help to disentangle the sulfur chemistry in these cores.

\begin{figure*}
    \centering
    \includegraphics[width=\textwidth]{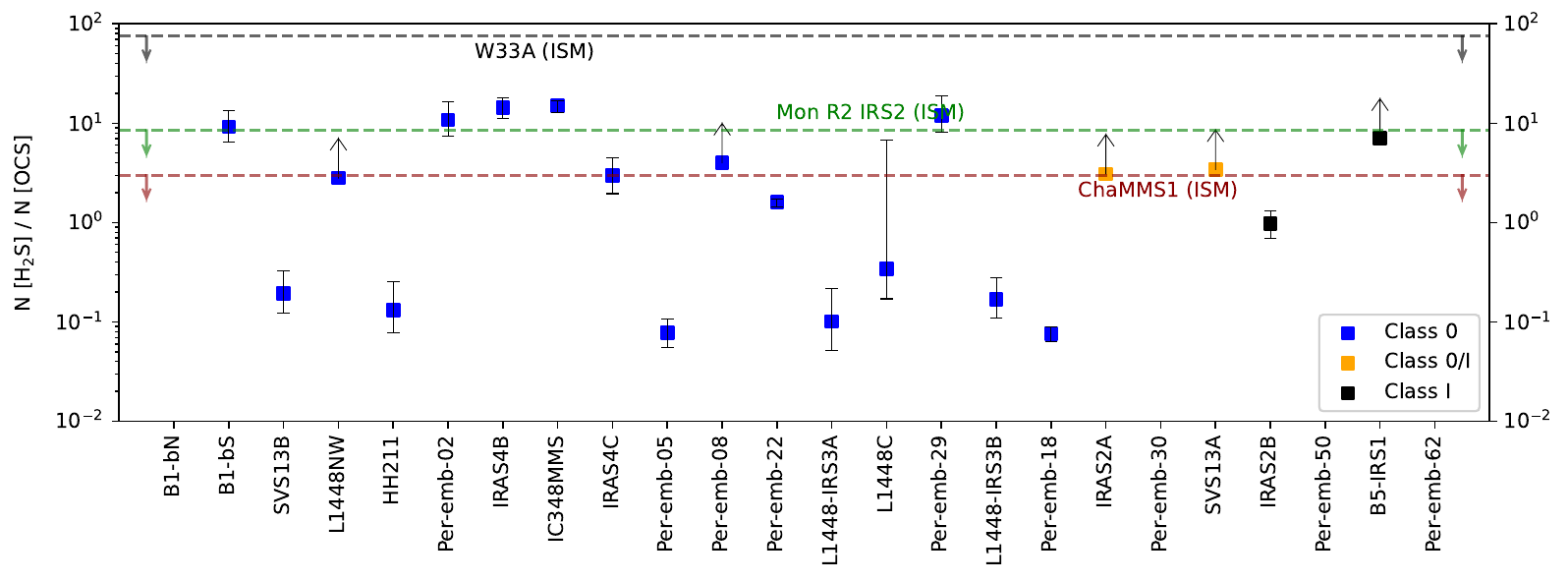}\caption{H$_2$S/OCS ratio in the 24 protostars of our sample. The class of each protostar has been represented using the color code in the legend. Sources are sorted by class, and then ordered by increasing bolometric temperature within each class. The discontinuous horizontal lines represent different upper limits to the H$_2$S/OCS ratio in ices from different regions: in black, W33A; in green, Mon R2 IRS2; in red, ChaMMS1. Only the OCS-rich sources are compatible with the more restrictive limit, ChaMMS1, which could give us some hints about the environment where the ice was formed.}
    \label{fig:ratios}
\end{figure*}

Our observations and simulations show that the composition of sulfur species in the ice formed during the pre-stellar and protostellar phase, and also later, in the warm gas enriched in sulfur compounds because of ice evaporation, are extremely sensitive to the thermal history of the gas and dust. We notice variations of more than one order of magnitude in the abundances of routinely observed species such a H$_2$S and OCS. This challenges the estimation of sulfur depletion based on the observations of only these molecules. The knowledge of the ice composition at the end of the pre-stellar phase would be essential to obtain accurate predictions of the abundances of all sulfur compounds in the warm phase. Observation of the chemical composition of ices would be a valuable input to constrain chemical predictions.

The detection of sulfur species in solid phase remains challenging even in the James Webb Space Telescope (JWST) era. Thus far, only OCS and SO$_2$ have been detected in interstellar ices  \citep{Geballe1985, Palumbo1997, Boogert1997, Boogert2022, McClure2023} , while solid H$_2$S remains undetected. Recent measurements carried out with the JWST have improved previous upper limits of the  H$_2$S/OCS in the ice. Yet, the limits are far from constraining our knowledge of sulfur chemistry. In Fig.~\ref{fig:ratios}, we compare the values of H$_2$S/OCS obtained towards the warm protostellar cores using millimeter interferometers with the upper limits obtained for the solid phase using infrared telescopes. Some of the gas-phase measurements are higher than the H$_2$S/OCS ratio obtained by \citet{McClure2023} for the ice towards ChaMMS1, but most of them remain compatible with it. Taking into account the great sensitivity if the icy H$_2$S/OCS ratio to the local physical conditions and past thermal history of the interstellar grains, current observations are far to be conclusive and new searches of solid H$_2$S are necessary to increase our knowledge of sulfur chemistry and further contrast our models.

\section{Summary and conclusions}

In this work, we studied several sulfur-bearing species ---H$_2$S, H$_2^{33}$S, OCS, OC$^{33}$S, OC$^{34}$S, SO and SO$_2$--- in the warm inner core of 24 Class 0/I protostars in the Perseus Molecular Cloud. We derived the column densities of the first five by fitting the gas emission lines in the NOEMA Band 3 receiver and the PolyFix correlator spectra from the PRODIGE large program observations, and we used the SO and SO$_2$ data from \citet{ArturDeLaVillarmois2023}. We also estimated the abundances of all seven molecules using our observations of the continuum, and analysed the variations of the H$_2$S/OCS ratio over the 24 sources. Our main conclusions are the following:
\begin{itemize}
    \item We calculated the gas-phase column densities of H$_2$S, H$_2^{33}$S, OCS, OC$^{33}$S and OC$^{34}$S NOEMA Band 3 data, and the N(H$_2$S) ratio over N(OCS). We detected H$_2$S in 20 of the 24 sources in our sample, and OCS in 17 of them.
    \item Protostars can be characterized by their H$_2$S and OCS composition. We differentiate two kinds of objects: OCS-poor protostars, with H$_2$S/OCS$\sim$7-10; OCS-rich protostars, with H$_2$S/OCS$\sim$0.1-1.
    \item Total sulfur abundance grows with evolution until the Class 0/I stage, where the minumum depletion is achieved: $D_S$$\sim$8, with a total S abundance of (1.7$\pm$0.1)$\times$10$^{-6}$. This value is in accordance with the H$_2$S abundance measured in Orion KL: (3.1$\pm^{1.1}_{1.9}\,)\times$10$^{-6}$ \citep{Crockett2014}. In more evolved stages (Class I), the total sulfur abundance decreases, though more observations of Class I/II sources would be needed to confirm this. The total sulfur abundance is usually dominated by the H$_2$S molecule, followed by OCS. SO and SO$_2$ abundances are typically two orders of magnitude lower than H$_2$S and OCS; however, SO and SO$_2$ abundances grow slowly towards the Class I stage, gaining relevance in the sulfur budget in the more evolved sources.
    \item Theoretical numerical simulations show that H$_2$S and OCS gas-phase and ice abundances are very responsive to temperature changes both in the pre-stellar phase and during the collapse. H$_2$S is less abundant when the initial temperature of the cloud is warmer, driving to lower amounts of OCS and, eventually, greater H$_2$S/OCS ratios after the collapse. Also, a lower temperature of the inner core during the collapse shows a fall of H$_2$S abundance, leading to lower H$_2$S/OCS ratios. Furthermore, the 7.1 and 7.2 (15 K) models (Table~\ref{tab:models}) return significant variations of the H$_2$S and OCS in ices after the pre-phase and a factor $\sim$2 difference of the H$_2$S/OCS ratio after the collapse, with only slightly different conditions of the initial cloud. Results suggest that, in some scenarios, H$_2$S$_x$ molecules, for $x\geq3$, might be a significant sulfur reservoir.
    \item We found a wide range of values for the depletion of sulfur (8 to $>$ 100) in our sample of protostars. This behaviour only seems to be explainable with differences in the initial conditions and in the first evolutionary stages. We consider that the details of the ice composition after the pre-stellar phase (before the collapse) are essential in order to understand the chemical composition of a protostar after the collapse. Also, higher angular resolution observations are required to resolve the circumstellar disk to probe the innermost and warmest region.
\end{itemize}
The NOEMA interferometer observations have allowed us to study the sulfur chemistry of the warm inner core of 24 protostars in Perseus. Our results show that the evolution history of a protostar plays a decisive role in the chemistry of its later stages. However, the low spatial resolution of the images and the low spectral resolution of the H$_2$S, H$_2^{33}$S and OC$^{34}$S bands, invites for higher resolution observations in order to better resolve the inner core and to better estimate the column densities of some species. Nevertheless, the PRODIGE data reveals that the sulfur chemistry of young protostars is still not well understood and needs of further study to comprehend how the different environmental and evolutionary factors affect the composition of these young stellar objects.

\begin{acknowledgements}
This work is supported by ERC grant SUL4LIFE, GA No. 101096293. Funded by the European Union. Views and opinions expressed are however those of the author(s) only and do not necessarily reflect those of the European Union or the European Research Council Executive Agency. Neither the European Union nor the granting authority can be held responsible for them. J.E.P., D.M.S.-C., P.C., M.T.V.-M., T.-H.H., L.B., C.G. Y.-R.C. and M.M. are grateful for support from the Max Planck Society. This work is based on observations carried out under project number L19MB with the IRAM NOEMA Interferometer. IRAM is supported by INSU/CNRS (France), MPG (Germany) and IGN (Spain). This paper makes use of the following ALMA data: ADS/JAO.ALMA\#2016.0.00391.S. ALMA is a partnership of ESO (representing its member states), NSF (USA) and NINS (Japan), together with NRC (Canada), NSTC and ASIAA (Taiwan), and KASI (Republic of Korea), in cooperation with the Republic of Chile. The Joint ALMA Observatory is operated by ESO, AUI/NRAO and NAOJ.

\end{acknowledgements}

\bibliography{biblio}

\begin{thebibliography}{151}
\expandafter\ifx\csname natexlab\endcsname\relax\def\natexlab#1{#1}\fi

\bibitem[{{Altwegg} {et~al.}(2022){Altwegg}, {Combi}, {Fuselier}, {H{\"a}nni}, {De Keyser}, {Mahjoub}, {M{\"u}ller}, {Pestoni}, {Rubin}, \& {Wampfler}}]{Altwegg2022}
{Altwegg}, K., {Combi}, M., {Fuselier}, S.~A., {et~al.} 2022, \mnras, 516, 3900

\bibitem[{{Anders} \& {Grevesse}(1989)}]{AndersGravesse1989}
{Anders}, E. \& {Grevesse}, N. 1989, \gca, 53, 197

\bibitem[{{Andr{\'e}} {et~al.}(2010){Andr{\'e}}, {Men'shchikov}, {Bontemps}, {K{\"o}nyves}, {Motte}, {Schneider}, {Didelon}, {Minier}, {Saraceno}, {Ward-Thompson}, {di Francesco}, {White}, {Molinari}, {Testi}, {Abergel}, {Griffin}, {Henning}, {Royer}, {Mer{\'\i}n}, {Vavrek}, {Attard}, {Arzoumanian}, {Wilson}, {Ade}, {Aussel}, {Baluteau}, {Benedettini}, {Bernard}, {Blommaert}, {Cambr{\'e}sy}, {Cox}, {di Giorgio}, {Hargrave}, {Hennemann}, {Huang}, {Kirk}, {Krause}, {Launhardt}, {Leeks}, {Le Pennec}, {Li}, {Martin}, {Maury}, {Olofsson}, {Omont}, {Peretto}, {Pezzuto}, {Prusti}, {Roussel}, {Russeil}, {Sauvage}, {Sibthorpe}, {Sicilia-Aguilar}, {Spinoglio}, {Waelkens}, {Woodcraft}, \& {Zavagno}}]{Andre2010}
{Andr{\'e}}, P., {Men'shchikov}, A., {Bontemps}, S., {et~al.} 2010, \aap, 518, L102

\bibitem[{{Aponte} {et~al.}(2023){Aponte}, {Dworkin}, {Glavin}, {Elsila}, {Parker}, {McLain}, {Naraoka}, {Okazaki}, {Takano}, {Tachibana}, {Dong}, {Zeichner}, {Eiler}, {Yurimoto}, {Nakamura}, {Yabuta}, {Terui}, {Noguchi}, {Sakamoto}, {Yada}, {Nishimura}, {Nakato}, {Miyazaki}, {Yogata}, {Abe}, T., {Usui}, {Yoshikawa}, {Tanaka}, {Nakazawa}, {Tsuda}, {Watanabe}, \& initial-analysis SOM team & The Hayabusa2-initial-analysis~core team}]{Aponte2023}
{Aponte}, J.~C., {Dworkin}, J.~P., {Glavin}, D.~P., {et~al.} 2023, Earth, Planets and Space, 75, 148

\bibitem[{{Artur de la Villarmois} {et~al.}(2022){Artur de la Villarmois}, {Guzm{\'a}n}, {J{\o}rgensen}, {Kristensen}, {Bergin}, {Harsono}, {Sakai}, {van Dishoeck}, \& {Yamamoto}}]{Villarmois2022}
{Artur de la Villarmois}, E., {Guzm{\'a}n}, V.~V., {J{\o}rgensen}, J.~K., {et~al.} 2022, \aap, 667, A20

\bibitem[{{Artur de la Villarmois} {et~al.}(2023){Artur de la Villarmois}, {Guzm{\'a}n}, {Yang}, {Zhang}, \& {Sakai}}]{ArturDeLaVillarmois2023}
{Artur de la Villarmois}, E., {Guzm{\'a}n}, V.~V., {Yang}, Y.~L., {Zhang}, Y., \& {Sakai}, N. 2023, \aap, 678, A124

\bibitem[{{Asplund} {et~al.}(2009){Asplund}, {Grevesse}, {Sauval}, \& {Scott}}]{Asplund2009}
{Asplund}, M., {Grevesse}, N., {Sauval}, A.~J., \& {Scott}, P. 2009, \araa, 47, 481

\bibitem[{{Bachiller} {et~al.}(1990){Bachiller}, {Cernicharo}, {Martin-Pintado}, {Tafalla}, \& {Lazareff}}]{Bachiller1990}
{Bachiller}, R., {Cernicharo}, J., {Martin-Pintado}, J., {Tafalla}, M., \& {Lazareff}, B. 1990, \aap, 231, 174

\bibitem[{{Bachiller} {et~al.}(1998){Bachiller}, {Guilloteau}, {Gueth}, {Tafalla}, {Dutrey}, {Codella}, \& {Castets}}]{Bachiller1998}
{Bachiller}, R., {Guilloteau}, S., {Gueth}, F., {et~al.} 1998, \aap, 339, L49

\bibitem[{{Bachiller} {et~al.}(1991){Bachiller}, {Martin-Pintado}, \& {Fuente}}]{Bachiller1991}
{Bachiller}, R., {Martin-Pintado}, J., \& {Fuente}, A. 1991, \aap, 243, L21

\bibitem[{{Bariosco} {et~al.}(2024){Bariosco}, {Pantaleone}, {Ceccarelli}, {Rimola}, {Balucani}, {Corno}, \& {Ugliengo}}]{Bariosco2024}
{Bariosco}, V., {Pantaleone}, S., {Ceccarelli}, C., {et~al.} 2024, \mnras, 531, 1371

\bibitem[{{Barsony} {et~al.}(1998){Barsony}, {Ward-Thompson}, {Andr{\'e}}, \& {O'Linger}}]{Barsony1998}
{Barsony}, M., {Ward-Thompson}, D., {Andr{\'e}}, P., \& {O'Linger}, J. 1998, \apj, 509, 733

\bibitem[{{Bianchi} {et~al.}(2022){Bianchi}, {Ceccarelli}, {Codella}, {L{\'o}pez-Sepulcre}, {Yamamoto}, {Balucani}, {Caselli}, {Podio}, {Neri}, {Bachiller}, {Favre}, {Fontani}, {Lefloch}, {Sakai}, \& {Segura-Cox}}]{Bianchi2022}
{Bianchi}, E., {Ceccarelli}, C., {Codella}, C., {et~al.} 2022, \aap, 662, A103

\bibitem[{{Bianchi} {et~al.}(2019){Bianchi}, {Codella}, {Ceccarelli}, {Vazart}, {Bachiller}, {Balucani}, {Bouvier}, {De Simone}, {Enrique-Romero}, {Kahane}, {Lefloch}, {L{\'o}pez-Sepulcre}, {Ospina-Zamudio}, {Podio}, \& {Taquet}}]{Bianchi2019}
{Bianchi}, E., {Codella}, C., {Ceccarelli}, C., {et~al.} 2019, \mnras, 483, 1850

\bibitem[{{Boogert} {et~al.}(2022){Boogert}, {Brewer}, {Brittain}, \& {Emerson}}]{Boogert2022}
{Boogert}, A.~C.~A., {Brewer}, K., {Brittain}, A., \& {Emerson}, K.~S. 2022, \apj, 941, 32

\bibitem[{{Boogert} {et~al.}(1997){Boogert}, {Schutte}, {Helmich}, {Tielens}, \& {Wooden}}]{Boogert1997}
{Boogert}, A.~C.~A., {Schutte}, W.~A., {Helmich}, F.~P., {Tielens}, A.~G.~G.~M., \& {Wooden}, D.~H. 1997, \aap, 317, 929

\bibitem[{{Brinch} {et~al.}(2009){Brinch}, {J{\o}rgensen}, \& {Hogerheijde}}]{Brinch2009}
{Brinch}, C., {J{\o}rgensen}, J.~K., \& {Hogerheijde}, M.~R. 2009, \aap, 502, 199

\bibitem[{{Calmonte} {et~al.}(2016){Calmonte}, {Altwegg}, {Balsiger}, {Berthelier}, {Bieler}, {Cessateur}, {Dhooghe}, {van Dishoeck}, {Fiethe}, {Fuselier}, {Gasc}, {Gombosi}, {H{\"a}ssig}, {Le Roy}, {Rubin}, {S{\'e}mon}, {Tzou}, \& {Wampfler}}]{Calmonte2016}
{Calmonte}, U., {Altwegg}, K., {Balsiger}, H., {et~al.} 2016, \mnras, 462, S253

\bibitem[{{Calmonte} {et~al.}(2017){Calmonte}, {Altwegg}, {Balsiger}, {Berthelier}, {Bieler}, {De Keyser}, {Fiethe}, {Fuselier}, {Gasc}, {Gombosi}, {Le Roy}, {Rubin}, {S{\'e}mon}, {Tzou}, \& {Wampfler}}]{Calmonte2017}
{Calmonte}, U., {Altwegg}, K., {Balsiger}, H., {et~al.} 2017, \mnras, 469, S787

\bibitem[{{Caratti o Garatti} {et~al.}(2024){Caratti o Garatti}, {Ray}, {Kavanagh}, {McCaughrean}, {Gieser}, {Giannini}, {van Dishoeck}, {Justtanont}, {van Gelder}, {Francis}, {Beuther}, {Tychoniec}, {Nisini}, {Navarro}, {Devaraj}, {Reyes}, {Nazari}, {Klaassen}, {G{\"u}del}, {Henning}, {Lagage}, {{\"O}stlin}, {Vandenbussche}, {Waelkens}, \& {Wright}}]{CarattioGaratti2024}
{Caratti o Garatti}, A., {Ray}, T.~P., {Kavanagh}, P.~J., {et~al.} 2024, \aap, 691, A134

\bibitem[{{Carrascosa} {et~al.}(2024){Carrascosa}, {Mu{\~n}oz Caro}, {Mart{\'\i}n-Dom{\'e}nech}, {Cazaux}, {Chen}, \& {Fuente}}]{Carrascosa2024}
{Carrascosa}, H., {Mu{\~n}oz Caro}, G.~M., {Mart{\'\i}n-Dom{\'e}nech}, R., {et~al.} 2024, \mnras, 533, 967

\bibitem[{{Caselli} {et~al.}(1997){Caselli}, {Hartquist}, \& {Havnes}}]{Caselli1997}
{Caselli}, P., {Hartquist}, T.~W., \& {Havnes}, O. 1997, \aap, 322, 296

\bibitem[{{Caselli} {et~al.}(2022){Caselli}, {Pineda}, {Sipil{\"a}}, {Zhao}, {Redaelli}, {Spezzano}, {Maureira}, {Alves}, {Bizzocchi}, {Bourke}, {Chac{\'o}n-Tanarro}, {Friesen}, {Galli}, {Harju}, {Jim{\'e}nez-Serra}, {Keto}, {Li}, {Padovani}, {Schmiedeke}, {Tafalla}, \& {Vastel}}]{Caselli2022}
{Caselli}, P., {Pineda}, J.~E., {Sipil{\"a}}, O., {et~al.} 2022, \apj, 929, 13

\bibitem[{{Cazaux} {et~al.}(2022){Cazaux}, {Carrascosa}, {Mu{\~n}oz Caro}, {Caselli}, {Fuente}, {Navarro-Almaida}, \& {Rivi{\'e}re-Marichalar}}]{Cazaux2022}
{Cazaux}, S., {Carrascosa}, H., {Mu{\~n}oz Caro}, G.~M., {et~al.} 2022, \aap, 657, A100

\bibitem[{{Ceccarelli}(2007)}]{Ceccarelli2007}
{Ceccarelli}, C. 2007, in Molecules in Space and Laboratory, ed. J.~L. {Lemaire} \& F.~{Combes}, 1

\bibitem[{Chen \& Yu(2019)}]{Chen2019}
Chen, M. \& Yu, X. 2019, in Carbonyl Sulfide-Mediated Synthesis of Peptides with Amino Acid Ionic Liquids

\bibitem[{{Chen} {et~al.}(2024){Chen}, {Rocha}, {van Dishoeck}, {van Gelder}, {Nazari}, {Slavicinska}, {Francis}, {Tabone}, {Ressler}, {Klaassen}, {Beuther}, {Boogert}, {Gieser}, {Kavanagh}, {Perotti}, {Le Gouellec}, {Majumdar}, {G{\"u}del}, \& {Henning}}]{Chen2024}
{Chen}, Y., {Rocha}, W.~R.~M., {van Dishoeck}, E.~F., {et~al.} 2024, \aap, 690, A205

\bibitem[{{Chin} {et~al.}(1996){Chin}, {Henkel}, {Whiteoak}, {Langer}, \& {Churchwell}}]{Chin1996}
{Chin}, Y.~N., {Henkel}, C., {Whiteoak}, J.~B., {Langer}, N., \& {Churchwell}, E.~B. 1996, \aap, 305, 960

\bibitem[{{Codella} {et~al.}(2021){Codella}, {Bianchi}, {Podio}, {Mercimek}, {Ceccarelli}, {L{\'o}pez-Sepulcre}, {Bachiller}, {Caselli}, {Sakai}, {Neri}, {Fontani}, {Favre}, {Balucani}, {Lefloch}, {Viti}, \& {Yamamoto}}]{Codella2021}
{Codella}, C., {Bianchi}, E., {Podio}, L., {et~al.} 2021, \aap, 654, A52

\bibitem[{{Codella} {et~al.}(2016){Codella}, {Ceccarelli}, {Bianchi}, {Podio}, {Bachiller}, {Lefloch}, {Fontani}, {Taquet}, \& {Testi}}]{Codella2016}
{Codella}, C., {Ceccarelli}, C., {Bianchi}, E., {et~al.} 2016, \mnras, 462, L75

\bibitem[{{Collings} {et~al.}(2003){Collings}, {Dever}, {Fraser}, {McCoustra}, \& {Williams}}]{Collings2003}
{Collings}, M.~P., {Dever}, J.~W., {Fraser}, H.~J., {McCoustra}, M.~R.~S., \& {Williams}, D.~A. 2003, \apj, 583, 1058

\bibitem[{{Crockett} {et~al.}(2014){Crockett}, {Bergin}, {Neill}, {Black}, {Blake}, \& {Kleshcheva}}]{Crockett2014}
{Crockett}, N.~R., {Bergin}, E.~A., {Neill}, J.~L., {et~al.} 2014, \apj, 781, 114

\bibitem[{{Daflon} {et~al.}(2009){Daflon}, {Cunha}, {de la Reza}, {Holtzman}, \& {Chiappini}}]{Daflon2009}
{Daflon}, S., {Cunha}, K., {de la Reza}, R., {Holtzman}, J., \& {Chiappini}, C. 2009, \aj, 138, 1577

\bibitem[{{Dionatos} {et~al.}(2010){Dionatos}, {Nisini}, {Cabrit}, {Kristensen}, \& {Pineau Des For{\^e}ts}}]{Dionatos2010}
{Dionatos}, O., {Nisini}, B., {Cabrit}, S., {Kristensen}, L., \& {Pineau Des For{\^e}ts}, G. 2010, \aap, 521, A7

\bibitem[{{Dionatos} {et~al.}(2018){Dionatos}, {Ray}, \& {G{\"u}del}}]{Dionatos2018}
{Dionatos}, O., {Ray}, T., \& {G{\"u}del}, M. 2018, \aap, 616, A84

\bibitem[{{Drozdovskaya} {et~al.}(2018){Drozdovskaya}, {van Dishoeck}, {J{\o}rgensen}, {Calmonte}, {van der Wiel}, {Coutens}, {Calcutt}, {M{\"u}ller}, {Bjerkeli}, {Persson}, {Wampfler}, \& {Altwegg}}]{Drozdovskaya2018}
{Drozdovskaya}, M.~N., {van Dishoeck}, E.~F., {J{\o}rgensen}, J.~K., {et~al.} 2018, \mnras, 476, 4949

\bibitem[{{Dutrey} {et~al.}(1997){Dutrey}, {Guilloteau}, \& {Bachiller}}]{Dutrey1997}
{Dutrey}, A., {Guilloteau}, S., \& {Bachiller}, R. 1997, \aap, 325, 758

\bibitem[{{el Akel} {et~al.}(2022){el Akel}, {Kristensen}, {Le Gal}, {van der Walt}, {Pitts}, \& {Dulieu}}]{elakel2022}
{el Akel}, M., {Kristensen}, L.~E., {Le Gal}, R., {et~al.} 2022, \aap, 659, A100

\bibitem[{{Enoch} {et~al.}(2009){Enoch}, {Evans}, {Sargent}, \& {Glenn}}]{Enoch2009}
{Enoch}, M.~L., {Evans}, Neal~J., I., {Sargent}, A.~I., \& {Glenn}, J. 2009, \apj, 692, 973

\bibitem[{{Esplugues} {et~al.}(2013){Esplugues}, {Tercero}, {Cernicharo}, {Goicoechea}, {Palau}, {Marcelino}, \& {Bell}}]{Esplugues2013}
{Esplugues}, G.~B., {Tercero}, B., {Cernicharo}, J., {et~al.} 2013, \aap, 556, A143

\bibitem[{{Fiorellino} {et~al.}(2021){Fiorellino}, {Manara}, {Nisini}, {Ramsay}, {Antoniucci}, {Giannini}, {Biazzo}, {Alcal{\`a}}, \& {Fedele}}]{Fiorellino2021}
{Fiorellino}, E., {Manara}, C.~F., {Nisini}, B., {et~al.} 2021, \aap, 650, A43

\bibitem[{{Froebrich}(2005)}]{Froebrich2005}
{Froebrich}, D. 2005, \apjs, 156, 169

\bibitem[{{Fuente} {et~al.}(2016){Fuente}, {Cernicharo}, {Roueff}, {Gerin}, {Pety}, {Marcelino}, {Bachiller}, {Lefloch}, {Roncero}, \& {Aguado}}]{Fuente2016}
{Fuente}, A., {Cernicharo}, J., {Roueff}, E., {et~al.} 2016, \aap, 593, A94

\bibitem[{{Fuente} {et~al.}(2017{\natexlab{a}}){Fuente}, {Gerin}, {Pety}, {Commer{\c{c}}on}, {Ag{\'u}ndez}, {Cernicharo}, {Marcelino}, {Roueff}, {Lis}, \& {Wootten}}]{Fuente2017}
{Fuente}, A., {Gerin}, M., {Pety}, J., {et~al.} 2017{\natexlab{a}}, \aap, 606, L3

\bibitem[{{Fuente} {et~al.}(2017{\natexlab{b}}){Fuente}, {Goicoechea}, {Pety}, {Le Gal}, {Mart{\'\i}n-Dom{\'e}nech}, {Gratier}, {Guzm{\'a}n}, {Roueff}, {Loison}, {Mu{\~n}oz Caro}, {Wakelam}, {Gerin}, {Riviere-Marichalar}, \& {Vidal}}]{Fuente2017-s2h}
{Fuente}, A., {Goicoechea}, J.~R., {Pety}, J., {et~al.} 2017{\natexlab{b}}, \apjl, 851, L49

\bibitem[{{Fuente} {et~al.}(2023){Fuente}, {Rivi{\`e}re-Marichalar}, {Beitia-Antero}, {Caselli}, {Wakelam}, {Esplugues}, {Rodr{\'\i}guez-Baras}, {Navarro-Almaida}, {Gerin}, {Kramer}, {Bachiller}, {Goicoechea}, {Jim{\'e}nez-Serra}, {Loison}, {Ivlev}, {Mart{\'\i}n-Dom{\'e}nech}, {Spezzano}, {Roncero}, {Mu{\~n}oz-Caro}, {Cazaux}, \& {Marcelino}}]{Fuente2023}
{Fuente}, A., {Rivi{\`e}re-Marichalar}, P., {Beitia-Antero}, L., {et~al.} 2023, \aap, 670, A114

\bibitem[{{Fuente} {et~al.}(2024){Fuente}, {Roueff}, {Le Petit}, {Le Bourlot}, {Bron}, {Wolfire}, {Babb}, {Yan}, {Onaka}, {Black}, {Schroetter}, {Van De Putte}, {Sidhu}, {Canin}, {Trahin}, {Alarc{\'o}n}, {Chown}, {Kannavou}, {Bern{\'e}}, {Habart}, {Peeters}, {Goicoechea}, {Zannese}, {Meshaka}, {Okada}, {R{\"o}llig}, {Le Gal}, {Sales}, {Palumbo}, {Baratta}, {Madden}, {Neelamkodan}, {Zhang}, \& {Stancil}}]{Fuente2024}
{Fuente}, A., {Roueff}, E., {Le Petit}, F., {et~al.} 2024, \aap, 687, A87

\bibitem[{{Geballe} {et~al.}(1985){Geballe}, {Baas}, {Greenberg}, \& {Schutte}}]{Geballe1985}
{Geballe}, T.~R., {Baas}, F., {Greenberg}, J.~M., \& {Schutte}, W. 1985, \aap, 146, L6

\bibitem[{{Gerin} {et~al.}(2017){Gerin}, {Pety}, {Commer{\c{c}}on}, {Fuente}, {Cernicharo}, {Marcelino}, {Ciardi}, {Lis}, {Roueff}, {Wootten}, \& {Chapillon}}]{Gerin2017}
{Gerin}, M., {Pety}, J., {Commer{\c{c}}on}, B., {et~al.} 2017, \aap, 606, A35

\bibitem[{{Gerin} {et~al.}(2015){Gerin}, {Pety}, {Fuente}, {Cernicharo}, {Commer{\c{c}}on}, \& {Marcelino}}]{Gerin2015}
{Gerin}, M., {Pety}, J., {Fuente}, A., {et~al.} 2015, \aap, 577, L2

\bibitem[{{Gieser} {et~al.}(2024){Gieser}, {Pineda}, {Segura-Cox}, {Caselli}, {Valdivia-Mena}, {Maureira}, {Hsieh}, {Busch}, {Bouscasse}, {Lopez-Sepulcre}, {Neri}, {Kuffmeier}, {Henning}, {Semenov}, {Cunningham}, \& {Jimenez-Serra}}]{Gieser2024}
{Gieser}, C., {Pineda}, J.~E., {Segura-Cox}, D.~M., {et~al.} 2024, \aap, 692, A55

\bibitem[{{Goicoechea} \& {Cuadrado}(2021)}]{Goicoechea2021}
{Goicoechea}, J.~R. \& {Cuadrado}, S. 2021, \aap, 647, L7

\bibitem[{{Goldsmith} \& {Langer}(1999)}]{Goldsmith1999}
{Goldsmith}, P.~F. \& {Langer}, W.~D. 1999, \apj, 517, 209

\bibitem[{{Gratier} {et~al.}(2016){Gratier}, {Majumdar}, {Ohishi}, {Roueff}, {Loison}, {Hickson}, \& {Wakelam}}]{Gratier2016}
{Gratier}, P., {Majumdar}, L., {Ohishi}, M., {et~al.} 2016, \apjs, 225, 25

\bibitem[{{Grossman} {et~al.}(1987){Grossman}, {Masson}, {Sargent}, {Scoville}, {Scott}, \& {Woody}}]{Grossman1987}
{Grossman}, E.~N., {Masson}, C.~R., {Sargent}, A.~I., {et~al.} 1987, \apj, 320, 356

\bibitem[{{Gueth} \& {Guilloteau}(1999)}]{Gueth1999}
{Gueth}, F. \& {Guilloteau}, S. 1999, \aap, 343, 571

\bibitem[{{Hatchell} \& {Dunham}(2009)}]{Hatchell2009}
{Hatchell}, J. \& {Dunham}, M.~M. 2009, \aap, 502, 139

\bibitem[{{Hatchell} \& {Fuller}(2008)}]{Hatchell2008}
{Hatchell}, J. \& {Fuller}, G.~A. 2008, \aap, 482, 855

\bibitem[{{Hatchell} {et~al.}(2013){Hatchell}, {Wilson}, {Drabek}, {Curtis}, {Richer}, {Nutter}, {Di Francesco}, {Ward-Thompson}, \& {JCMT GBS Consortium}}]{Hatchell2013}
{Hatchell}, J., {Wilson}, T., {Drabek}, E., {et~al.} 2013, \mnras, 429, L10

\bibitem[{{Hennebelle} {et~al.}(2016){Hennebelle}, {Commer{\c{c}}on}, {Chabrier}, \& {Marchand}}]{Hennebelle2016}
{Hennebelle}, P., {Commer{\c{c}}on}, B., {Chabrier}, G., \& {Marchand}, P. 2016, \apjl, 830, L8

\bibitem[{{Herczeg} {et~al.}(2012){Herczeg}, {Karska}, {Bruderer}, {Kristensen}, {van Dishoeck}, {J{\o}rgensen}, {Visser}, {Wampfler}, {Bergin}, {Y{\i}ld{\i}z}, {Pontoppidan}, \& {Gracia-Carpio}}]{Herczeg2012}
{Herczeg}, G.~J., {Karska}, A., {Bruderer}, S., {et~al.} 2012, \aap, 540, A84

\bibitem[{{HGBS team}(2020)}]{HGBS2020}
{HGBS team}. 2020, Herschel Gould Belt Survey, 10.26131/IRSA72

\bibitem[{{Hily-Blant} {et~al.}(2022){Hily-Blant}, {Pineau des For{\^e}ts}, {Faure}, \& {Lique}}]{Hily-Blant2022}
{Hily-Blant}, P., {Pineau des For{\^e}ts}, G., {Faure}, A., \& {Lique}, F. 2022, \aap, 658, A168

\bibitem[{{Hirano} {et~al.}(2010){Hirano}, {Ho}, {Liu}, {Shang}, {Lee}, \& {Bourke}}]{Hirano2010}
{Hirano}, N., {Ho}, P. P.~T., {Liu}, S.-Y., {et~al.} 2010, \apj, 717, 58

\bibitem[{{Hirano} {et~al.}(1999){Hirano}, {Kamazaki}, {Mikami}, {Ohashi}, \& {Umemoto}}]{Hirano1999}
{Hirano}, N., {Kamazaki}, T., {Mikami}, H., {Ohashi}, N., \& {Umemoto}, T. 1999, in Star Formation 1999, ed. T.~{Nakamoto}, 181--182

\bibitem[{{Hirano} \& {Liu}(2014)}]{Hirano2014}
{Hirano}, N. \& {Liu}, F.-c. 2014, \apj, 789, 50

\bibitem[{{Hirota} {et~al.}(2008){Hirota}, {Bushimata}, {Choi}, {Honma}, {Imai}, {Iwadate}, {Jike}, {Kameya}, {Kamohara}, {Kan-Ya}, {Kawaguchi}, {Kijima}, {Kobayashi}, {Kuji}, {Kurayama}, {Manabe}, {Miyaji}, {Nagayama}, {Nakagawa}, {Oh}, {Omodaka}, {Oyama}, {Sakai}, {Sasao}, {Sato}, {Shibata}, {Tamura}, \& {Yamashita}}]{Hirota2008}
{Hirota}, T., {Bushimata}, T., {Choi}, Y.~K., {et~al.} 2008, \pasj, 60, 37

\bibitem[{{Holdship} {et~al.}(2016){Holdship}, {Viti}, {Jimenez-Serra}, {Lefloch}, {Codella}, {Podio}, {Benedettini}, {Fontani}, {Bachiller}, {Tafalla}, \& {Ceccarelli}}]{Holdship2016}
{Holdship}, J., {Viti}, S., {Jimenez-Serra}, I., {et~al.} 2016, \mnras, 463, 802

\bibitem[{{Hsieh} {et~al.}(2024){Hsieh}, {Pineda}, {Segura-Cox}, {Caselli}, {Valdivia-Mena}, {Gieser}, {Maureira}, {Lopez-Sepulcre}, {Bouscasse}, {Neri}, {M{\"o}ller}, {Dutrey}, {Fuente}, {Semenov}, {Chapillon}, {Cunningham}, {Henning}, {Pi{\'e}tu}, {Jimenez-Serra}, {Marino}, \& {Ceccarelli}}]{Hsieh2024}
{Hsieh}, T.~H., {Pineda}, J.~E., {Segura-Cox}, D.~M., {et~al.} 2024, \aap, 686, A289

\bibitem[{{Hsieh} {et~al.}(2023){Hsieh}, {Segura-Cox}, {Pineda}, {Caselli}, {Bouscasse}, {Neri}, {Lopez-Sepulcre}, {Valdivia-Mena}, {Maureira}, {Henning}, {Smirnov-Pinchukov}, {Semenov}, {M{\"o}ller}, {Cunningham}, {Fuente}, {Marino}, {Dutrey}, {Tafalla}, {Chapillon}, {Ceccarelli}, \& {Zhao}}]{Hsieh2023}
{Hsieh}, T.~H., {Segura-Cox}, D.~M., {Pineda}, J.~E., {et~al.} 2023, \aap, 669, A137

\bibitem[{{Jennings} {et~al.}(1987){Jennings}, {Cameron}, {Cudlip}, \& {Hirst}}]{Jennings1987}
{Jennings}, R.~E., {Cameron}, D.~H.~M., {Cudlip}, W., \& {Hirst}, C.~J. 1987, \mnras, 226, 461

\bibitem[{{Jim{\'e}nez-Escobar} \& {Mu{\~n}oz Caro}(2011)}]{Jimenez-Escobar2011}
{Jim{\'e}nez-Escobar}, A. \& {Mu{\~n}oz Caro}, G.~M. 2011, \aap, 536, A91

\bibitem[{{Jim{\'e}nez-Escobar} {et~al.}(2012){Jim{\'e}nez-Escobar}, {Mu{\~n}oz Caro}, {Ciaravella}, {Cecchi-Pestellini}, {Candia}, \& {Micela}}]{Jimenez-Escobar2012}
{Jim{\'e}nez-Escobar}, A., {Mu{\~n}oz Caro}, G.~M., {Ciaravella}, A., {et~al.} 2012, \apjl, 751, L40

\bibitem[{{Jim{\'e}nez-Serra} {et~al.}(2008){Jim{\'e}nez-Serra}, {Caselli}, {Mart{\'\i}n-Pintado}, \& {Hartquist}}]{Jimenez-Serra2008}
{Jim{\'e}nez-Serra}, I., {Caselli}, P., {Mart{\'\i}n-Pintado}, J., \& {Hartquist}, T.~W. 2008, \aap, 482, 549

\bibitem[{{J{\o}rgensen} {et~al.}(2020){J{\o}rgensen}, {Belloche}, \& {Garrod}}]{Jorgensen2020}
{J{\o}rgensen}, J.~K., {Belloche}, A., \& {Garrod}, R.~T. 2020, \araa, 58, 727

\bibitem[{{J{\o}rgensen} {et~al.}(2005){J{\o}rgensen}, {Bourke}, {Myers}, {Sch{\"o}ier}, {van Dishoeck}, \& {Wilner}}]{Jorgensen2005}
{J{\o}rgensen}, J.~K., {Bourke}, T.~L., {Myers}, P.~C., {et~al.} 2005, \apj, 632, 973

\bibitem[{{J{\o}rgensen} {et~al.}(2006){J{\o}rgensen}, {Harvey}, {Evans}, {Huard}, {Allen}, {Porras}, {Blake}, {Bourke}, {Chapman}, {Cieza}, {Koerner}, {Lai}, {Mundy}, {Myers}, {Padgett}, {Rebull}, {Sargent}, {Spiesman}, {Stapelfeldt}, {van Dishoeck}, {Wahhaj}, \& {Young}}]{Jorgensen2006}
{J{\o}rgensen}, J.~K., {Harvey}, P.~M., {Evans}, Neal~J., I., {et~al.} 2006, \apj, 645, 1246

\bibitem[{{J{\o}rgensen} {et~al.}(2004){J{\o}rgensen}, {Hogerheijde}, {van Dishoeck}, {Blake}, \& {Sch{\"o}ier}}]{Jorgensen2004}
{J{\o}rgensen}, J.~K., {Hogerheijde}, M.~R., {van Dishoeck}, E.~F., {Blake}, G.~A., \& {Sch{\"o}ier}, F.~L. 2004, \aap, 413, 993

\bibitem[{{J{\o}rgensen} {et~al.}(2002){J{\o}rgensen}, {Sch{\"o}ier}, \& {van Dishoeck}}]{Jorgensen2002}
{J{\o}rgensen}, J.~K., {Sch{\"o}ier}, F.~L., \& {van Dishoeck}, E.~F. 2002, \aap, 389, 908

\bibitem[{{J{\o}rgensen} {et~al.}(2009){J{\o}rgensen}, {van Dishoeck}, {Visser}, {Bourke}, {Wilner}, {Lommen}, {Hogerheijde}, \& {Myers}}]{Jorgensen2009}
{J{\o}rgensen}, J.~K., {van Dishoeck}, E.~F., {Visser}, R., {et~al.} 2009, \aap, 507, 861

\bibitem[{{Kauffmann} {et~al.}(2008){Kauffmann}, {Bertoldi}, {Bourke}, {Evans}, \& {Lee}}]{Kauffmann2008}
{Kauffmann}, J., {Bertoldi}, F., {Bourke}, T.~L., {Evans}, N.~J., I., \& {Lee}, C.~W. 2008, \aap, 487, 993

\bibitem[{{Keller} {et~al.}(2002){Keller}, {Hony}, {Bradley}, {Molster}, {Waters}, {Bouwman}, {de Koter}, {Brownlee}, {Flynn}, {Henning}, \& {Mutschke}}]{Keller2002}
{Keller}, L.~P., {Hony}, S., {Bradley}, J.~P., {et~al.} 2002, \nat, 417, 148

\bibitem[{{Laas} \& {Caselli}(2019)}]{Laas2019}
{Laas}, J.~C. \& {Caselli}, P. 2019, \aap, 624, A108

\bibitem[{{Le Gal} {et~al.}(2019){Le Gal}, {{\"O}berg}, {Loomis}, {Pegues}, \& {Bergner}}]{LeGal2019}
{Le Gal}, R., {{\"O}berg}, K.~I., {Loomis}, R.~A., {Pegues}, J., \& {Bergner}, J.~B. 2019, \apj, 876, 72

\bibitem[{{Le Roy} {et~al.}(2015){Le Roy}, {Altwegg}, {Balsiger}, {Berthelier}, {Bieler}, {Briois}, {Calmonte}, {Combi}, {De Keyser}, {Dhooghe}, {Fiethe}, {Fuselier}, {Gasc}, {Gombosi}, {H{\"a}ssig}, {J{\"a}ckel}, {Rubin}, \& {Tzou}}]{LeRoy2015}
{Le Roy}, L., {Altwegg}, K., {Balsiger}, H., {et~al.} 2015, \aap, 583, A1

\bibitem[{{Lee} {et~al.}(2009){Lee}, {Hirano}, {Palau}, {Ho}, {Bourke}, {Zhang}, \& {Shang}}]{Lee2009}
{Lee}, C.-F., {Hirano}, N., {Palau}, A., {et~al.} 2009, \apj, 699, 1584

\bibitem[{{Lee} {et~al.}(2015){Lee}, {Dunham}, {Myers}, {Tobin}, {Kristensen}, {Pineda}, {Vorobyov}, {Offner}, {Arce}, {Li}, {Bourke}, {J{\o}rgensen}, {Goodman}, {Sadavoy}, {Chandler}, {Harris}, {Kratter}, {Looney}, {Melis}, {Perez}, \& {Segura-Cox}}]{Lee2015}
{Lee}, K.~I., {Dunham}, M.~M., {Myers}, P.~C., {et~al.} 2015, \apj, 814, 114

\bibitem[{{Lef{\`e}vre} {et~al.}(2017){Lef{\`e}vre}, {Cabrit}, {Maury}, {Gueth}, {Tabone}, {Podio}, {Belloche}, {Codella}, {Maret}, {Anderl}, {Andr{\'e}}, \& {Hennebelle}}]{Lefevre2017}
{Lef{\`e}vre}, C., {Cabrit}, S., {Maury}, A.~J., {et~al.} 2017, \aap, 604, L1

\bibitem[{{Lefloch} {et~al.}(1998){Lefloch}, {Castets}, {Cernicharo}, {Langer}, \& {Zylka}}]{Lefloch1998}
{Lefloch}, B., {Castets}, A., {Cernicharo}, J., {Langer}, W.~D., \& {Zylka}, R. 1998, \aap, 334, 269

\bibitem[{Leman {et~al.}(2004)Leman, Orgel, \& Ghadiri}]{Leman2004}
Leman, L.~J., Orgel, L.~E., \& Ghadiri, M.~R. 2004, Science, 306, 283

\bibitem[{{Lin} {et~al.}(2024){Lin}, {Yen}, \& {Lai}}]{Lin2024}
{Lin}, S.-J., {Yen}, H.-W., \& {Lai}, S.-P. 2024, \aj, 168, 107

\bibitem[{{Looney} {et~al.}(2000){Looney}, {Mundy}, \& {Welch}}]{Looney2000}
{Looney}, L.~W., {Mundy}, L.~G., \& {Welch}, W.~J. 2000, \apj, 529, 477

\bibitem[{{Machida} {et~al.}(2006){Machida}, {Inutsuka}, \& {Matsumoto}}]{Machida2006}
{Machida}, M.~N., {Inutsuka}, S.-i., \& {Matsumoto}, T. 2006, \apjl, 647, L151

\bibitem[{{Marcelino} {et~al.}(2018){Marcelino}, {Gerin}, {Cernicharo}, {Fuente}, {Wootten}, {Chapillon}, {Pety}, {Lis}, {Roueff}, {Commer{\c{c}}on}, \& {Ciardi}}]{Marcelino2018}
{Marcelino}, N., {Gerin}, M., {Cernicharo}, J., {et~al.} 2018, \aap, 620, A80

\bibitem[{{Mart{\'\i}n-Dom{\'e}nech} {et~al.}(2021){Mart{\'\i}n-Dom{\'e}nech}, {Bergner}, {{\"O}berg}, {Carpenter}, {Law}, {Huang}, {J{\o}rgensen}, {Schwarz}, \& {Wilner}}]{MartinDomenech2021}
{Mart{\'\i}n-Dom{\'e}nech}, R., {Bergner}, J.~B., {{\"O}berg}, K.~I., {et~al.} 2021, \apj, 923, 155

\bibitem[{{Mart{\'\i}n-Dom{\'e}nech} {et~al.}(2024){Mart{\'\i}n-Dom{\'e}nech}, {{\"O}berg}, {Mu{\~n}oz Caro}, {Carrascosa}, {Fuente}, \& {Rajappan}}]{Martin2024}
{Mart{\'\i}n-Dom{\'e}nech}, R., {{\"O}berg}, K.~I., {Mu{\~n}oz Caro}, G.~M., {et~al.} 2024, \mnras, 535, 807

\bibitem[{{Maury} {et~al.}(2014){Maury}, {Belloche}, {Andr{\'e}}, {Maret}, {Gueth}, {Codella}, {Cabrit}, {Testi}, \& {Bontemps}}]{Maury2014}
{Maury}, A.~J., {Belloche}, A., {Andr{\'e}}, P., {et~al.} 2014, \aap, 563, L2

\bibitem[{{McClure} {et~al.}(2023){McClure}, {Rocha}, {Pontoppidan}, {Crouzet}, {Chu}, {Dartois}, {Lamberts}, {Noble}, {Pendleton}, {Perotti}, {Qasim}, {Rachid}, {Smith}, {Sun}, {Beck}, {Boogert}, {Brown}, {Caselli}, {Charnley}, {Cuppen}, {Dickinson}, {Drozdovskaya}, {Egami}, {Erkal}, {Fraser}, {Garrod}, {Harsono}, {Ioppolo}, {Jim{\'e}nez-Serra}, {Jin}, {J{\o}rgensen}, {Kristensen}, {Lis}, {McCoustra}, {McGuire}, {Melnick}, {{\~A}-berg}, {Palumbo}, {Shimonishi}, {Sturm}, {van Dishoeck}, \& {Linnartz}}]{McClure2023}
{McClure}, M.~K., {Rocha}, W.~R.~M., {Pontoppidan}, K.~M., {et~al.} 2023, Nature Astronomy, 7, 431

\bibitem[{{McGuire}(2022)}]{McGuire2022}
{McGuire}, B.~A. 2022, \apjs, 259, 30

\bibitem[{{Mercimek} {et~al.}(2022){Mercimek}, {Codella}, {Podio}, {Bianchi}, {Chahine}, {Bouvier}, {L{\'o}pez-Sepulcre}, {Neri}, \& {Ceccarelli}}]{Mercimek2022}
{Mercimek}, S., {Codella}, C., {Podio}, L., {et~al.} 2022, \aap, 659, A67

\bibitem[{{M{\"u}ller} {et~al.}(2005){M{\"u}ller}, {Schl{\"o}der}, {Stutzki}, \& {Winnewisser}}]{Muller2005}
{M{\"u}ller}, H. S.~P., {Schl{\"o}der}, F., {Stutzki}, J., \& {Winnewisser}, G. 2005, Journal of Molecular Structure, 742, 215

\bibitem[{{Navarro-Almaida} {et~al.}(2020){Navarro-Almaida}, {Le Gal}, {Fuente}, {Rivi{\`e}re-Marichalar}, {Wakelam}, {Cazaux}, {Caselli}, {Laas}, {Alonso-Albi}, {Loison}, {Gerin}, {Kramer}, {Roueff}, {Bachiller}, {Commer{\c{c}}on}, {Friesen}, {Garc{\'\i}a-Burillo}, {Goicoechea}, {Giuliano}, {Jim{\'e}nez-Serra}, {Kirk}, {Lattanzi}, {Malinen}, {Marcelino}, {Mart{\'\i}n-Dom{\`e}nech}, {Mu{\~n}oz Caro}, {Pineda}, {Tercero}, {Trevi{\~n}o-Morales}, {Roncero}, {Hacar}, {Tafalla}, \& {Ward-Thompson}}]{NavarroAlmaida2020}
{Navarro-Almaida}, D., {Le Gal}, R., {Fuente}, A., {et~al.} 2020, \aap, 637, A39

\bibitem[{{Navarro-Almaida} {et~al.}(2024){Navarro-Almaida}, {Lebreuilly}, {Hennebelle}, {Fuente}, {Commer{\c{c}}on}, {Le Gal}, {Wakelam}, {Gerin}, {Rivi{\'e}re-Marichalar}, {Beitia-Antero}, \& {Ascasibar}}]{NavarroAlmaida2024}
{Navarro-Almaida}, D., {Lebreuilly}, U., {Hennebelle}, P., {et~al.} 2024, \aap, 685, A112

\bibitem[{{Neufeld} {et~al.}(2015){Neufeld}, {Godard}, {Gerin}, {Pineau des For{\^e}ts}, {Bernier}, {Falgarone}, {Graf}, {G{\"u}sten}, {Herbst}, {Lesaffre}, {Schilke}, {Sonnentrucker}, \& {Wiesemeyer}}]{Neufeld2015}
{Neufeld}, D.~A., {Godard}, B., {Gerin}, M., {et~al.} 2015, \aap, 577, A49

\bibitem[{Olson \& Straub(2016)}]{Olson2016}
Olson, K.~R. \& Straub, K.~D. 2016, Physiology, 31, 60, pMID: 26674552

\bibitem[{{Ossenkopf} \& {Henning}(1994)}]{OssenkopfHenning1994}
{Ossenkopf}, V. \& {Henning}, T. 1994, \aap, 291, 943

\bibitem[{{Palau} {et~al.}(2014){Palau}, {Zapata}, {Rodr{\'\i}guez}, {Bouy}, {Barrado}, {Morales-Calder{\'o}n}, {Myers}, {Chapman}, {Ju{\'a}rez}, \& {Li}}]{Palau2014}
{Palau}, A., {Zapata}, L.~A., {Rodr{\'\i}guez}, L.~F., {et~al.} 2014, \mnras, 444, 833

\bibitem[{{Palumbo} {et~al.}(1997){Palumbo}, {Geballe}, \& {Tielens}}]{Palumbo1997}
{Palumbo}, M.~E., {Geballe}, T.~R., \& {Tielens}, A.~G.~G.~M. 1997, \apj, 479, 839

\bibitem[{{Palumbo} {et~al.}(1995){Palumbo}, {Tielens}, \& {Tokunaga}}]{Palumbo1995}
{Palumbo}, M.~E., {Tielens}, A.~G.~G.~M., \& {Tokunaga}, A.~T. 1995, \apj, 449, 674

\bibitem[{{Perrero} {et~al.}(2024){Perrero}, {Beitia-Antero}, {Fuente}, {Ugliengo}, \& {Rimola}}]{Perrero2024}
{Perrero}, J., {Beitia-Antero}, L., {Fuente}, A., {Ugliengo}, P., \& {Rimola}, A. 2024, \apj, 971, 36

\bibitem[{{Pineda} {et~al.}(2022){Pineda}, {Harju}, {Caselli}, {Sipil{\"a}}, {Juvela}, {Vastel}, {Rosolowsky}, {Burkert}, {Friesen}, {Shirley}, {Maureira}, {Choudhury}, {Segura-Cox}, {G{\"u}sten}, {Punanova}, {Bizzocchi}, \& {Goodman}}]{Pineda2022}
{Pineda}, J.~E., {Harju}, J., {Caselli}, P., {et~al.} 2022, \aj, 163, 294

\bibitem[{{Pineda} {et~al.}(2020){Pineda}, {Segura-Cox}, {Caselli}, {Cunningham}, {Zhao}, {Schmiedeke}, {Maureira}, \& {Neri}}]{Pineda2020}
{Pineda}, J.~E., {Segura-Cox}, D., {Caselli}, P., {et~al.} 2020, Nature Astronomy, 4, 1158

\bibitem[{{Ray} {et~al.}(2023){Ray}, {McCaughrean}, {Caratti o Garatti}, {Kavanagh}, {Justtanont}, {van Dishoeck}, {Reitsma}, {Beuther}, {Francis}, {Gieser}, {Klaassen}, {Perotti}, {Tychoniec}, {van Gelder}, {Colina}, {Greve}, {G{\"u}del}, {Henning}, {Lagage}, {{\"O}stlin}, {Vandenbussche}, {Waelkens}, \& {Wright}}]{Ray2023}
{Ray}, T.~P., {McCaughrean}, M.~J., {Caratti o Garatti}, A., {et~al.} 2023, \nat, 622, 48

\bibitem[{{Reipurth} {et~al.}(1993){Reipurth}, {Heathcote}, {Roth}, {Noriega-Crespo}, \& {Raga}}]{Reipurth1993}
{Reipurth}, B., {Heathcote}, S., {Roth}, M., {Noriega-Crespo}, A., \& {Raga}, A.~C. 1993, \apjl, 408, L49

\bibitem[{{Reynolds} {et~al.}(2021){Reynolds}, {Tobin}, {Sheehan}, {Sadavoy}, {Kratter}, {Li}, {Chandler}, {Segura-Cox}, {Looney}, \& {Dunham}}]{Reynolds2021}
{Reynolds}, N.~K., {Tobin}, J.~J., {Sheehan}, P., {et~al.} 2021, \apjl, 907, L10

\bibitem[{{Rivi{\`e}re-Marichalar} {et~al.}(2020){Rivi{\`e}re-Marichalar}, {Fuente}, {Le Gal}, {Baruteau}, {Neri}, {Navarro-Almaida}, {Trevi{\~n}o-Morales}, {Mac{\'\i}as}, {Bachiller}, \& {Osorio}}]{Riviere-Marichalar2020}
{Rivi{\`e}re-Marichalar}, P., {Fuente}, A., {Le Gal}, R., {et~al.} 2020, \aap, 642, A32

\bibitem[{{Rocha} {et~al.}(2024){Rocha}, {van Dishoeck}, {Ressler}, {van Gelder}, {Slavicinska}, {Brunken}, {Linnartz}, {Ray}, {Beuther}, {Caratti o Garatti}, {Geers}, {Kavanagh}, {Klaassen}, {Justtanont}, {Chen}, {Francis}, {Gieser}, {Perotti}, {Tychoniec}, {Barsony}, {Majumdar}, {le Gouellec}, {Chu}, {Lew}, {Henning}, \& {Wright}}]{Rocha2024}
{Rocha}, W.~R.~M., {van Dishoeck}, E.~F., {Ressler}, M.~E., {et~al.} 2024, \aap, 683, A124

\bibitem[{{Rodr{\'\i}guez} {et~al.}(2014){Rodr{\'\i}guez}, {Zapata}, \& {Palau}}]{Rodriguez2014}
{Rodr{\'\i}guez}, L.~F., {Zapata}, L.~A., \& {Palau}, A. 2014, \apj, 790, 80

\bibitem[{{Rodr{\'\i}guez-Baras} {et~al.}(2021){Rodr{\'\i}guez-Baras}, {Fuente}, {Rivi{\'e}re-Marichalar}, {Navarro-Almaida}, {Caselli}, {Gerin}, {Kramer}, {Roueff}, {Wakelam}, {Esplugues}, {Garc{\'\i}a-Burillo}, {Le Gal}, {Spezzano}, {Alonso-Albi}, {Bachiller}, {Cazaux}, {Commercon}, {Goicoechea}, {Loison}, {Trevi{\~n}o-Morales}, {Roncero}, {Jim{\'e}nez-Serra}, {Laas}, {Hacar}, {Kirk}, {Lattanzi}, {Mart{\'\i}n-Dom{\'e}nech}, {Mu{\~n}oz-Caro}, {Pineda}, {Tercero}, {Ward-Thompson}, {Tafalla}, {Marcelino}, {Malinen}, {Friesen}, \& {Giuliano}}]{Rodriguez-Baras2021}
{Rodr{\'\i}guez-Baras}, M., {Fuente}, A., {Rivi{\'e}re-Marichalar}, P., {et~al.} 2021, \aap, 648, A120

\bibitem[{{Ruaud} {et~al.}(2016){Ruaud}, {Wakelam}, \& {Hersant}}]{Ruaud2016}
{Ruaud}, M., {Wakelam}, V., \& {Hersant}, F. 2016, \mnras, 459, 3756

\bibitem[{{Ruffle} {et~al.}(1999){Ruffle}, {Hartquist}, {Caselli}, \& {Williams}}]{Ruffle1999}
{Ruffle}, D.~P., {Hartquist}, T.~W., {Caselli}, P., \& {Williams}, D.~A. 1999, \mnras, 306, 691

\bibitem[{{Sadavoy} {et~al.}(2014){Sadavoy}, {Di Francesco}, {Andr{\'e}}, {Pezzuto}, {Bernard}, {Maury}, {Men'shchikov}, {Motte}, {Nguyen-Lu'o'ng}, {Schneider}, {Arzoumanian}, {Benedettini}, {Bontemps}, {Elia}, {Hennemann}, {Hill}, {K{\"o}nyves}, {Louvet}, {Peretto}, {Roy}, \& {White}}]{Sadavoy2014}
{Sadavoy}, S.~I., {Di Francesco}, J., {Andr{\'e}}, P., {et~al.} 2014, \apjl, 787, L18

\bibitem[{{Sandell} {et~al.}(1991){Sandell}, {Aspin}, {Duncan}, {Russell}, \& {Robson}}]{Sandell1991}
{Sandell}, G., {Aspin}, C., {Duncan}, W.~D., {Russell}, A. P.~G., \& {Robson}, E.~I. 1991, \apjl, 376, L17

\bibitem[{{Santos} {et~al.}(2024){Santos}, {van Gelder}, {Nazari}, {Ahmadi}, \& {van Dishoeck}}]{Santos2024}
{Santos}, J.~C., {van Gelder}, M.~L., {Nazari}, P., {Ahmadi}, A., \& {van Dishoeck}, E.~F. 2024, \aap, 689, A248

\bibitem[{{Schilke} {et~al.}(1997){Schilke}, {Walmsley}, {Pineau des Forets}, \& {Flower}}]{Schilke1997}
{Schilke}, P., {Walmsley}, C.~M., {Pineau des Forets}, G., \& {Flower}, D.~R. 1997, \aap, 321, 293

\bibitem[{{Semenov} {et~al.}(2018){Semenov}, {Favre}, {Fedele}, {Guilloteau}, {Teague}, {Henning}, {Dutrey}, {Chapillon}, {Hersant}, \& {Pi{\'e}tu}}]{Semenov2018}
{Semenov}, D., {Favre}, C., {Fedele}, D., {et~al.} 2018, \aap, 617, A28

\bibitem[{{Shingledecker} {et~al.}(2020){Shingledecker}, {Lamberts}, {Laas}, {Vasyunin}, {Herbst}, {K{\"a}stner}, \& {Caselli}}]{Shingledecker2020}
{Shingledecker}, C.~N., {Lamberts}, T., {Laas}, J.~C., {et~al.} 2020, \apj, 888, 52

\bibitem[{{Stephens} {et~al.}(2019){Stephens}, {Bourke}, {Dunham}, {Myers}, {Pokhrel}, {Tobin}, {Arce}, {Sadavoy}, {Vorobyov}, {Pineda}, {Offner}, {Lee}, {Kristensen}, {J{\o}rgensen}, {Gurwell}, \& {Goodman}}]{Stephens2019}
{Stephens}, I.~W., {Bourke}, T.~L., {Dunham}, M.~M., {et~al.} 2019, \apjs, 245, 21

\bibitem[{{Taillard} {et~al.}(2025){Taillard}, {Mart{\'\i}n-Dom{\'e}nech}, {Carrascosa}, {Noble}, {Mu{\~n}oz Caro}, {Dartois}, {Navarro-Almaida}, {Escribano}, {S{\'a}nchez-Monge}, \& {Fuente}}]{Taillard2025}
{Taillard}, A., {Mart{\'\i}n-Dom{\'e}nech}, R., {Carrascosa}, H., {et~al.} 2025, \aap, 694, A263

\bibitem[{{Taniguchi} {et~al.}(2024){Taniguchi}, {Pineda}, {Caselli}, {Shimoikura}, {Friesen}, {Segura-Cox}, \& {Schmiedeke}}]{Taniguchi2024}
{Taniguchi}, K., {Pineda}, J.~E., {Caselli}, P., {et~al.} 2024, \apj, 965, 162

\bibitem[{{Tobin} {et~al.}(2015{\natexlab{a}}){Tobin}, {Dunham}, {Looney}, {Li}, {Chandler}, {Segura-Cox}, {Sadavoy}, {Melis}, {Harris}, {Perez}, {Kratter}, {J{\o}rgensen}, {Plunkett}, \& {Hull}}]{Tobin2015}
{Tobin}, J.~J., {Dunham}, M.~M., {Looney}, L.~W., {et~al.} 2015{\natexlab{a}}, \apj, 798, 61

\bibitem[{{Tobin} {et~al.}(2016{\natexlab{a}}){Tobin}, {Kratter}, {Persson}, {Looney}, {Dunham}, {Segura-Cox}, {Li}, {Chandler}, {Sadavoy}, {Harris}, {Melis}, \& {P{\'e}rez}}]{Tobin2016-L1448IRS3B}
{Tobin}, J.~J., {Kratter}, K.~M., {Persson}, M.~V., {et~al.} 2016{\natexlab{a}}, \nat, 538, 483

\bibitem[{{Tobin} {et~al.}(2016{\natexlab{b}}){Tobin}, {Looney}, {Li}, {Chandler}, {Dunham}, {Segura-Cox}, {Sadavoy}, {Melis}, {Harris}, {Kratter}, \& {Perez}}]{Tobin2016}
{Tobin}, J.~J., {Looney}, L.~W., {Li}, Z.-Y., {et~al.} 2016{\natexlab{b}}, \apj, 818, 73

\bibitem[{{Tobin} {et~al.}(2018){Tobin}, {Looney}, {Li}, {Sadavoy}, {Dunham}, {Segura-Cox}, {Kratter}, {Chandler}, {Melis}, {Harris}, \& {Perez}}]{Tobin2018}
{Tobin}, J.~J., {Looney}, L.~W., {Li}, Z.-Y., {et~al.} 2018, \apj, 867, 43

\bibitem[{{Tobin} {et~al.}(2007){Tobin}, {Looney}, {Mundy}, {Kwon}, \& {Hamidouche}}]{Tobin2007}
{Tobin}, J.~J., {Looney}, L.~W., {Mundy}, L.~G., {Kwon}, W., \& {Hamidouche}, M. 2007, \apj, 659, 1404

\bibitem[{{Tobin} {et~al.}(2015{\natexlab{b}}){Tobin}, {Looney}, {Wilner}, {Kwon}, {Chandler}, {Bourke}, {Loinard}, {Chiang}, {Schnee}, \& {Chen}}]{Tobin2015-2}
{Tobin}, J.~J., {Looney}, L.~W., {Wilner}, D.~J., {et~al.} 2015{\natexlab{b}}, \apj, 805, 125

\bibitem[{{Tobin} \& {Sheehan}(2024)}]{Tobin2024}
{Tobin}, J.~J. \& {Sheehan}, P.~D. 2024, \araa, 62, 203

\bibitem[{{Valdivia-Mena} {et~al.}(2022){Valdivia-Mena}, {Pineda}, {Segura-Cox}, {Caselli}, {Neri}, {L{\'o}pez-Sepulcre}, {Cunningham}, {Bouscasse}, {Semenov}, {Henning}, {Pi{\'e}tu}, {Chapillon}, {Dutrey}, {Fuente}, {Guilloteau}, {Hsieh}, {Jim{\'e}nez-Serra}, {Marino}, {Maureira}, {Smirnov-Pinchukov}, {Tafalla}, \& {Zhao}}]{Valdivia-Mena2022}
{Valdivia-Mena}, M.~T., {Pineda}, J.~E., {Segura-Cox}, D.~M., {et~al.} 2022, \aap, 667, A12

\bibitem[{{Valdivia-Mena} {et~al.}(2023){Valdivia-Mena}, {Pineda}, {Segura-Cox}, {Caselli}, {Schmiedeke}, {Choudhury}, {Offner}, {Neri}, {Goodman}, \& {Fuller}}]{Valdivia-Mena2023}
{Valdivia-Mena}, M.~T., {Pineda}, J.~E., {Segura-Cox}, D.~M., {et~al.} 2023, \aap, 677, A92

\bibitem[{{van Gelder} {et~al.}(2022){van Gelder}, {Nazari}, {Tabone}, {Ahmadi}, {van Dishoeck}, {Beltr{\'a}n}, {Fuller}, {Sakai}, {S{\'a}nchez-Monge}, {Schilke}, {Yang}, \& {Zhang}}]{vanGelder2022}
{van Gelder}, M.~L., {Nazari}, P., {Tabone}, B., {et~al.} 2022, \aap, 662, A67

\bibitem[{{van Gelder} {et~al.}(2020){van Gelder}, {Tabone}, {Tychoniec}, {van Dishoeck}, {Beuther}, {Boogert}, {Caratti o Garatti}, {Klaassen}, {Linnartz}, {M{\"u}ller}, \& {Taquet}}]{vanGelder2020}
{van Gelder}, M.~L., {Tabone}, B., {Tychoniec}, {\L}., {et~al.} 2020, \aap, 639, A87

\bibitem[{{Vastel} {et~al.}(2018){Vastel}, {Qu{\'e}nard}, {Le Gal}, {Wakelam}, {Andrianasolo}, {Caselli}, {Vidal}, {Ceccarelli}, {Lefloch}, \& {Bachiller}}]{Vastel2018}
{Vastel}, C., {Qu{\'e}nard}, D., {Le Gal}, R., {et~al.} 2018, \mnras, 478, 5514

\bibitem[{{Vidal} \& {Wakelam}(2018)}]{Vidal2018}
{Vidal}, T. H.~G. \& {Wakelam}, V. 2018, \mnras, 474, 5575

\bibitem[{{Vitorino} {et~al.}(2024){Vitorino}, {Loison}, {Wakelam}, {Congiu}, \& {Dulieu}}]{Vitorino2024}
{Vitorino}, J., {Loison}, J.~C., {Wakelam}, V., {Congiu}, E., \& {Dulieu}, F. 2024, \mnras, 533, 52

\bibitem[{{Wakelam} {et~al.}(2024){Wakelam}, {Gratier}, {Loison}, {Hickson}, {Penguen}, \& {Mechineau}}]{Wakelam2024}
{Wakelam}, V., {Gratier}, P., {Loison}, J.~C., {et~al.} 2024, \aap, 689, A63

\bibitem[{{Wakelam} {et~al.}(2011){Wakelam}, {Hersant}, \& {Herpin}}]{Wakelam2011}
{Wakelam}, V., {Hersant}, F., \& {Herpin}, F. 2011, \aap, 529, A112

\bibitem[{{Wakelam} {et~al.}(2017){Wakelam}, {Loison}, {Mereau}, \& {Ruaud}}]{Wakelam2017}
{Wakelam}, V., {Loison}, J.~C., {Mereau}, R., \& {Ruaud}, M. 2017, Molecular Astrophysics, 6, 22

\bibitem[{{Walmsley} {et~al.}(2004){Walmsley}, {Flower}, \& {Pineau des For{\^e}ts}}]{Walmsley2004}
{Walmsley}, C.~M., {Flower}, D.~R., \& {Pineau des For{\^e}ts}, G. 2004, \aap, 418, 1035

\bibitem[{{Yang} {et~al.}(2021){Yang}, {Sakai}, {Zhang}, {Murillo}, {Zhang}, {Higuchi}, {Zeng}, {L{\'o}pez-Sepulcre}, {Yamamoto}, {Lefloch}, {Bouvier}, {Ceccarelli}, {Hirota}, {Imai}, {Oya}, {Sakai}, \& {Watanabe}}]{Yang2021}
{Yang}, Y.-L., {Sakai}, N., {Zhang}, Y., {et~al.} 2021, \apj, 910, 20

\bibitem[{{Zapata} {et~al.}(2014){Zapata}, {Arce}, {Brassfield}, {Palau}, {Patel}, \& {Pineda}}]{Zapata2014}
{Zapata}, L.~A., {Arce}, H.~G., {Brassfield}, E., {et~al.} 2014, \mnras, 441, 3696

\bibitem[{{Zhang} {et~al.}(2023){Zhang}, {Yang}, {Zhang}, {Cox}, {Zeng}, {Murillo}, {Ohashi}, \& {Sakai}}]{Zhang2023}
{Zhang}, Z.~E., {Yang}, Y.-l., {Zhang}, Y., {et~al.} 2023, \apj, 946, 113

\end{thebibliography}
\bibliographystyle{aa}

\appendix

\section{The H$_2^{33}$S isotopologue}
\label{sec:apendix_h233s}

H$_2$S is a very abundant molecule in the core of YSOs, showing high column densities too. Observations of the lower excitation temperature lines can result in optically thick lines that do not represent the actual H$_2$S budget, and, sometimes, the observation of the lesser abundant isotopologues is needed in order to achieve an accurate column density estimation.

\begin{figure}[h]
    \includegraphics[width=0.49\textwidth]{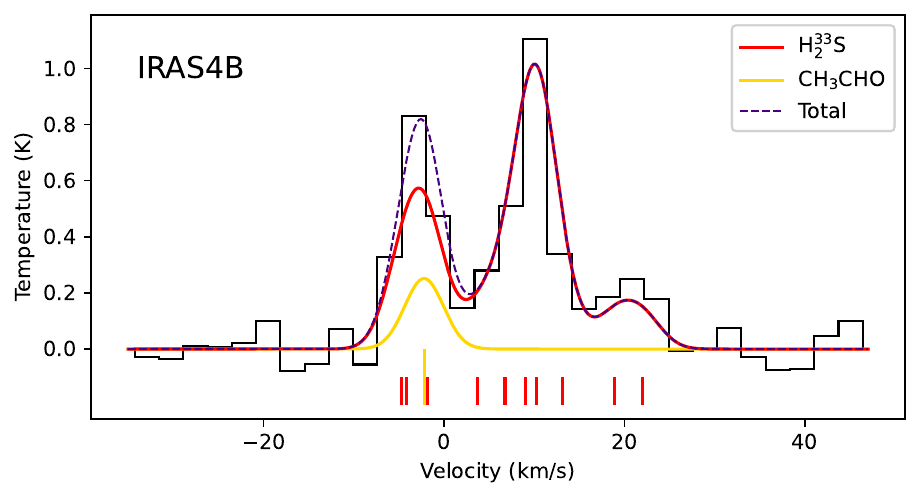}
    \caption{IRAS4B spectrum around the 215 GHz H$_2^{33}$S line. In black, the integrated emission in a 1.5$\arcsec$ circle around the protostar's position. In red, the fit to the H$_2^{33}$S emission. In yellow, the emission of CH$_3$CHO, assuming its column density is three times greater than that of H$_2^{33}$S. The purple dashed line represents the sum of the emission of both species. The different lines from Table~\ref{tab:h233s} appear marked in the bottom side of the figure with small vertical lines. For reference, the most intense H$_2^{33}$S line (the fourth starting from the left) is at a velocity of -3.1 km/s.}
    \label{fig:H233S}
\end{figure}

The H$_2^{33}$S molecule presents two important structural behaviors. On the one hand, the molecule has a para and an orto variants, with a distribution of 1:3 respectively, as in the case of H$_2$S. On the other hand, this species exhibits hyperfine structure, which is discernible at lower excitation levels. The transition observed in this work is actually a low-level transition ---H$_2^{33}$S: 2$_{2,0}$--2$_{1,1}$---, therefore, we took into account the line splitting when we calculated the column density. Table~\ref{tab:h233s} displays the hyperfine structure of the molecule for the 2$_{2,0}$--2$_{1,1}$ transition. Furthermore, there is an additional factor to take into account, as Acetaldehyde ---CH$_3$CHO--- has a relatively bright transition ($\sim$33\% of the maximum H$_2^{33}$S intensity) in the same range of frequencies. To deal with this part, we masked the CH$_3$CHO peak before the fitting with \texttt{CLASS} by fittin only the not contaminated hyperfine components, in order to avoid misrepresentations of the total emission or the opacity of the line. In Fig.~\ref{fig:H233S} we show the spectrum of IRAS4B in the spectral window of H$_2^{33}$S and CH$_3$CHO. When fitting a line with hyperfine splitting with \texttt{CLASS}, the program also estimates the opacity for the whole transition. The least value possible for the opacity is 0.1, so when the transition is not optically thick, it will return this result. The opacities estimated with \texttt{CLASS} were very low in all cases (0.1) except for IRAS2A and SVS13A, for which we obtained higher opacities. In these two cases, our estimates should be understood only as a lower limit to the real column density.

\setlength{\tabcolsep}{3.5pt}
\renewcommand{\arraystretch}{1.5}
\begin{table}[h]
\caption[H$_2^{33}$S 2$_{2,0}$--2$_{1,1}$ hyperfine structure and nearby CH$_3$CHO transition.]{H$_2^{33}$S 2$_{2,0}$--2$_{1,1}$ hyperfine structure and nearby CH$_3$CHO transition. Data was retrieved or calculated from the CDMS\tablefootnote{\url{https://cdms.astro.uni-koeln.de/}} database.}
\label{tab:h233s}
\centerline{
\begin{tabular}{lcccl}
\hline\hline
Molecule & $\nu$ (GHz) & Transition & I (log$_{10}$) & Norm. I \\ 
\hline
H$_2^{33}$S & 215.4945 & 2$_{2,0,1}$--2$_{1,1,2}$   & -3.7205 & 0.1458 \\
            & 215.4967 & 2$_{2,0,1}$--2$_{1,1,1}$   & -3.7205 & 0.1458 \\
            & 215.5008 & 2$_{2,0,4}$--2$_{1,1,3}$   & -3.6625 & 0.1666 \\
            & 215.5029 & 2$_{2,0,4}$--2$_{1,1,4}$   & -2.8843 & 1.0    \\
            & 215.5037 & 2$_{2,0,2}$--2$_{1,1,3}$   & -3.5743 & 0.2042 \\
            & 215.5054 & 2$_{2,0,2}$--2$_{1,1,2}$   & -3.5163 & 0.2333 \\
            & 215.5076 & 2$_{2,0,2}$--2$_{1,1,1}$   & -3.7204 & 0.1458 \\
            & 215.5116 & 2$_{2,0,3}$--2$_{1,1,3}$   & -3.1817 & 0.5042 \\
            & 215.5132 & 2$_{2,0.3}$--2$_{1,1,2}$   & -3.5743 & 0.2042 \\
            & 215.5136 & 2$_{2,0,3}$--2$_{1,1,4}$   & -3.6624 & 0.1666 \\
CH$_3$CHO   & 215.5118 & 11(2 9)6--10(2 8)6 & -3.9744 & 0.3251 \\
\hline 
\end{tabular}
}
\begin{flushleft}
    \tiny\textbf{Notes:} The normalised intensity is the result of dividing each individual intensity (without the log$_{10}$) by the more intense of the lines of the hyperfine splitting, which in this case is 2$_{2,0}$4--2$_{1,1}$4.
\end{flushleft}
\end{table}

\section{IRAS2A and IRAS4B extended emission integrated maps}

The following figures show the extended emission of IRAS2A and IRAS4B as a complement of Fig.~\ref{fig:moment-0-maps}. The extended emission is found in the north-south outflows direction, and is specially noticeable in IRAS4B.

\begin{figure}[h]
    \centering
    \includegraphics[width=\linewidth]{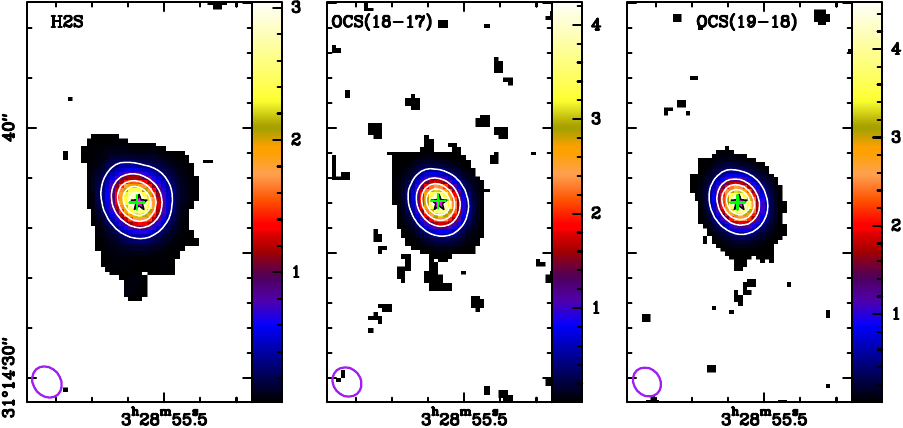}
    \caption{Integrated emission of H$_2$S and OCS in IRAS2A in a 8$\arcsec$$\times$16$\arcsec$ region. See Fig.~\ref{fig:moment-0-maps} caption for more details.}
    \label{fig:map_IRAS2A_BIG}
\end{figure}

\begin{figure}[h]
    \centering
    \includegraphics[width=\linewidth]{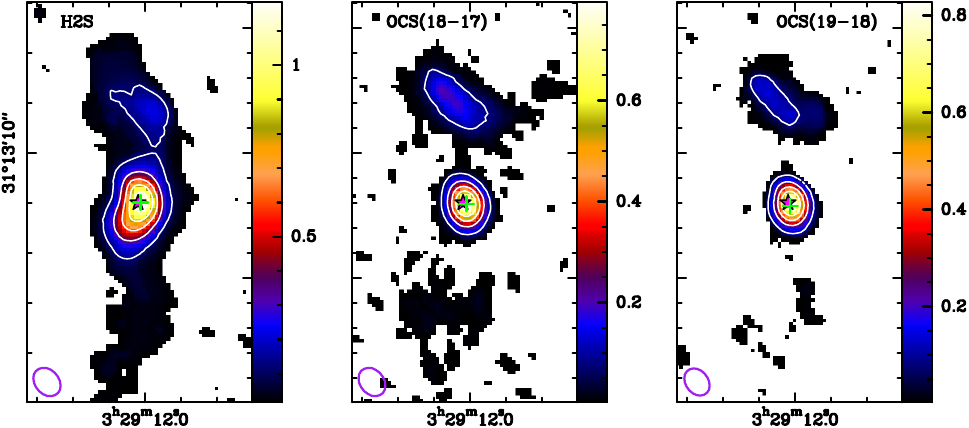}
    \caption{Integrated emission of H$_2$S and OCS in IRAS4B in a 8$\arcsec$$\times$16$\arcsec$ region. See Fig.~\ref{fig:moment-0-maps} caption for more details.}
    \label{fig:map_IRAS4B_BIG}
\end{figure}

\onecolumn
\section{H$_2$S opacity worst-case scenario}
\label{sec:appendix_mod-ratios}

\begin{figure*}[h]
\begin{subfigure}{0.95\textwidth}
    \centering
    \includegraphics[width=\linewidth]{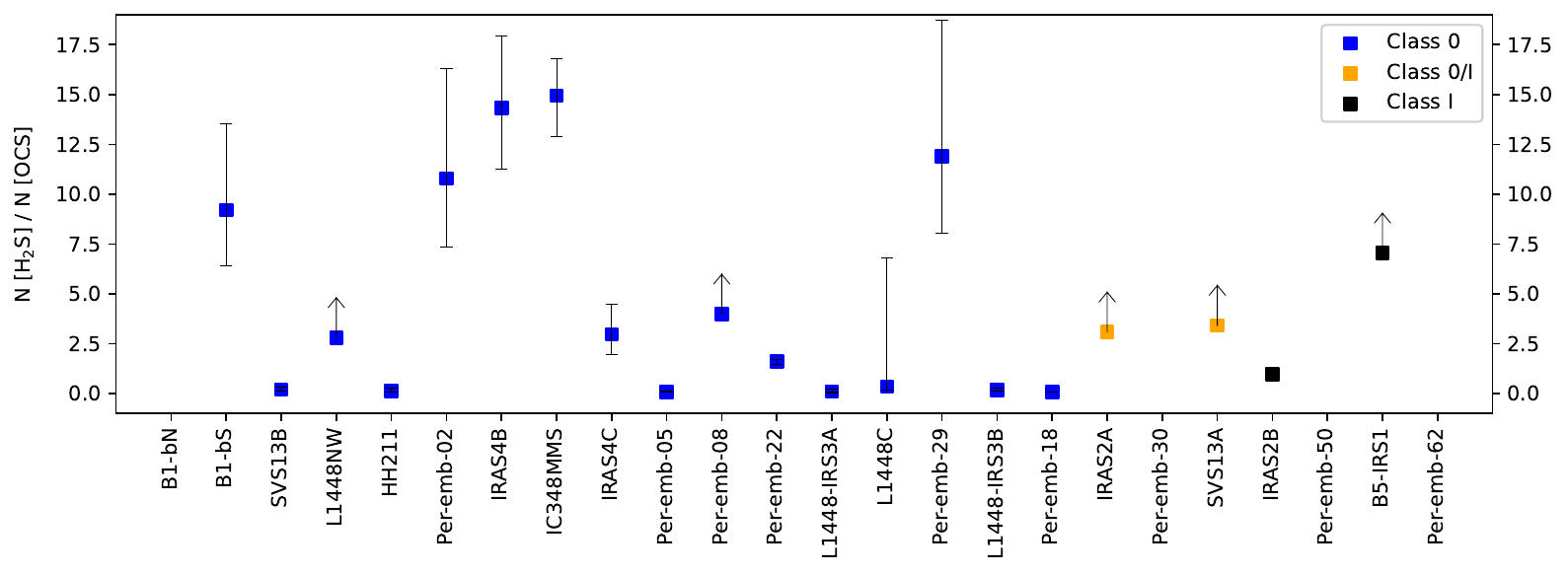}
    \caption{Measured ratios}
    \label{fig:ratio-measured}
\end{subfigure}

\vspace{0.5cm}

\begin{subfigure}{0.95\textwidth}
    \centering
    \includegraphics[width=\linewidth]{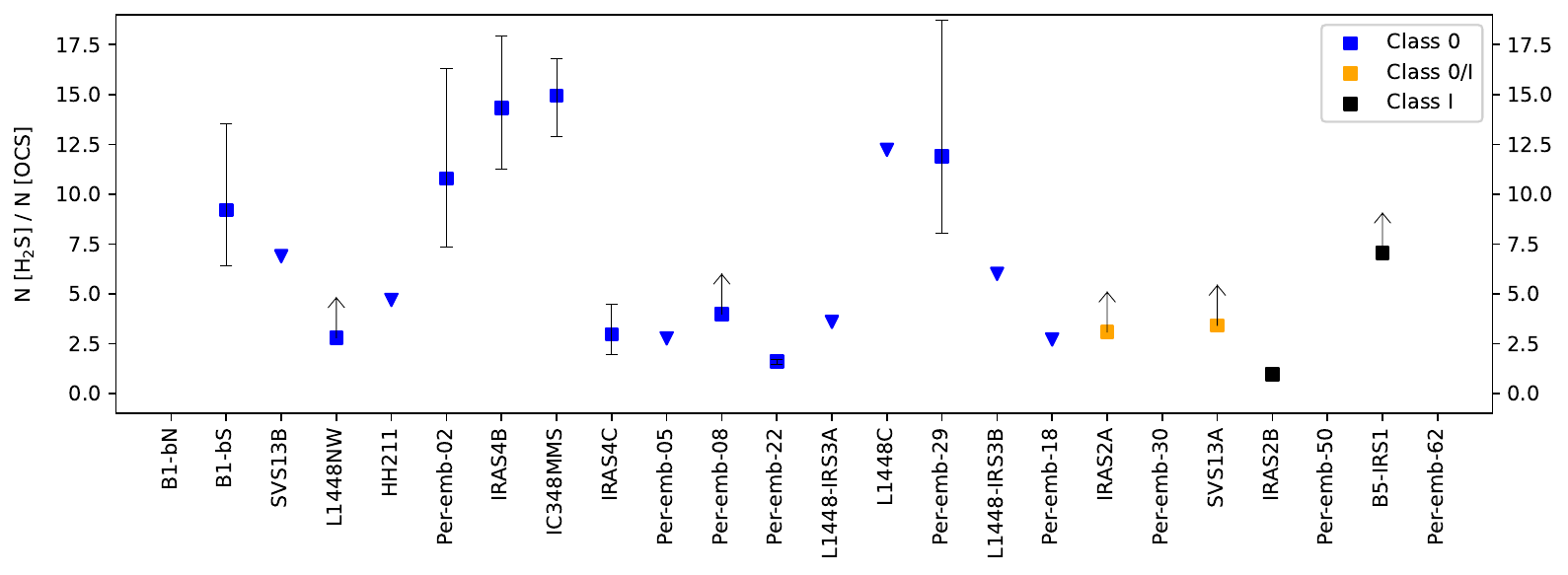}
    \caption{Worst case scenario (scaled $\times37$)}
    \label{fig:ratio-scaled}
\end{subfigure}
\caption{H$_2$S/OCS ratio in the 24 protostars of our sample. The class of each protostar has been represented using the color code in the legend. Sources are sorted by class, and then ordered by increasing bolometric temperature within each class. Panel a) shows the ratios calculated with the estimated column densities of H$_2$S and OCS. Panel b) shows the same data but assuming a worst-case scenario where all protostars with OC$^{34}$S detection but no H$_2^{33}$S detection have their H$_2$S column density scaled by a factor 37, to take into account possible optically thick emission in the H$_2$S line. The scaled ratios have been represented with a downwards triangle, which is a superior limit to the maximum expected H$_2$S/OCS ratio after accounting for the opacity. In this case, the two differentiated trends still appear.}
\label{fig:ratio_comparison}
\end{figure*}

Figure~\ref{fig:ratio_comparison} shows the comparison between the measured H$_2$S/OCS ratios in the sources from our sample, and the corresponding H$_2$S/OCS ratios in the worst case scenario. As explained at the end of Section~\ref{sec:ratios}, this scenario considers that, in the OCS-rich protostars were OC$^{34}$S was detected but H$_2^{33}$S was not, the correction for the H$_2$S opacity is of a factor of 30. We find that, with these assumptions, the two-family segregation into OCS-poor and OCS-rich sources remains valid.

\section{Complementary Tables}
\label{sec:appendix_tables}

In Tables \ref{tab:sources}, \ref{tab:detections}, \ref{tab:col_dens_and_ratios} and \ref{tab:abundances}, we present crucial information about the 24 sources that form our sample, including tabulated data from previous works and results from our observations. All values have been calculated with their corresponding uncertainties, but they have not been included in these tables for the sake of conciseness and clarity. The complete \ref{tab:col_dens_and_ratios} and \ref{tab:abundances} tables, as well as the spectra and an extra table regarding additional information about the moment-0 maps (see Section \ref{sec:results}), can be consulted at \url{https://github.com/JMiranzo/PRODIGE-MissingSulfurProblem}.

\setlength{\tabcolsep}{8pt}
\renewcommand{\arraystretch}{1.5}
\begin{sidewaystable*}[]
    \caption{General source information of the 24 protostars in our sample.}
    \label{tab:sources}
    \centering
    \begin{tabular}{lllllccccc}
    \hline\hline
        \multirow{2}{*}{Source} & \multirow{2}{*}{Other Names} & \multirow{2}{*}{Subregion} & RA (J2000) & DEC (J2000) & \multirow{2}{*}{Class} & $L_\textrm{bol}$$^{(a)}$ & $T_\textrm{bol}$$^{(a)}$ & $T_\textrm{kin}$$^{(b)}$ & Hot \\
        & & & \scriptsize (hh:mm:ss) & \scriptsize (dd:mm:ss) & & \scriptsize ($L_\odot$) & \scriptsize (K) & \scriptsize (K) & Corino \\
    \hline
    B1bN & & Barnard 1 & 03:33:21.21 & 31:07:43.6 & 0 & 0.32$\pm$0.10 & 14.7$\pm$1.0 & 100.0 & N \\
    B1bS & & Barnard 1 & 03:33:21.36 & 03:07:26.3 & 0 & 0.70$\pm$0.01 & 17.7$\pm$1.0 & 100.0 & Y \\
    B5-IRS1 & Per-emb-53 & Barnard 5 & 03:47:41.59 & 32:51:43.6 & I & 4.70$\pm$0.90 & 287.0$\pm$8.0 & 287.0 & N \\
    HH211MMS & Per-emb-01 & IC348 & 03:43:56.81 & 32:00:50 & 0 & 1.80$\pm$0.10 & 27.0$\pm$1.0 & 100.0 & Y \\
    IC348MMS & Per-emb-11 & IC348 & 03:43:57.07 & 32:03:04.8 & 0 & 1.50$\pm$0.10 & 30.0$\pm$2.0 & 100.0 & Y \\
    IRAS2A & Per-emb-27 & NGC1333 & 03:28:55.57 & 31:14:37.0 & 0/I & 19.00$\pm$0.40 & 69.0$\pm$1.0 & 100.0 & Y \\
    IRAS2B & Per-emb-36 & NGC1333 & 03:28:57.38 & 31:14:15.8 & I & 5.30$\pm$1.00 & 106.0$\pm$12.0 & 106.0 & N \\
    IRAS4B & Per-emb-13 & NGC1333 & 03:29:12.02 & 31:13:08.0 & 0 & 4.00$\pm$0.30 & 28.0$\pm$1.0 & 100.0 & Y \\
    IRAS4C & Per-emb-14 & NGC1333 & 03:29:13.55 & 31:13:58.1 & 0 & 0.70$\pm$0.08 & 31.0$\pm$2.0 & 100.0 & N \\
    L1448-IRS3A & & L1448 & 03:25:36:50 & 30:45:21.9 & 0 & 9.20$\pm$1.30 & 47.0$\pm$2.0 & 100.0 & N \\
    L1448-IRS3B & Per-emb-33 & L1448 & 03:25:36.38 & 30:45:14.7 & 0 & 8.30$\pm$0.80 & 57.0$\pm$3.0 & 100.0 & N \\
    L1448C & Per-emb-26 & L1448 & 03:25:38.88 & 30:44:05.3 & 0 & 8.40$\pm$1.50 & 47.0$\pm$7.0 & 100.0 & Y \\
    L1448NW & L1448-IRS3C & L1448 & 03:25:35.67 & 30:45:34.2 & 0 & 1.40$\pm$0.10 & 22.0$\pm$1.0 & 100.0 & N \\
    Per-emb-02 & & Barnard 1 & 03:32:17.92 & 30:49:47.8 & 0 & 0.90$\pm$0.07 & 27.0$\pm$1.0 & 100.0 & N \\
    Per-emb-05 & & ... $^{(c)}$ & 03:31:20.94 & 30:45:30.2 & 0 & 1.30$\pm$0.10 & 32.0$\pm$2.0 & 100.0 & N \\
    Per-emb-08 & & IC348 & 03:44:43.98 & 32:01:35.2 & 0 & 2.60$\pm$0.50 & 43.0$\pm$6.0 & 100.0 & N \\
    Per-emb-18 & IRAS7 & NGC1333 & 03:29:11.27 & 31:18:31.1 & 0 & 2.80$\pm$1.70 & 59.0$\pm$12.0 & 100.0 & N \\
    Per-emb-22 & L1448-IRS2 & L1448 & 03:25:22.41 & 30:45:13.3 & 0 & 3.60$\pm$0.50 & 43.0$\pm$2.0 & 100.0 & N \\
    Per-emb-29 & B1c & Barnard 1 & 03:33:17.88 & 31:09:31.7 & 0 & 3.70$\pm$0.40 & 48.0$\pm$1.0 & 100.0 & N \\
    Per-emb-30 & & Barnard 1 & 03:33:27.28 & 31:07:10.2 & 0/I & 1.70$\pm$0.01 & 78.0$\pm$6.0 & 100.0 & N \\
    Per-emb-50 & & NCG1333 & 03:29:07.77 & 31:21:57.1 & I & 23.20$\pm$3.00 & 128.0$\pm$23.0 & 128.0 & N \\
    Per-emb-62 & & IC348 & 03:44:12.98 & 32:01:35.4 & I & 1.80$\pm$0.40 & 378.0$\pm$29.0 & 378.0 & N \\
    SVS13A & Per-emb-44 & NGC1333 & 03:29:03.76 & 31:16:03.7 & 0/I & 32.50$\pm$7.10 & 188.0$\pm$9.0 & 188.0 & Y \\
    SVS13B & & NGC1333 & 03:29:03.08 & 31:15:51.7 & 0 & 1.00$\pm$1.00 & 20.0$\pm$20.0 & 100.0 & N \\
    \hline
    \end{tabular}
    \begin{flushleft}
    \tiny
    \textbf{Notes:} $^{(a)}$ Data extracted from \citet{Tobin2016}. $^{(b)}$ We have assumed the $T_\textrm{kin}$ of the warm inner core to be the maximum between 100K and $T_\textrm{kin}$ (which implies, $T\geq$100K).  $^{(c)}$ Close to the Barnard 1 region, but isolated.
\end{flushleft}
\end{sidewaystable*}

\renewcommand{\arraystretch}{1.5} 
\setlength{\tabcolsep}{8pt}
\begin{table*}[h]
\caption{Detected and undetected transitions in the warm inner cores of the 24 observed protostars.}
\label{tab:detections}
\centerline{
\begin{tabular}{lccccccccc}
    \hline\hline
    Protostar  &  Class  &  H$_2$S  &  H$_2^{33}$S  &  OCS  &  OCS  & OC$^{33}$S & OC$^{34}$S & SO & SO$_2$$^{(a)}$ \\
    & & \scriptsize{2$_{2,0}$--2$_{1,1}$} & \scriptsize{2(2,0,4)$-$2(1,1,4)} & \scriptsize{J=18$-$17} & \scriptsize{J=19$-$18} & \scriptsize{J=18$-$17} & \scriptsize{J=20$-$19} &  \scriptsize{14(0,14)$-$14(1,13)}\\
    \hline
    B1bN & 0 & $-$ & \checkmark & $-$ & $-$ & $-$ & $-$ & $-$ & $-$ \\     
    B1bS & 0 & \checkmark & \checkmark & \checkmark & \checkmark & \checkmark & \checkmark & $-$ & $-$ \\
    B5-IRS1 & I & \checkmark & $-$ & $-$ & $-$ & $-$ & $-$ & \checkmark & $-$ \\
    HH211MMS & 0 & \checkmark & $-$ & \checkmark & \checkmark & $-$ & \checkmark & \checkmark & \checkmark \\
    IC348MMS & 0 & \checkmark & \checkmark & \checkmark & \checkmark & \checkmark & \checkmark & \checkmark & \checkmark \\
    IRAS2A & 0/I & \checkmark & \checkmark & \checkmark & \checkmark & \checkmark & \checkmark & \checkmark & \checkmark \\
    IRAS2B & I & \checkmark & $-$ & \checkmark & \checkmark & $-$ & $-$ & \checkmark & \checkmark \\
    IRAS4B & 0 & \checkmark & \checkmark & \checkmark & \checkmark & \checkmark & \checkmark & \checkmark & \checkmark \\
    IRAS4C & 0 & \checkmark & $-$ & \checkmark & \checkmark & $-$ & $-$ & \checkmark & $-$ \\
    L1448-IRS3A & 0 & \checkmark & $-$ & \checkmark & \checkmark & $-$ & \checkmark & \checkmark & \checkmark \\
    L1448-IRS3B & 0 & \checkmark & $-$ & \checkmark & \checkmark & $-$ & \checkmark & \checkmark & $-$ \\
    L1448C & 0 & \checkmark & $-$ & \checkmark & \checkmark & \checkmark & \checkmark & \checkmark & \checkmark \\
    L1448NW & 0 & \checkmark & $-$ & $-$ & $-$ & $-$ & $-$ & \checkmark & \checkmark \\
    Per-emb-2 & 0 & \checkmark & $-$ & \checkmark & \checkmark & $-$ & $-$ & \checkmark & $-$ \\
    Per-emb-5 & 0 & \checkmark & $-$ & \checkmark & \checkmark & \checkmark & \checkmark & \checkmark & $-$ \\
    Per-emb-8 & 0 & \checkmark & $-$ & $-$ & $-$ & $-$ & $-$ & \checkmark & $-$ \\
    Per-emb-18 & 0 & \checkmark & $-$ & \checkmark & \checkmark & \checkmark & \checkmark & \checkmark & \checkmark \\
    Per-emb-22 & 0 & \checkmark & $-$ & \checkmark & \checkmark & $-$ & $-$ & \checkmark & \checkmark \\
    Per-emb-29 & 0 & \checkmark & \checkmark & \checkmark & \checkmark & \checkmark & \checkmark & \checkmark & \checkmark \\
    Per-emb-30 & 0/I & $-$ & $-$ & $-$ & $-$ & $-$ & $-$ & $-^{(b)}$ & $-^{(b)}$ \\
    Per-emb-50 & I & $-$ & $-$ & $-$ & $-$ & $-$ & $-$ & \checkmark & \checkmark \\
    Per-emb-62 & I & $-$ & $-$ & $-$ & $-$ & $-$ & $-$ & $-^{(b)}$ & $-^{(b)}$\\
    SVS13A & 0/I & \checkmark & \checkmark & \checkmark & \checkmark & \checkmark & \checkmark & \checkmark & \checkmark \\
    SVS13B & 0 & \checkmark & $-$ & \checkmark & \checkmark & $-$ & \checkmark & \checkmark & $-$ \\
    \hline
\end{tabular}
}
\begin{flushleft}
\tiny
    \textbf{Notes:} $^{(a)}$ Retrieved from \citet{ArturDeLaVillarmois2023}. $^{(b)}$ Objects not observed in \citet{ArturDeLaVillarmois2023}.
\end{flushleft}
\end{table*}

\setlength{\tabcolsep}{2.1pt}
\renewcommand{\arraystretch}{1.5} 
\begin{sidewaystable*}[]
    \caption{Column densities and specific ratios in Class 0/I protostars in Perseus.}
    \label{tab:col_dens_and_ratios}
    \centering
    \begin{tabular}{lrrlrrrlcc|ccc|c|}
    \hline\hline
        \multirow{2}{*}{Source} & N(H$_2$S) & N(H$_2^{33}$S) & Calculated H$_2$S$^{(a)}$ & N(OCS) & N(OC$^{33}$S) & N(OC$^{34}$S) & Calculated OCS$^{(b)}$ & N(SO)$^{(c)}$ & N(SO$_2$)$^{(c)}$ & \multirow{2}{*}{$^{34}$S/$^{33}$S} &  \multirow{2}{*}{H$_2$S|$_{33}$/H$_2$S}$^{(d)}$ & \multirow{2}{*}{OCS|$_{34}$/OCS}$^{(d)}$ & \multirow{2}{*}{H$_2$S/OCS} \\
        & \scriptsize(cm$^{-2}$)$\times$10$^{13}$ & \scriptsize(cm$^{-2}$)$\times$10$^{13}$ & \multicolumn{1}{c}{\scriptsize(cm$^{-2}$)$\times$10$^{13}$} & \scriptsize(cm$^{-2}$)$\times$10$^{13}$ & \scriptsize(cm$^{-2}$)$\times$10$^{13}$ & \scriptsize(cm$^{-2}$)$\times$10$^{13}$ & \multicolumn{1}{c}{\scriptsize(cm$^{-2}$)$\times$10$^{13}$} & \scriptsize(cm$^{-2}$)$\times$10$^{13}$ & \scriptsize(cm$^{-2}$)$\times$10$^{13}$ & & & & \\
    \hline
        B1bN & 0.0 & 2.890 & $\,\,$(N(H$_2$S)=154.0) & 0.0 & 0.0 & 0.0 & & <7 & <28 & - & - & - & - \\
        B1bS & 18.45 & 6.055 & $\,\,$(N(H$_2$S)=2659) & 13.37 & 5.099 & 12.83 & $\,\,$(N(OCS)=288.8) & <7 & <28 & 2.52 & 144.1 & 21.6 & 9.21 \\
        B5-IRS1 & 50.40 & 0.0 & & <7.154 & 0.0 & 0.0 & & [41.3-61.2] & <38 & - & - & - & >7.05\\
        HH211MMS & 21.01 & 0.0 & & 9.871 & 1.831 & 7.095 & $\,\,$(N(OCS)=159.6) & [48.3-71.6] & [50-200] & 3.88 & - & 16.2 & 0.13 \\
        IC348MMS & 33.96 & 4.092 & $\,\,$(N(H$_2$S)=2936) & 17.01 & 1.738 & 8.735 & $\,\,$(N(OCS)=196.5) & >40.9 & [20-60] & 5.02 & 86.5 & 11.6 & 14.9 \\
        IRAS2A & 446.3 & >20.05 & $\,\,$(N(H$_2$S)>5687) & 565.7 & 13.59 & 81.98 & $\,\,$(N(OCS)=1845) & >188.1 & >300 & 6.03 & >12.7 & 3.26 & >3.08 \\
        IRAS2B & 10.29 & 0.0 & & 10.58 & 0.0 & 0.0 & & >52.7 & [10-40] & - & - & - & 0.97 \\
        IRAS4B & 266.6 & 31.93 & $\,\,$(N(H$_2$S)=20600) & 145.9 & 15.20 & 63.89 & $\,\,$(N(OCS)=1437) & >11.9 & [20-70] & 4.20 & 77.3 & 9.85 & 14.3 \\
        IRAS4C & 11.54 & 0.0 & & 3.886 & 0.0 & 0.0 & & [33.2-49.2] & <18 & - & - & - & 2.97 \\
        L1448-IRS3A & 21.83 & 0.0 & & 37.36 & 2.360 & 9.640 & $\,\,$(N(OCS)=216.9) & >35.5 & [20-60] & 4.08 & - & 5.81 & 0.10 \\
        L1448-IRS3B & 27.05 & 0.0 & & 11.00 & 0.804 & 7.152 & $\,\,$(N(OCS)=160.9) & [10.8-16.0] & <15 & 8.90 & - & 14.6 & 0.17 \\
        L1448C & 130.9 & 0.0 & & 222.0 & 2.965 & 16.99 & $\,\,$(N(OCS)=382.3) & >59.0 & [60-400] & 5.73 & - & 1.70 & 0.34 \\
        L1448NW & 9.612 & 0.0 & & <3.426 & 0.0 & 0.0 & & >33.4 & [10-50] & - & - & - & >2.81 \\
        Per-emb-02 & 30.04 & 0.0 & & 2.785 & 0.0 & 0.0 & & [19.4-28.8] & <49 & - & - & - & 10.8 \\
        Per-emb-05 & 14.83 & 0.0 & & 12.84 & 0.695 & 8.509 & $\,\,$(N(OCS)=191.5) & [13.7-20.2] & <20 & 12.2 & - & 14.9 & 0.08 \\
        Per-emb-08 & 11.64 & 0.0 & & <2.923 & 0.0 & 0.0 & & [17.6-26.1] & <40 & - & - & - & >3.98 \\
        Per-emb-18 & 27.68 & 0.0 & & 74.03 & 2.348 & 16.20 & $\,\,$(N(OCS)=364.5) & >65.2 & [30-100] & 6.90 & - & 4.92 & 0.08 \\
        Per-emb-22 & 37.87 & 0.0 & & 23.66 & 0.0 & 0.0 & & [16.8-51.8] & [10-40] & - & - & - & 1.60 \\
        Per-emb-29 & 98.54 & 4.68 & $\,\,$(N(H$_2$S)=5364) & 36.83 & 5.763 & 20.04 & $\,\,$(N(OCS)=451.0) & >56.3 & [30-100] & 3.48 & 54.4 & 12.2 & 11.9 \\
        Per-emb-30 & 0.0 & 0.0 & & 0.0 & 0.0 & 0.0 & & - & - & - & - & - & - \\
        Per-emb-50 & 0.0 & 0.0 & & 0.0 & 0.0 & 0.0 & & >127.6 & [50-300] & - & - & - & - \\
        Per-emb-62 & 0.0 & 0.0 & & 0.0 & 0.0 & 0.0 & & - & - & - & - & - & - \\
        SVS13A & 974.7 & >67.76 & $\,\,$(N(H$_2$S)>14820) & 946.6 & 39.10 & 192.4 & $\,\,$(N(OCS)=4329) & >210.5 & >300 & 4.92 & >15.2 & 4.57 & >3.42 \\
        SVS13B & 26.76 & 0.0 & & 31.59 & 1.964 & 6.163 & $\,\,$(N(OCS)=138.7) & [5.0-7.4] & <18 & 3.14 & - & 4.39 & 0.19 \\
    \hline
    \end{tabular}
    \begin{flushleft}
    \tiny
    \textbf{Notes:} $^{(a)}$ Calculated using the solar isotopic ratio, $^{32}$S/$^{33}$S=126 \citep{AndersGravesse1989}. $^{(b)}$ Calculated using the solar isotopic ratio, $^{32}$S/$^{34}$S=22.5 \citep{AndersGravesse1989}. $^{(c)}$ Retrieved from \citet{ArturDeLaVillarmois2023}; no data from Per-emb-30 and Per-emb-62 (not observed). $^{(d)}$ A|$_x$/A notation stands for the ratio between the A species column density calculated with the corresponding $^x$S isotopologue, and the column density of A estimated without the isotopologue. 
\end{flushleft}
\end{sidewaystable*}

\setlength{\tabcolsep}{8pt}
\renewcommand{\arraystretch}{1.5}
\begin{sidewaystable*}[]
    \centering
    \caption{Abundances in Class 0/I protostars from Perseus.}
    \label{tab:abundances}
    \begin{tabular}{lrrrrrcc|cc|}
    \hline\hline
        \multirow{2}{*}{Source} & H$_2$S & H$_2^{33}$S & OCS & OC$^{33}$S & OC$^{34}$S & SO$^{(a)}$ & SO$_2$$^{(a)}$ & TOTAL$^{(b)}$ & \multirow{2}{*}{$D_S$} \\
        & \scriptsize$\times$10$^{-9}$ & \scriptsize$\times$10$^{-9}$ & \scriptsize$\times$10$^{-9}$ & \scriptsize$\times$10$^{-9}$ & \scriptsize$\times$10$^{-9}$ & \scriptsize$\times$10$^{-9}$ & \scriptsize$\times$10$^{-9}$ & \scriptsize$\times$10$^{-9}$ & \\
    \hline
        B1bN & 1.442 & 0.025 & 0.0 & 0.0 & 0.0 & <0.007 & <0.029 & 1.464 & 1$\times$10$^4$\\
        B1bS & 16.03 & 0.036 & 1.741 & 0.031 & 0.077 & <0.005 & <0.019 & 17.78 & 8$\times$10$^2$\\
        B5-IRS1 & 89.98 & 0.0 & 12.77 & 0.0 & 0.0 & [0.307-0.950] & <0.590 & 103.8 & 1$\times$10$^2$\\
        HH211MMS & 0.586 & 0.0 & 4.453 & 0.051 & 0.198 & [0.034-0.100] & [0.035-0.279] & 5.263 & 3$\times$10$^3$ \\
        IC348MMS & 41.88 & 0.083 & 2.803 & 0.025 & 0.125 & >0.011 & [0.005-0.030] & 44.72 & 3$\times$10$^2$ \\
        IRAS2A & >207.1 & >0.411 & 67.16 & 0.495 & 2.985 & >0.608 & >0.970 & >276.3 & < 5$\times$10$^1$ \\
        IRAS2B & 1.142 & 0.0 & 1.173 & 0.0 & 0.0 & >0.367 & [0.070-0.828] & 3.642 & 4$\times$10$^3$ \\
        IRAS4B & 84.13 & 0.167 & 5.871 & 0.062 & 0.261 & >0.013 & [0.021-0.196] & 90.12 & 2$\times$10$^2$ \\
        IRAS4C & 0.379 & 0.0 & <0.128 & 0.0 & 0.0 & [0.059-0.179] & <0.066 & 0.688 & 2$\times$10$^4$ \\
        L1448-IRS3A & 1.383 & 0.0 & 13.74 & 0.149 & 0.611 & >0.198 & [0.111-1.003] & 16.12 & 9$\times$10$^2$ \\
        L1448-IRS3B & 0.537 & 0.0 & 3.195 & 0.016 & 0.142 & [0.015-0.048] & <0.045 & 3.793 & 4$\times$10$^3$ \\
        L1448C & 4.113 & 0.0 & 12.01 & 0.093 & 0.534 & >0.198 & [0.201-3.133] & 18.12 & 8$\times$10$^2$ \\
        L1448NW & 0.275 & 0.0 & <0.097 & 0.0 & 0.0 & >0.045 & [0.013-0.120] & 0.518 & 3$\times$10$^4$ \\
        Per-emb-02 & 0.240 & 0.0 & 0.022 & 0.0 & 0.0 & [0.015-0.040] & <0.068 & 0.330 & 5$\times$10$^4$ \\
        Per-emb-05 & 0.206 & 0.0 & 2.658 & 0.010 & 0.118 & [0.031-0.091] & <0.090 & 2.986 & 5$\times$10$^3$ \\
        Per-emb-08 & 0.501 & 0.0 & <0.126 & 0.0 & 0.0 & [0.016-0.042] & <0.064 & 0.697 & 2$\times$10$^4$ \\
        Per-emb-18 & 1.271 & 0.0 & 16.73 & 0.108 & 0.744 & >0.201 & [0.093-0.681] & 18.76 & 8$\times$10$^2$ \\
        Per-emb-22 & 2.683 & 0.0 & 1.676 & 0.0 & 0.0 & [0.024-0.164] & [0.014-0.173] & 4.527 & 3$\times$10$^3$ \\
        Per-emb-29 & 65.55 & 0.209 & 20.18 & 0.258 & 0.897 & >0.063 & [0.033-0.744] & 85.93 & 6$\times$10$^1$  \\
        Per-emb-30 & 0.0 & 0.0 & 0.0 & 0.0 & 0.0 & - & - & 0.0 & - \\
        Per-emb-50 & 0.0 & 0.0 & 0.0 & 0.0 & 0.0 & >1.108 & [0.434-7.437] & >6.408 & <2$\times$10$^3$ \\
        Per-emb-62 & 0.0 & 0.0 & 0.0 & 0.0 & 0.0 & - & - & 0.0 & - \\
        SVS13A & >1343 & >6.139 & 392.2 & 3.542 & 17.43 & >1.885 & >2.686 & >1742 & <8.6 \\
        SVS13B & 0.232 & 0.0 & 1.204 & 0.017 & 0.053 & [0.019-0.078] & <0.190 & 1.589 & 9$\times$10$^3$ \\
    \hline
        
    \hline
    \end{tabular}
    \begin{flushleft}
    \tiny
    \textbf{Notes:} $^{(a)}$ Retrieved from \citet{ArturDeLaVillarmois2023}; no data from Per-emb-30 and Per-emb-62 (not observed). $^{(b)}$ Total abundance has been calculated using all lower and upper limits, and using the mean value of SO and SO$_2$ in the cases a range is given. Lower and upper limits to the total sulfur abundance have only been explicited when a limit was given for the most abundant species in the source.
\end{flushleft}
\end{sidewaystable*}

\onecolumn
\section{Complementary Figures}

In Figs.~\ref{fig:moment-0-maps-appendix} and \ref{fig:specs-appendix}, we show the moment-0 maps and the spectra, respectively, of the 20 sources not shown in the main text. The moment-0 maps cover the H$_2$S and both OCS lines, while in the spectra we display the emission of the H$_2$S line, one of the OCS lines and the OC$^{34}$S line.

\begin{figure*}[h!]
    \centering
    \begin{subfigure}{0.47\textwidth}
        \includegraphics[width=\linewidth]{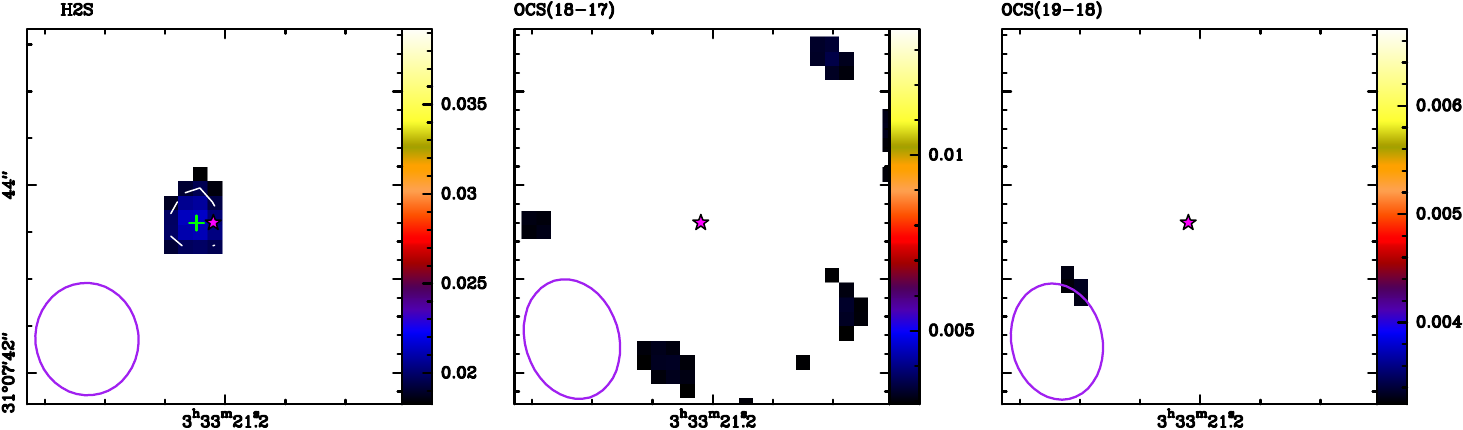}
        \caption{B1bN}
        \label{fig:map_B1bN}
    \end{subfigure}
    \hspace{0.9cm}
    \begin{subfigure}{0.47\textwidth}
        \centering
        \includegraphics[width=\linewidth]{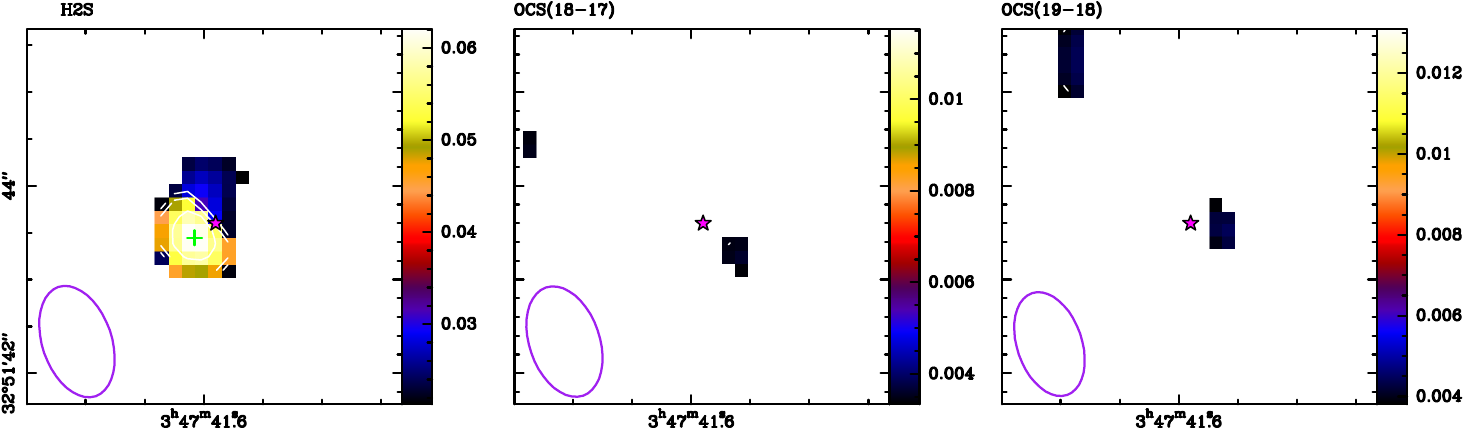}
        \caption{B5-IRS1}
        \label{fig:map_B5-IRS1}
    \end{subfigure}

    \vspace{1cm}
    \begin{subfigure}{0.47\textwidth}
        \centering
        \includegraphics[width=\linewidth]{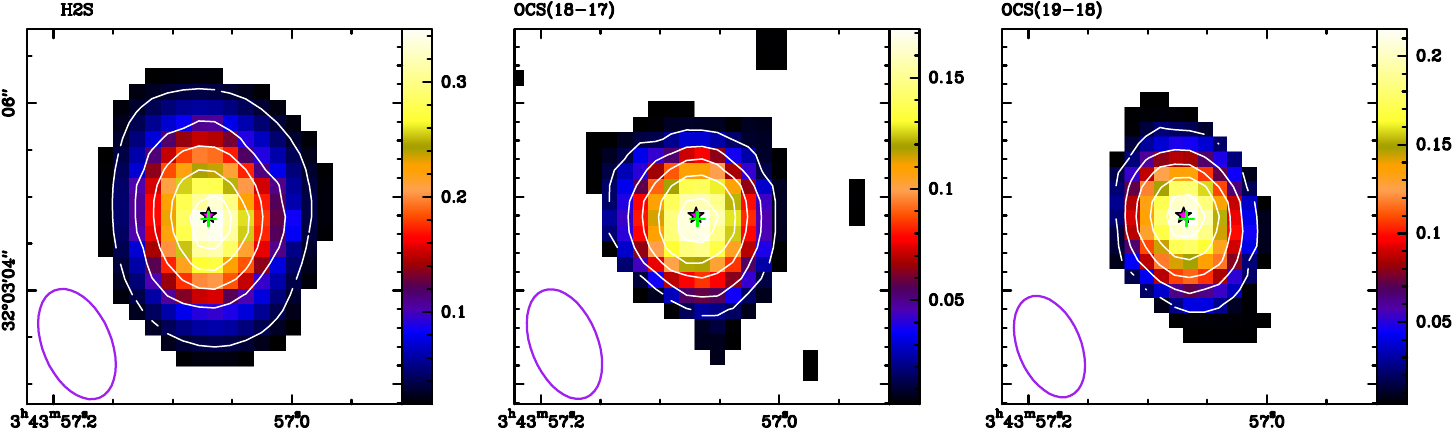}
        \caption{IC348MMS}
        \label{fig:map_IC348MMS}
    \end{subfigure}
    \hspace{0.9cm}
    \begin{subfigure}{0.47\textwidth}
        \centering
        \includegraphics[width=\linewidth]{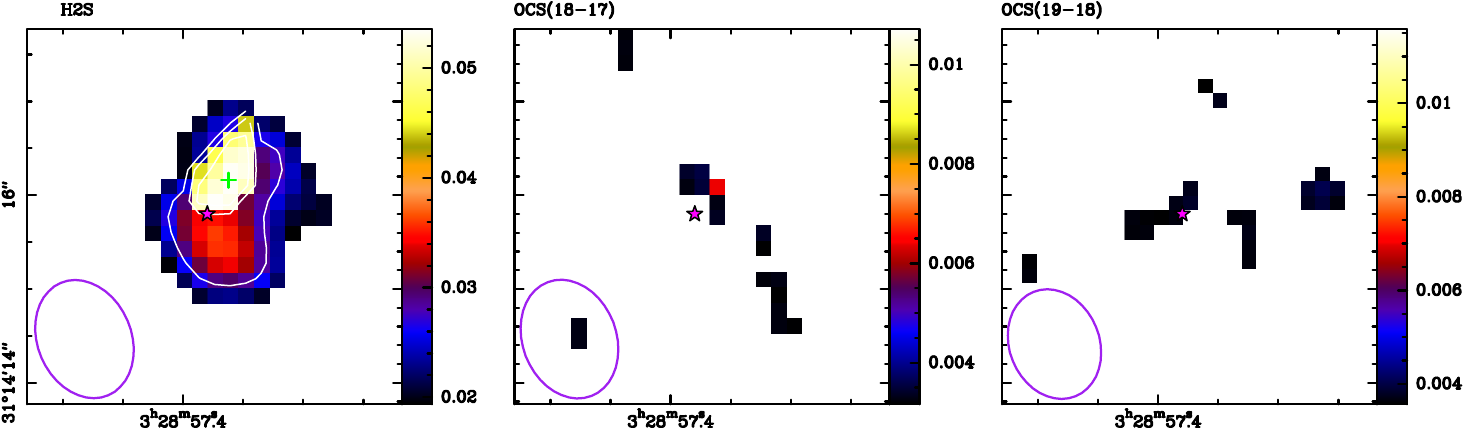}
        \caption{IRAS2B}
        \label{fig:map_IRAS2B}
    \end{subfigure}

    \vspace{1cm}
    \begin{subfigure}{0.47\textwidth}
        \centering
        \includegraphics[width=\linewidth]{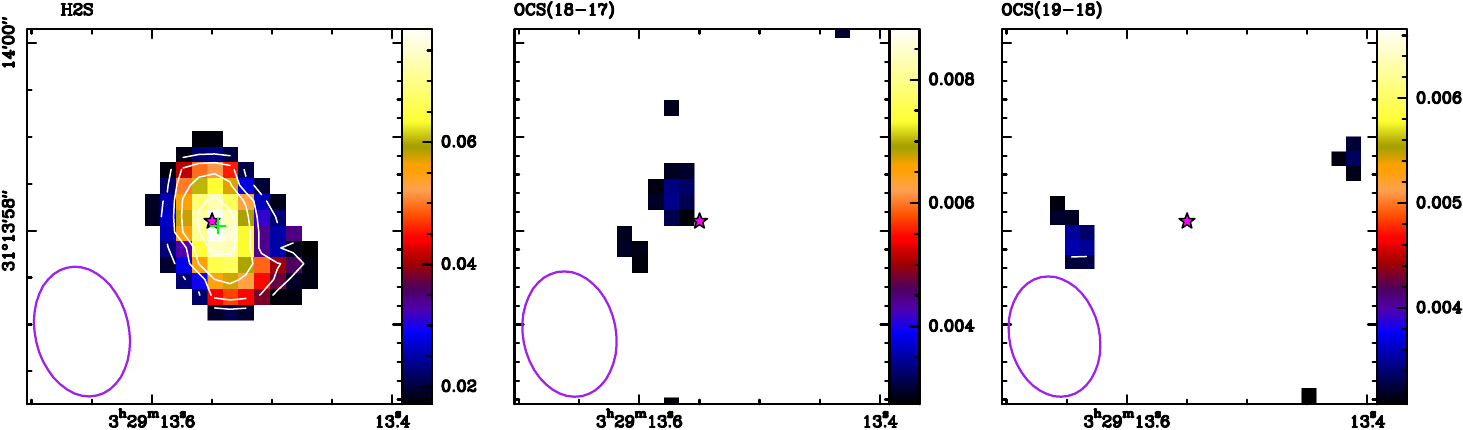}
        \caption{IRAS4C}
        \label{fig:map_IRAS4C}
    \end{subfigure}
    \hspace{0.9cm}
    \begin{subfigure}{0.47\textwidth}
        \centering
        \includegraphics[width=\linewidth]{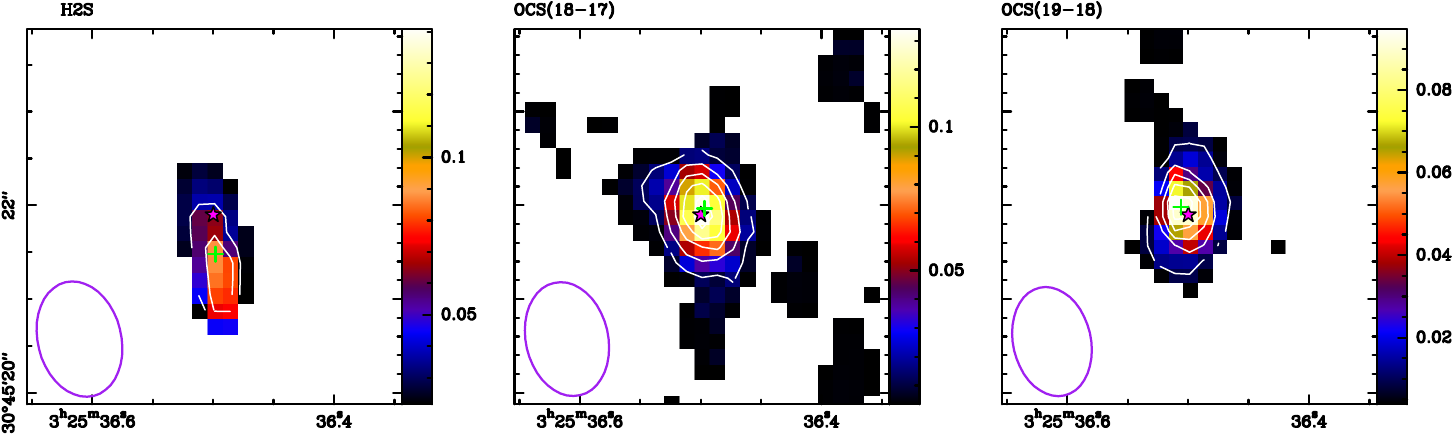}
        \caption{L1448-IRS3A}
        \label{fig:map_L1448-IRS3A}
    \end{subfigure}

    \vspace{1cm}
    \begin{subfigure}{0.47\textwidth}
        \centering
        \includegraphics[width=\linewidth]{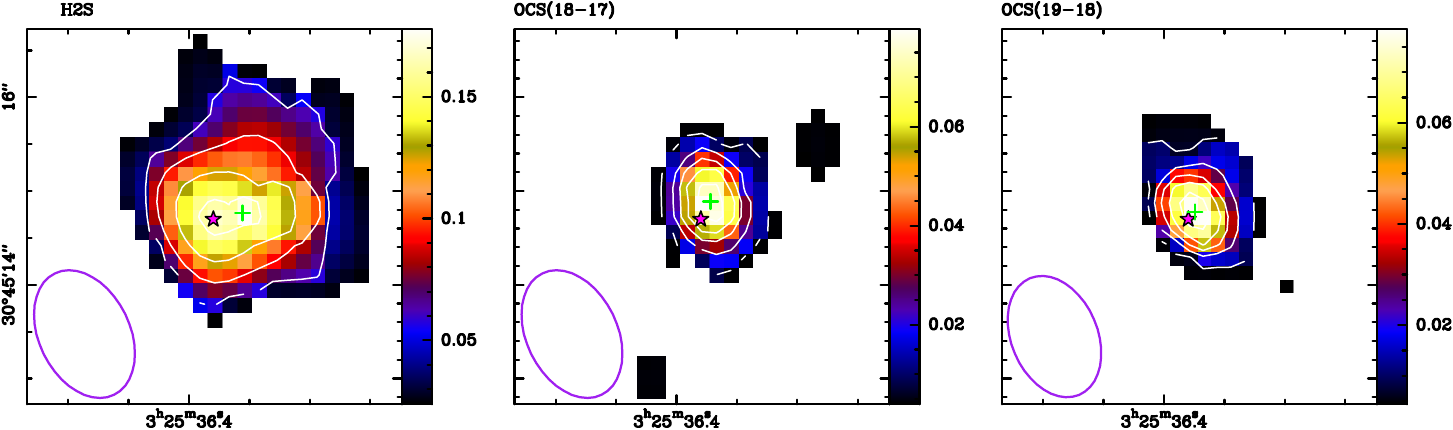}
        \caption{L1448-IRS3B}
        \label{fig:map_L1448-IRS3B}
    \end{subfigure}
    \hspace{0.9cm}
    \begin{subfigure}{0.47\textwidth}
        \centering
        \includegraphics[width=\linewidth]{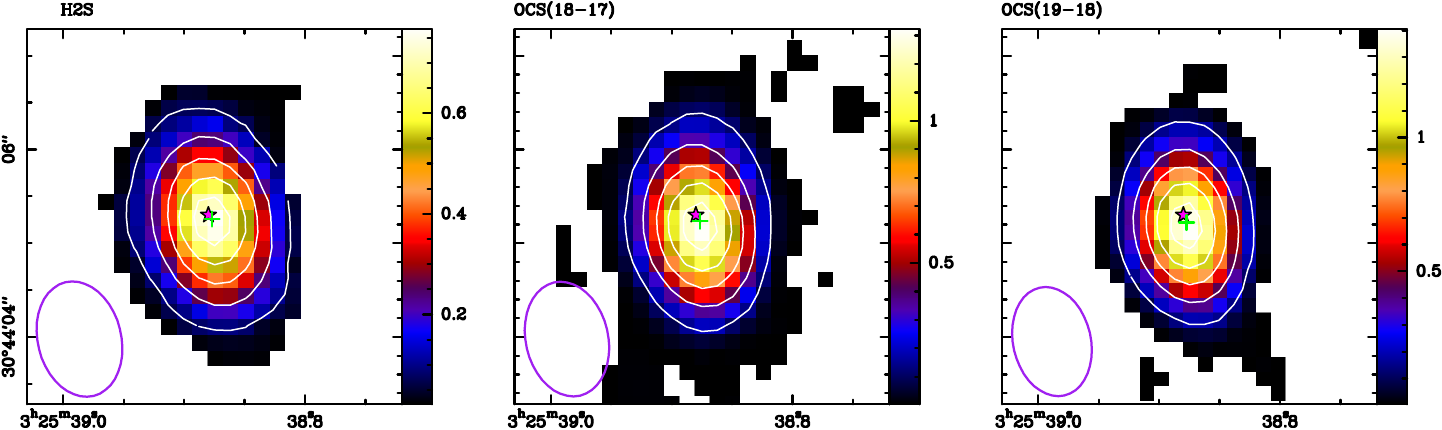}
        \caption{L1448C}
        \label{fig:map_L1448C}
    \end{subfigure}

    \vspace{1cm}
    \centering
    \begin{subfigure}{0.47\textwidth}
        \includegraphics[width=\linewidth]{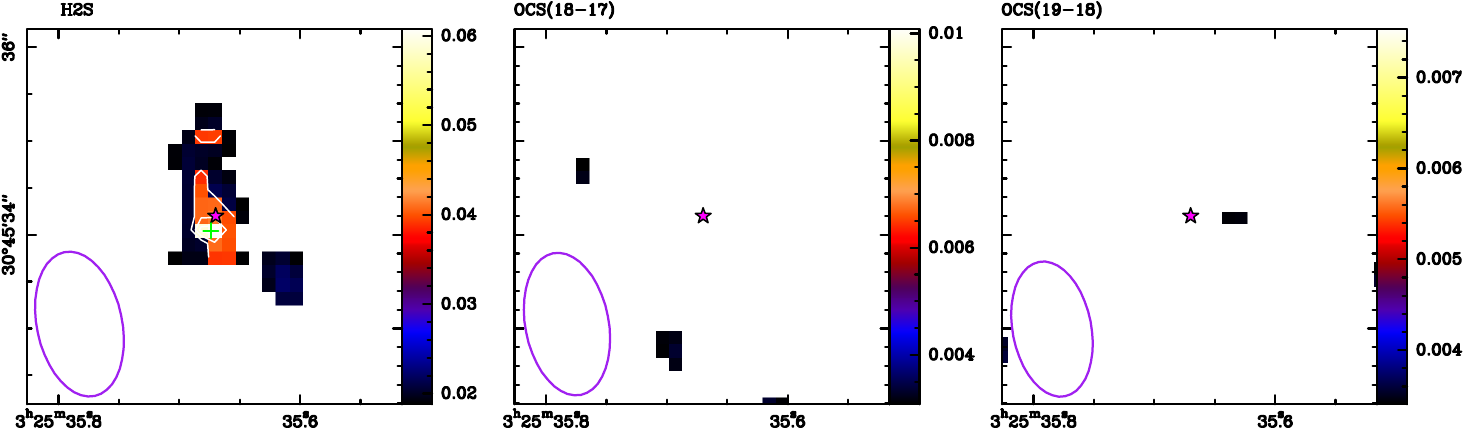}
        \caption{L1448NW}
        \label{fig:map_L1448NW}
    \end{subfigure}
    \hspace{0.9cm}
    \begin{subfigure}{0.47\textwidth}
    \centering
        \includegraphics[width=\linewidth]{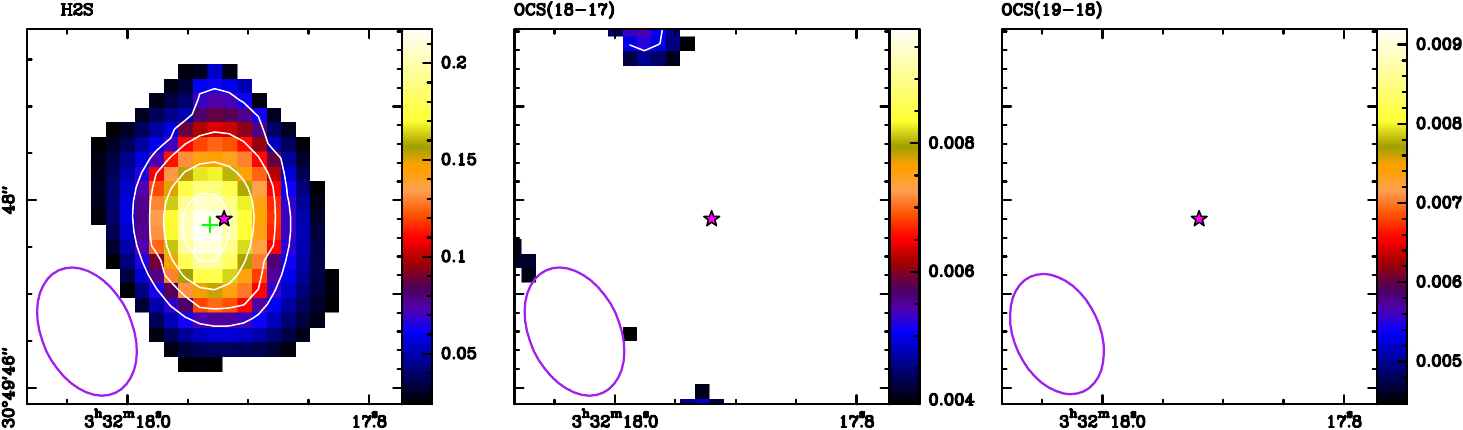}
        \caption{Per-emb-02}
        \label{fig:map_Per-emb-2}
    \end{subfigure}

    \caption{Emission of the main species (H$_2$S, and both OCS lines) in the warm inner core of the 20 protostars from our sample not shown in Section~\ref{sec:results}. The colormap represents the >3$\sigma$ emission integrated images in a 4$\arcsec$$\times$4$\arcsec$ square region. The color scale, shown at the right of each map, is the brightness temperature in K. The white contours represent 10\%, 30\%, 50\%, 70\% and 90\% of the peak temperature. The pink star shows the position of the protostar, determined by the position of the maximum emission in the continuum \citep{Tobin2016}. The green cross marks the point with maximum emission of the line.}
    \label{fig:moment-0-maps-appendix}
\end{figure*}

\begin{figure*}\ContinuedFloat

    \vspace{1cm}
    \begin{subfigure}{0.47\textwidth}
        \centering
        \includegraphics[width=\linewidth]{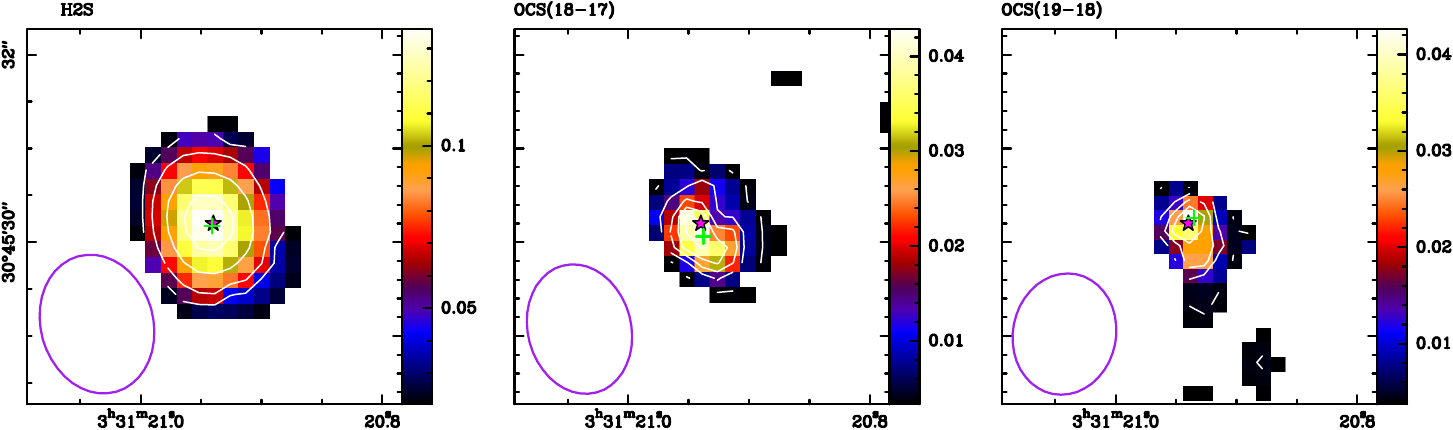}
        \caption{Per-emb-05}
        \label{fig:map_Per-emb-5}
    \end{subfigure}
    \hspace{0.9cm}
    \begin{subfigure}{0.47\textwidth}
        \centering
        \includegraphics[width=\linewidth]{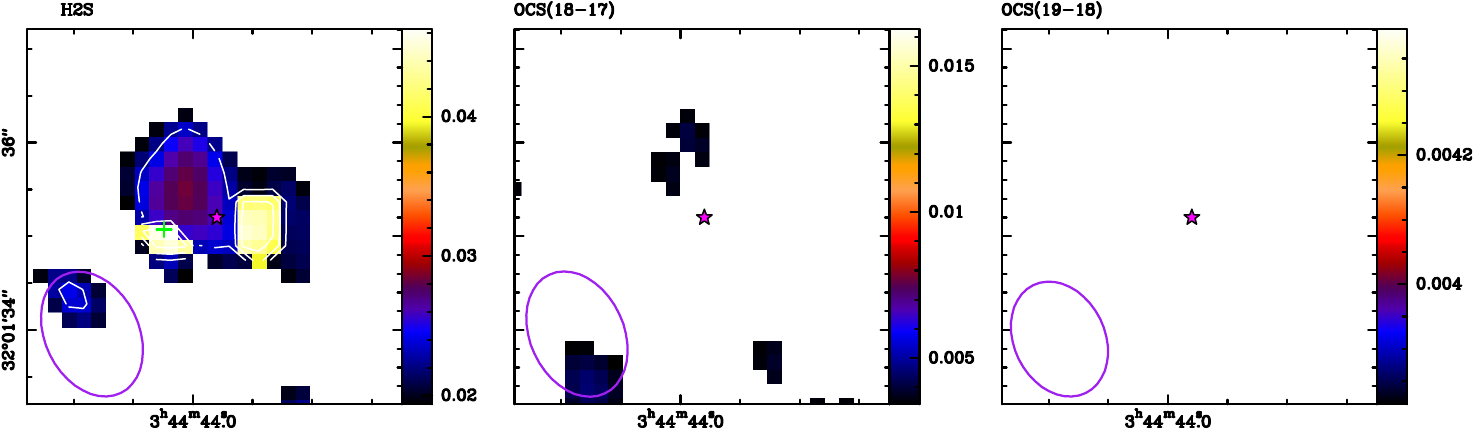}
        \caption{Per-emb-08}
        \label{fig:map_Per-emb-8}
    \end{subfigure}

    \vspace{1cm}
    \begin{subfigure}{0.47\textwidth}
        \centering
        \includegraphics[width=\linewidth]{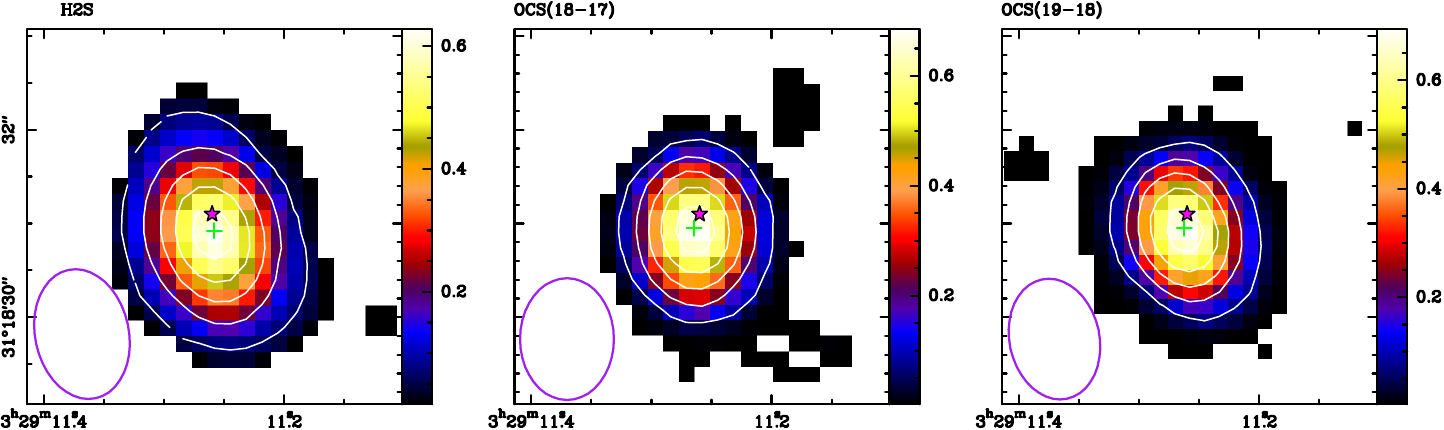}
        \caption{Per-emb-18}
        \label{fig:map_Per-emb-18}
    \end{subfigure}
    \hspace{0.9cm}
    \begin{subfigure}{0.47\textwidth}
        \centering
        \includegraphics[width=\linewidth]{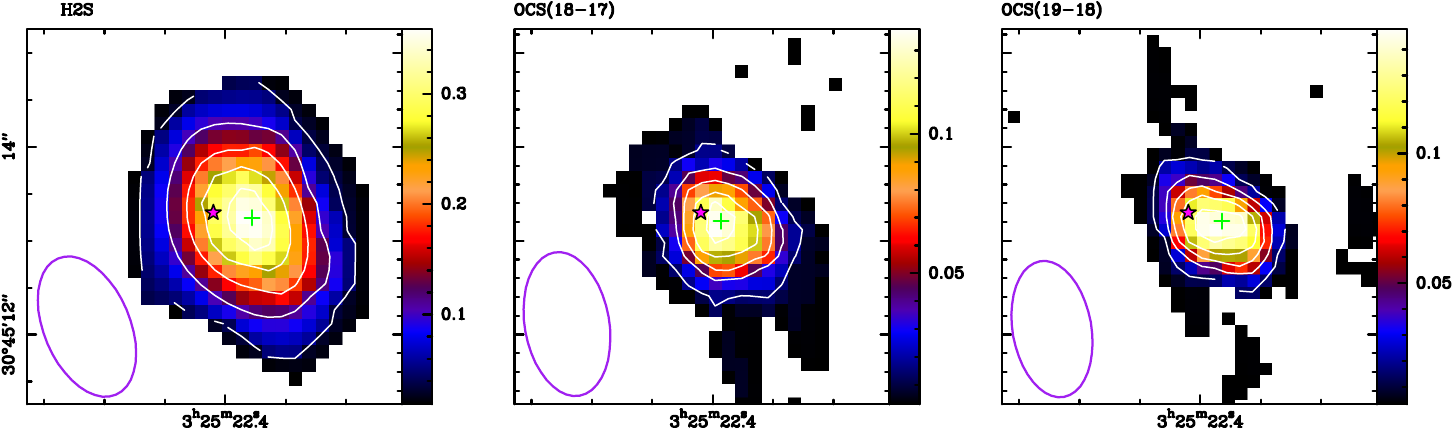}
        \caption{Per-emb-22}
        \label{fig:map_Per-emb-22}
    \end{subfigure}

    \vspace{1cm}
    \begin{subfigure}{0.47\textwidth}
        \centering
        \includegraphics[width=\linewidth]{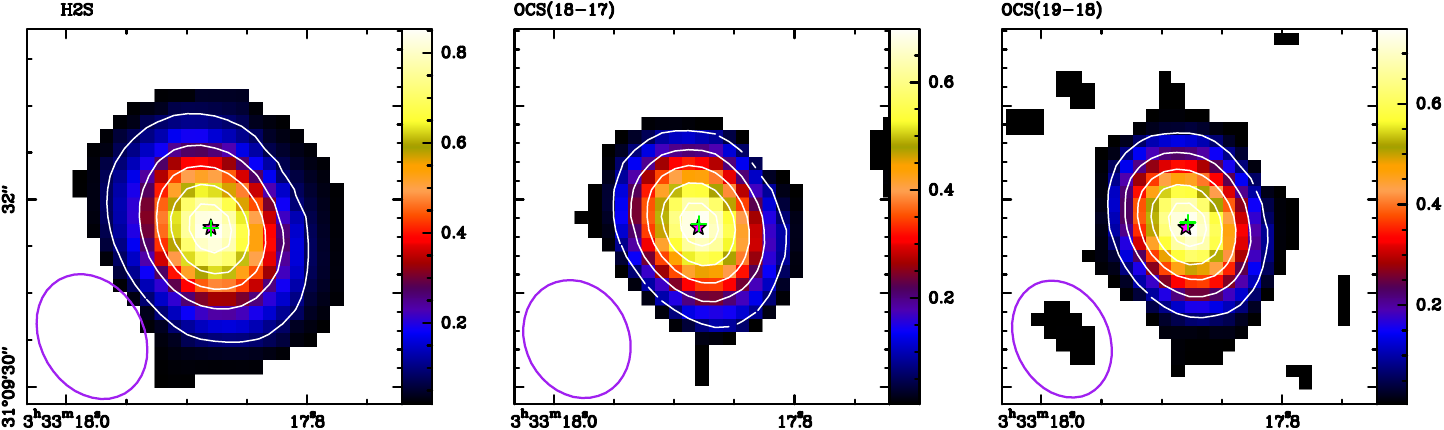}
        \caption{Per-emb-29}
        \label{fig:map_Per-emb-29}
    \end{subfigure}
    \hspace{0.9cm}
    \begin{subfigure}{0.47\textwidth}
        \centering
        \includegraphics[width=\linewidth]{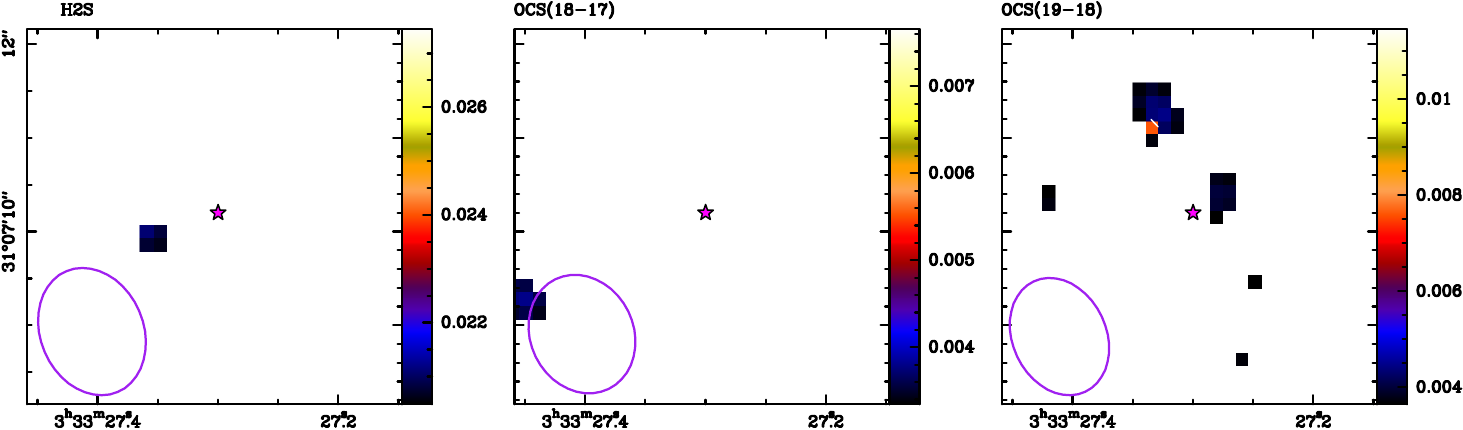}
        \caption{Per-emb-30}
        \label{fig:map_Per-emb-30}
    \end{subfigure}

    \vspace{1cm}
    \begin{subfigure}{0.47\textwidth}
        \centering
        \includegraphics[width=\linewidth]{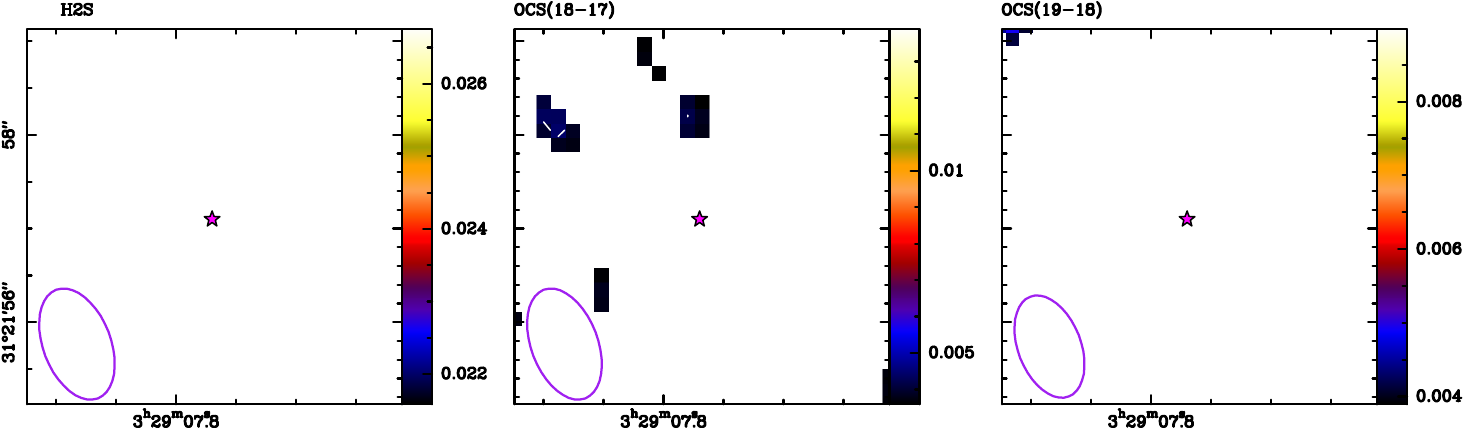}
        \caption{Per-emb-50}
        \label{fig:map_Per-emb-50}
    \end{subfigure}
    \hspace{0.9cm}
    \begin{subfigure}{0.47\textwidth}
        \centering
        \includegraphics[width=\linewidth]{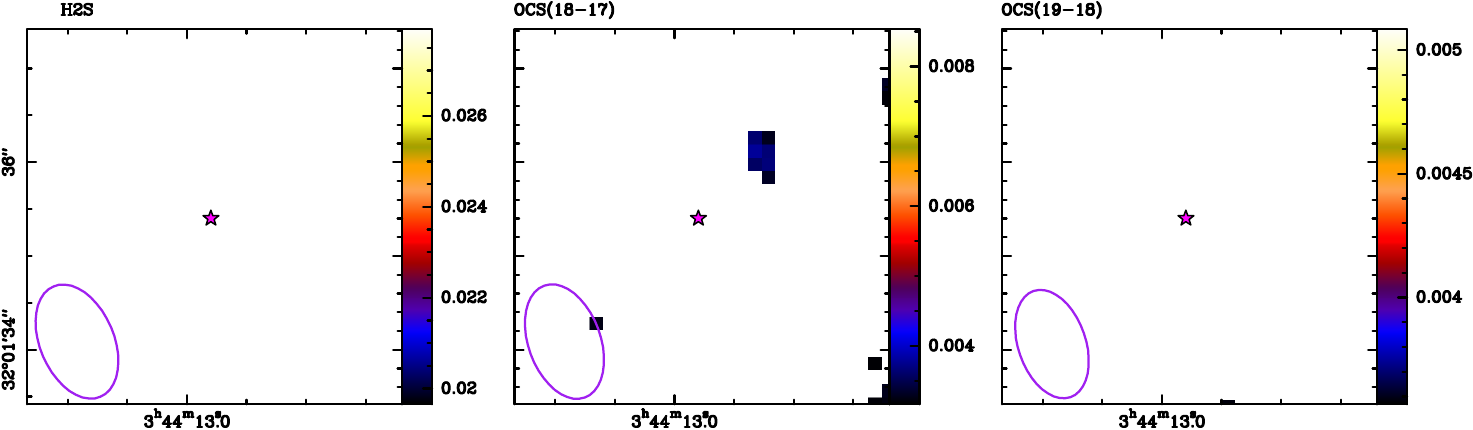}
        \caption{Per-emb-62}
        \label{fig:map_Per-emb-62}
    \end{subfigure}

    \vspace{1cm}
    \begin{subfigure}{0.47\textwidth}
        \centering
        \includegraphics[width=\linewidth]{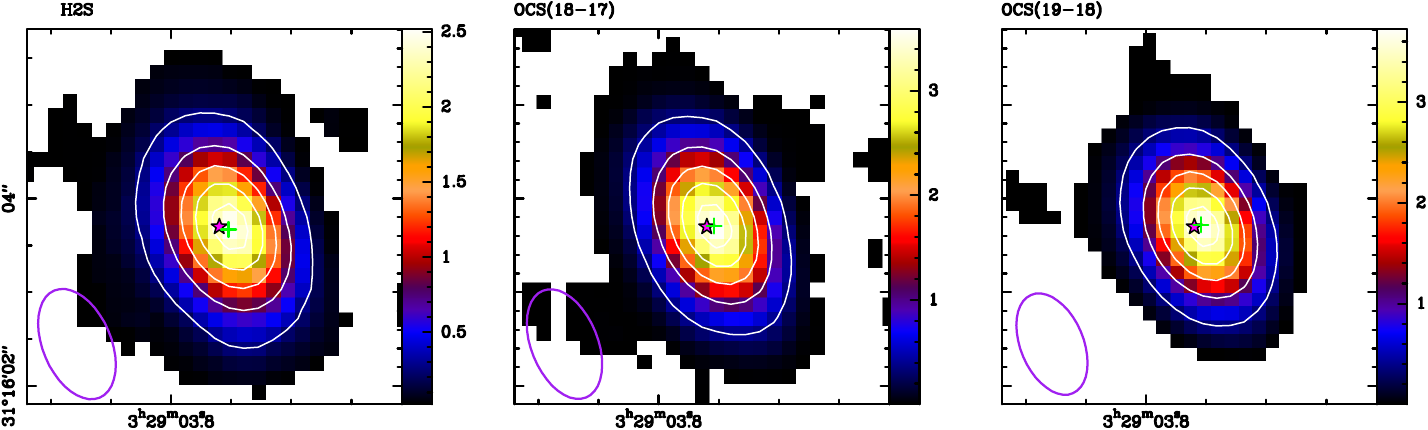}
        \caption{SVS13A}
        \label{fig:map_SVS13A}
    \end{subfigure}
    \hspace{0.9cm}
    \begin{subfigure}{0.47\textwidth}
        \centering
        \includegraphics[width=\linewidth]{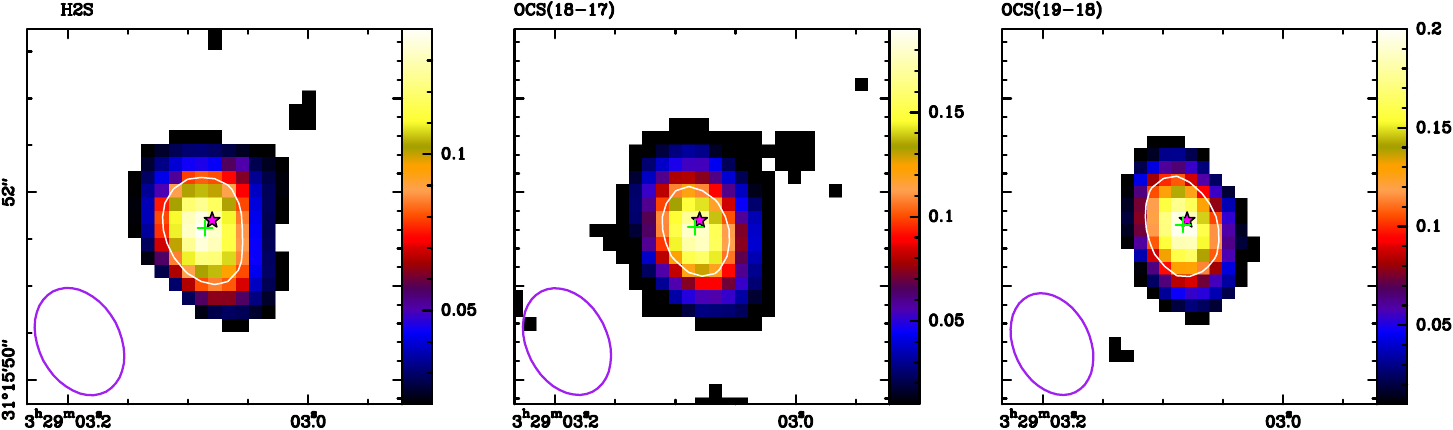}
        \caption{SVS13B}
        \label{fig:map_SVS13B}
    \end{subfigure}

    \caption{(Continuation)}
\end{figure*}

\begin{figure*}
    \centering

    \vspace{1cm}
    \begin{subfigure}{0.47\textwidth}
        \includegraphics[width=\linewidth]{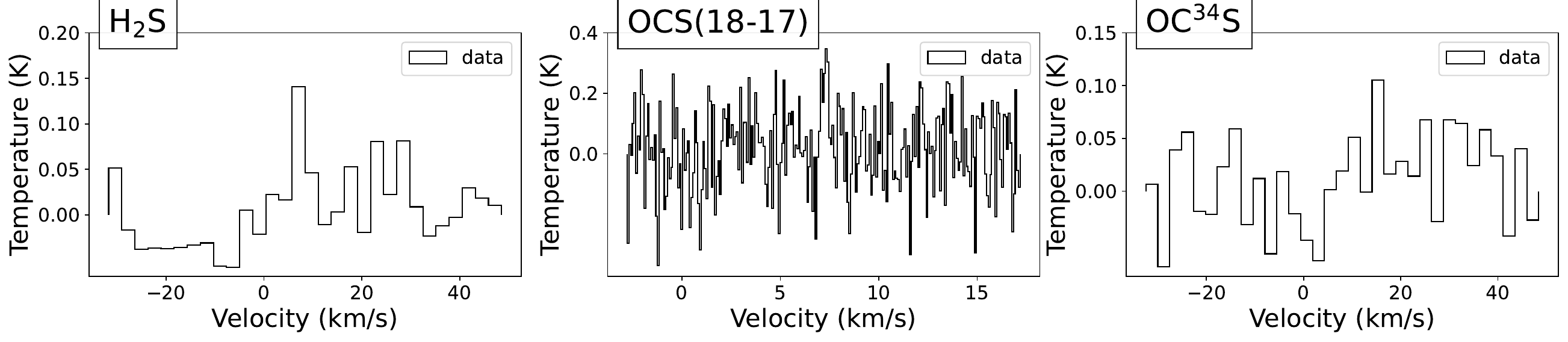}
        \caption{B1bN}
        \label{fig:spec_B1bN}
    \end{subfigure}
    \hspace{0.9cm}
    \begin{subfigure}{0.47\textwidth}
        \centering
        \includegraphics[width=\linewidth]{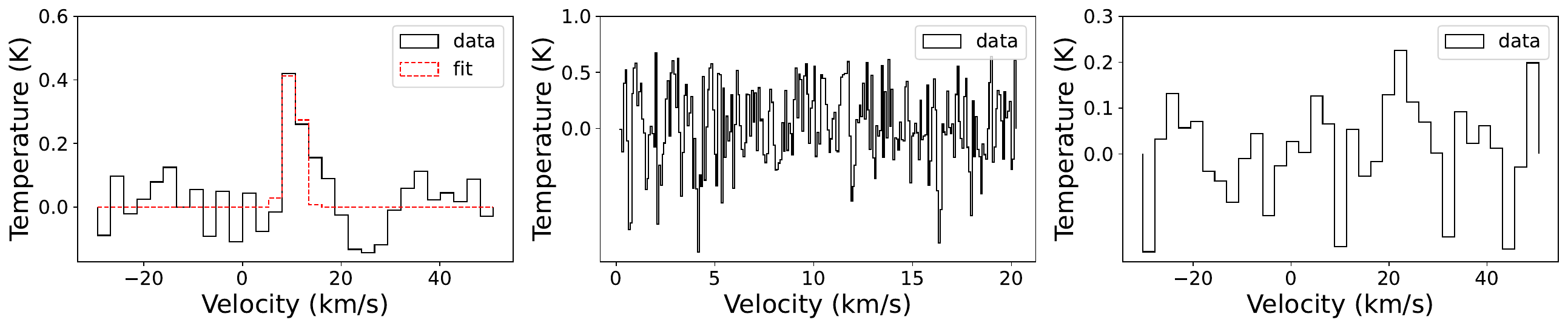}
        \caption{B5-IRS1}
        \label{fig:spec_B5-IRS1}
    \end{subfigure}

    \vspace{1cm}
    \begin{subfigure}{0.47\textwidth}
        \centering
        \includegraphics[width=\linewidth]{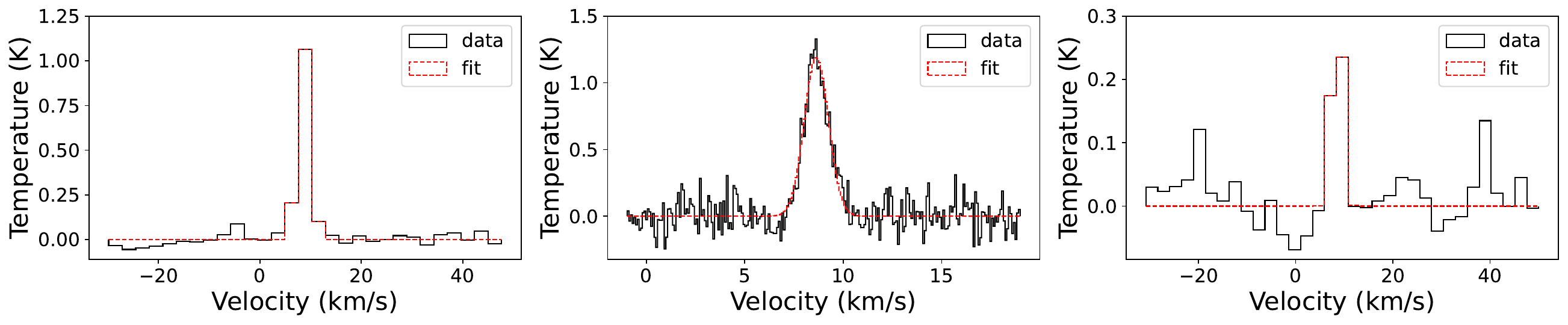}
        \caption{IC348MMS}
        \label{fig:spec_IC348MMS}
    \end{subfigure}
    \hspace{0.9cm}
    \begin{subfigure}{0.47\textwidth}
        \centering
        \includegraphics[width=\linewidth]{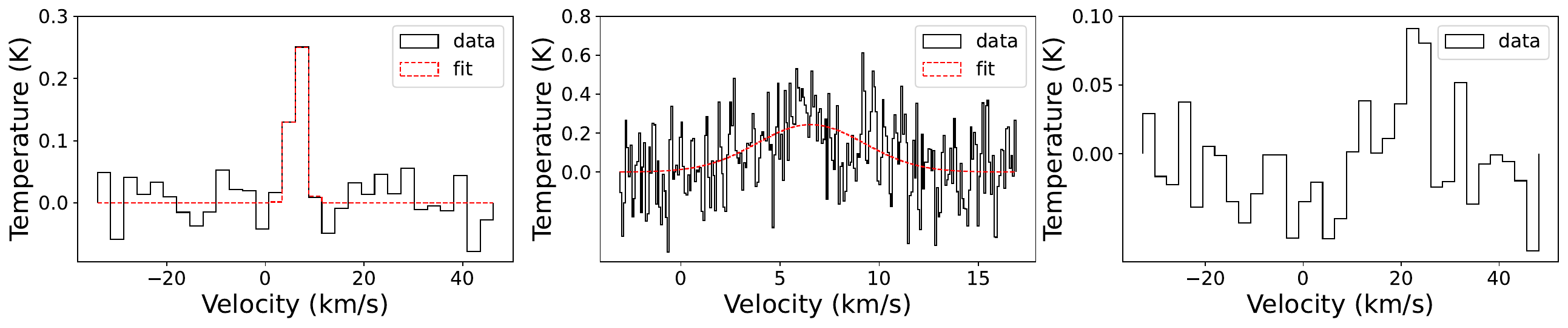}
        \caption{IRAS2B}
        \label{fig:spec_IRAS2B}
    \end{subfigure}

    \vspace{1cm}
    \begin{subfigure}{0.47\textwidth}
        \centering
        \includegraphics[width=\linewidth]{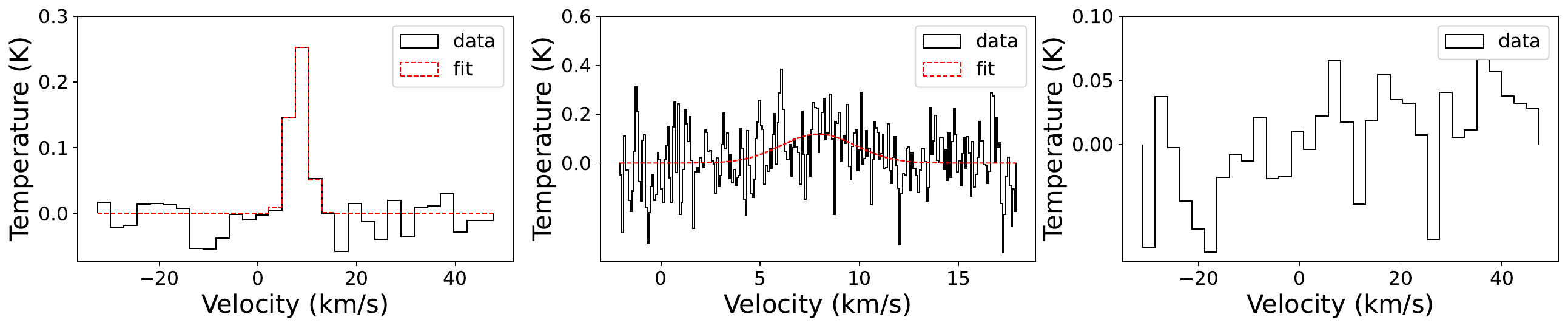}
        \caption{IRAS4C}
        \label{fig:spec_IRAS4C}
    \end{subfigure}
    \hspace{0.9cm}
    \begin{subfigure}{0.47\textwidth}
        \centering
        \includegraphics[width=\linewidth]{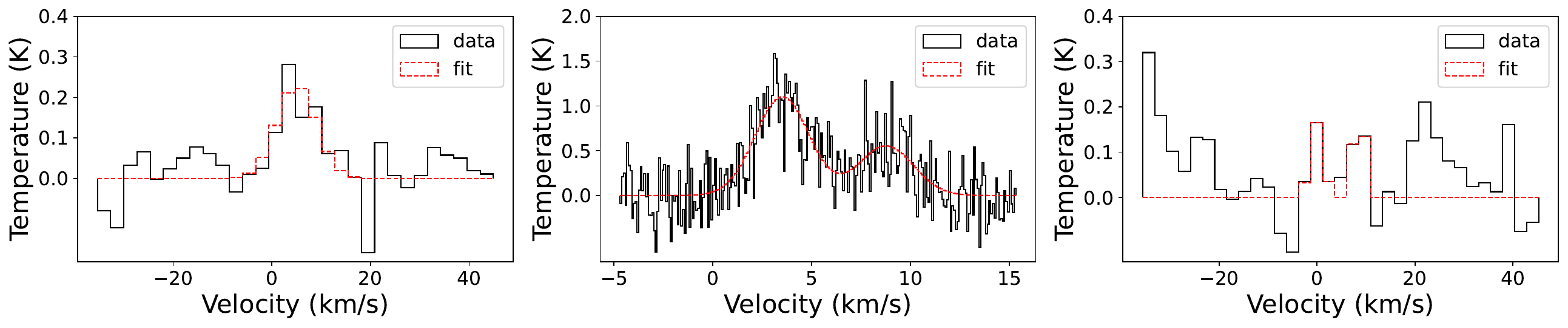}
        \caption{L1448-IRS3A}
        \label{fig:spec_L1448-IRS3A}
    \end{subfigure}

    \vspace{1cm}
    \begin{subfigure}{0.47\textwidth}
        \centering
        \includegraphics[width=\linewidth]{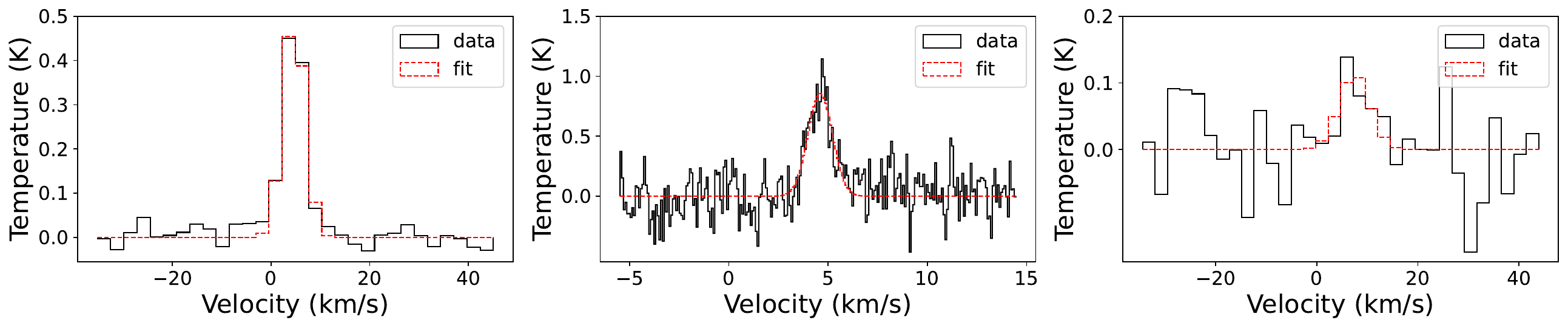}
        \caption{L1448-IRS3B}
        \label{fig:spec_L1448-IRS3B}
    \end{subfigure}
    \hspace{0.9cm}
    \begin{subfigure}{0.47\textwidth}
        \centering
        \includegraphics[width=\linewidth]{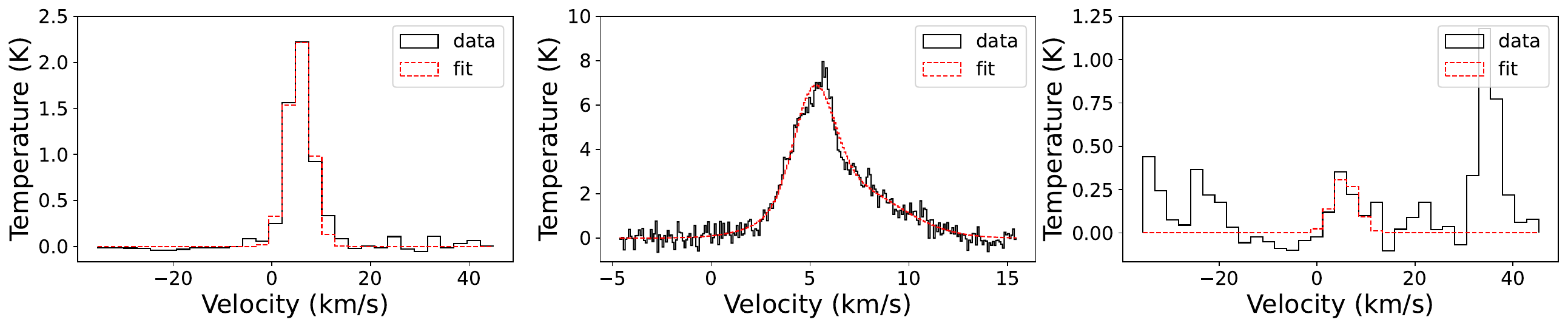}
        \caption{L1448C}
        \label{fig:spec_L1448C}
    \end{subfigure}

    \vspace{1cm}
    \begin{subfigure}{0.47\textwidth}
        \centering
        \includegraphics[width=\linewidth]{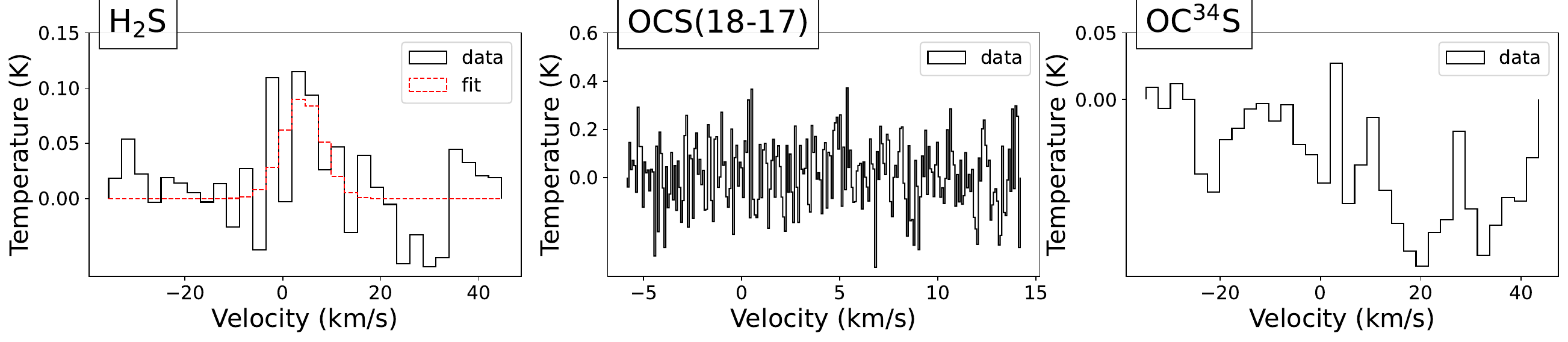}
        \caption{L1448NW}
        \label{fig:spec_L1448NW}
    \end{subfigure}
    \hspace{0.9cm}
    \begin{subfigure}{0.47\textwidth}
    \centering
        \includegraphics[width=\linewidth]{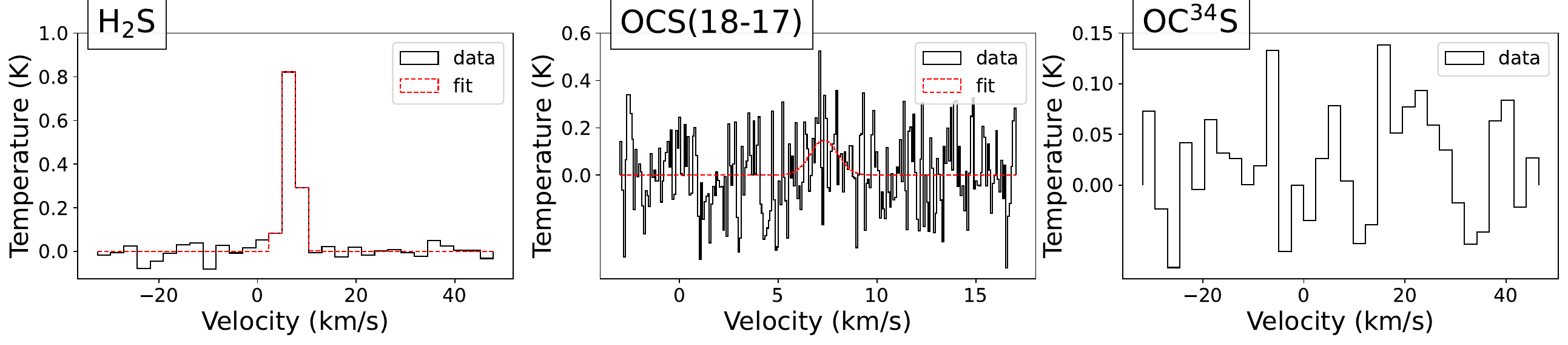}
        \caption{Per-emb-02}
        \label{fig:spec_Per-emb-2}
    \end{subfigure}

    \caption{Spectra of the H$_2$S 
    (2$_{2,0}$--2$_{1,1}$), OCS (18--17) and OC$^{34}$S (20--19) lines in the warm inner core of the 20 protostars from our sample not shown in Section~\ref{sec:results}. The red dotted lines represent the values of a Gaussian fit to each of the detected lines. In some cases, the combination of two Gaussian fits was necessary for the fit. The OCS (18-17) line was observed with a spectral resolution of 62.5 kHz while the H$_2$S and OC$^{34}$S lines were observed with a spectral resolution of 2 MHz. Gaussian fits have been plotted for the >3$\sigma$ detections, except for the OC$^{34}$S were we also show the <3$\sigma$ fits when the OC$^{33}$S counterpart was detected with >3$\sigma$.}
    \label{fig:specs-appendix}
\end{figure*}

\begin{figure*}\ContinuedFloat
    \centering

    \vspace{1cm}
    \begin{subfigure}{0.47\textwidth}
        \includegraphics[width=\linewidth]{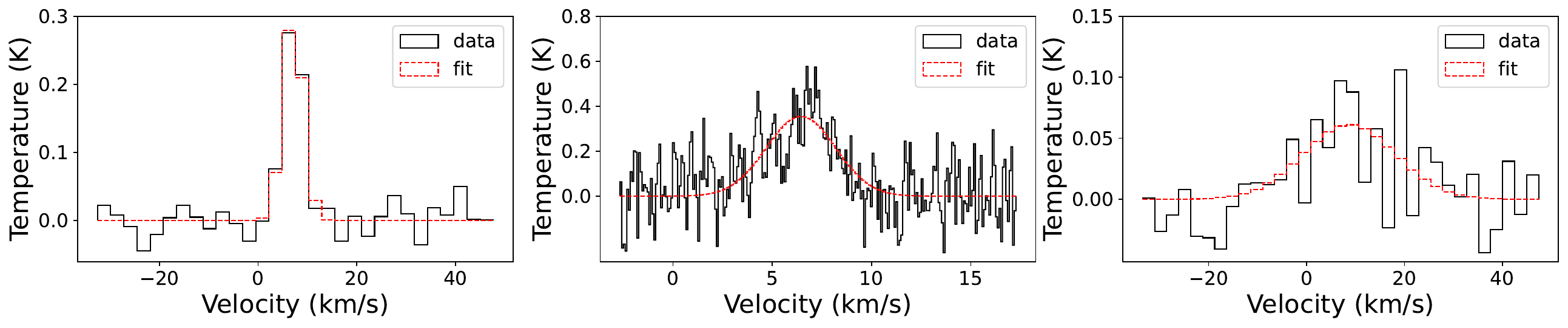}
        \caption{Per-emb-05}
        \label{fig:spec_Per-emb-5}
    \end{subfigure}
    \hspace{0.9cm}
    \begin{subfigure}{0.47\textwidth}
        \centering
        \includegraphics[width=\linewidth]{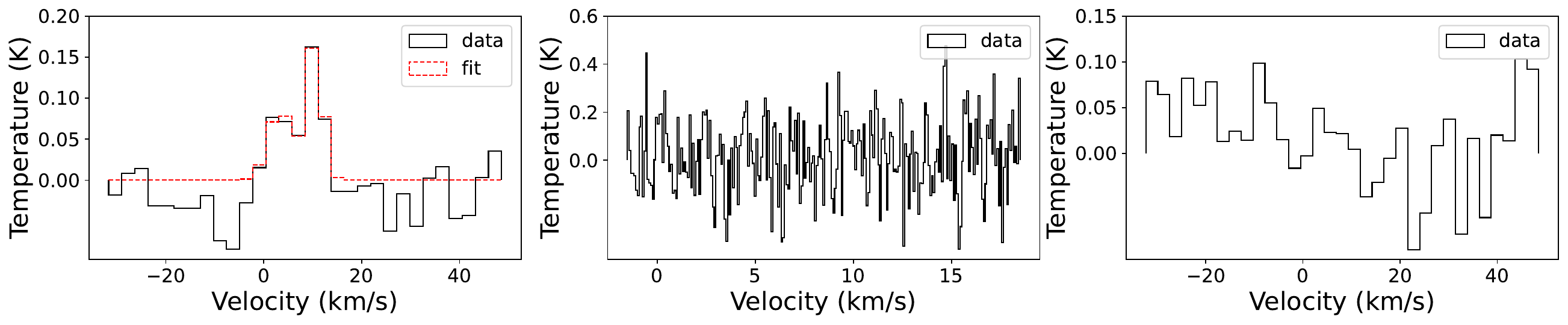}
        \caption{Per-emb-08}
        \label{fig:spec_Per-emb-8}
    \end{subfigure}

    \vspace{1cm}
    \begin{subfigure}{0.47\textwidth}
        \centering
        \includegraphics[width=\linewidth]{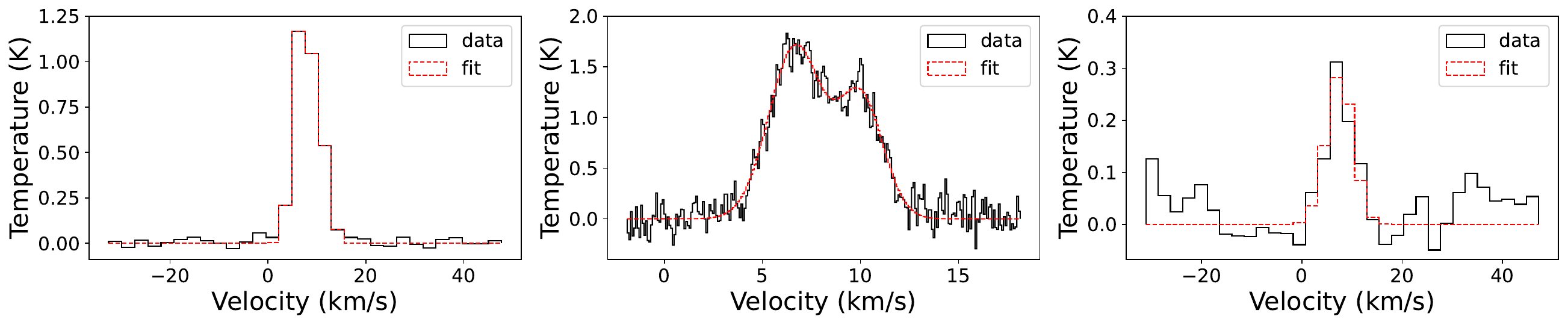}
        \caption{Per-emb-18}
        \label{fig:spec_Per-emb-18}
    \end{subfigure}
    \hspace{0.9cm}
    \begin{subfigure}{0.47\textwidth}
        \centering
        \includegraphics[width=\linewidth]{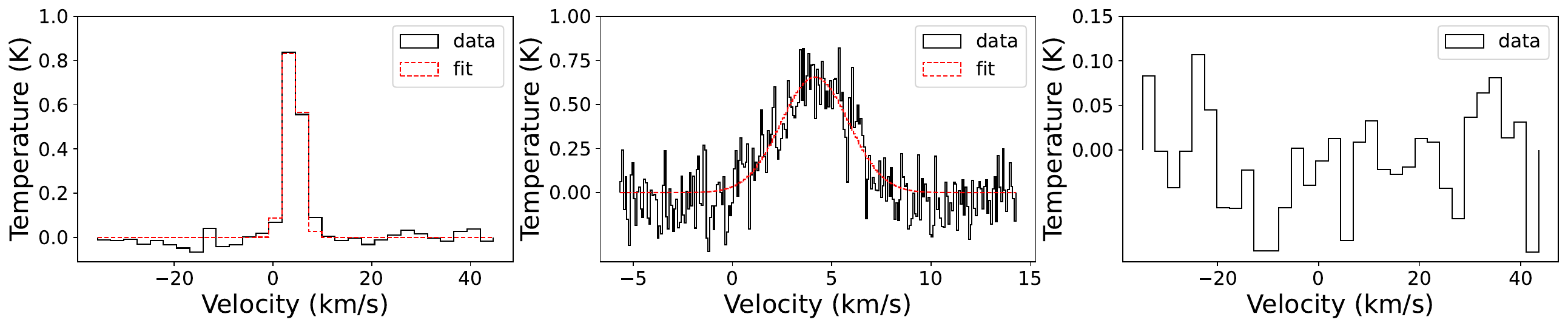}
        \caption{Per-emb-22}
        \label{fig:spec_Per-emb-22}
    \end{subfigure}

    \vspace{1cm}
    \begin{subfigure}{0.47\textwidth}
        \centering
        \includegraphics[width=\linewidth]{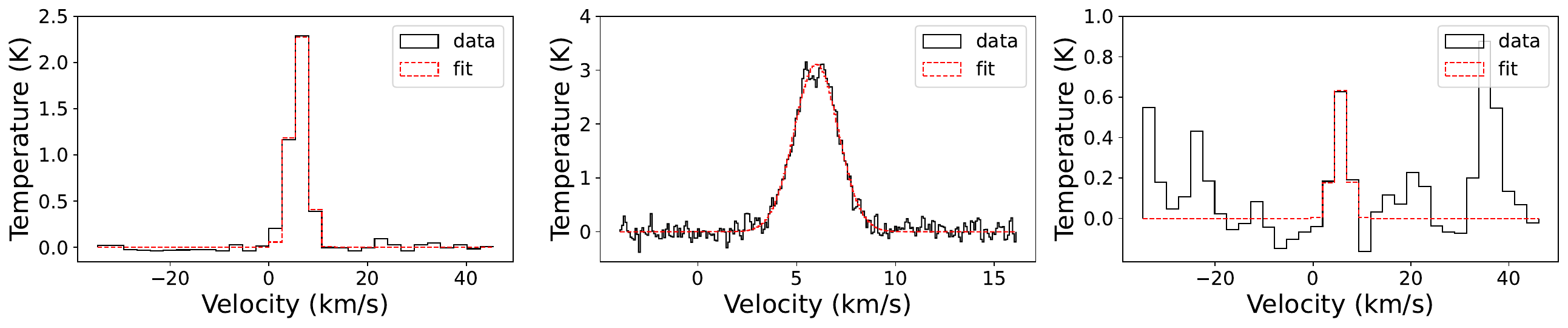}
        \caption{Per-emb-29}
        \label{fig:spec_Per-emb-29}
    \end{subfigure}
    \hspace{0.9cm}
    \begin{subfigure}{0.47\textwidth}
        \centering
        \includegraphics[width=\linewidth]{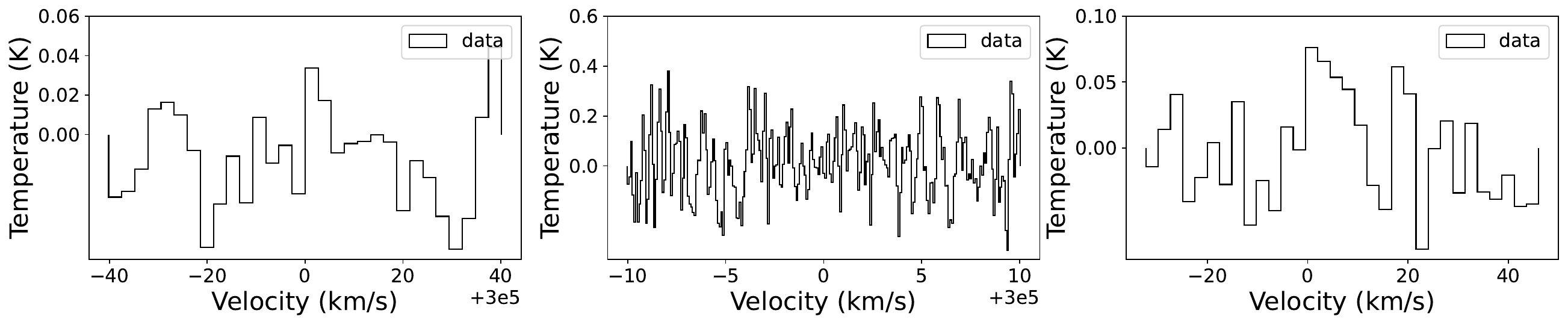}
        \caption{Per-emb-30}
        \label{fig:spec_Per-emb-30}
    \end{subfigure}

    \vspace{1cm}
    \begin{subfigure}{0.47\textwidth}
        \centering
        \includegraphics[width=\linewidth]{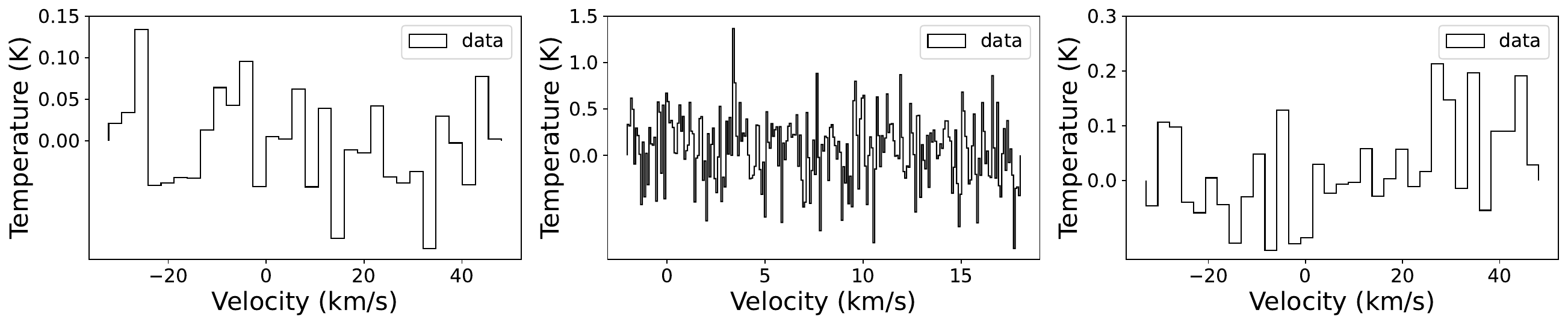}
        \caption{Per-emb-50}
        \label{fig:spec_Per-emb-50}
    \end{subfigure}
    \hspace{0.9cm}
    \begin{subfigure}{0.47\textwidth}
        \centering
        \includegraphics[width=\linewidth]{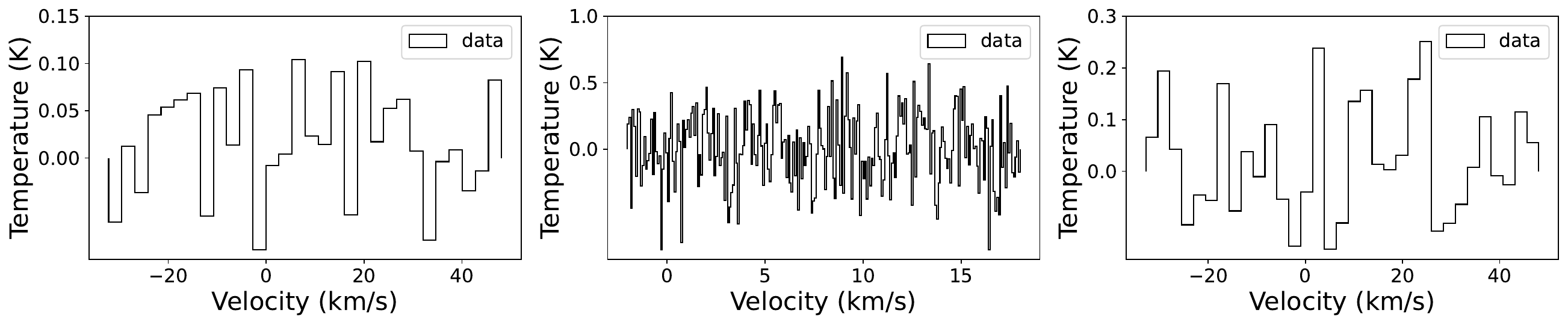}
        \caption{Per-emb-62}
        \label{fig:spec_Per-emb-62}
    \end{subfigure}

    \vspace{1cm}
    \begin{subfigure}{0.47\textwidth}
        \centering
        \includegraphics[width=\linewidth]{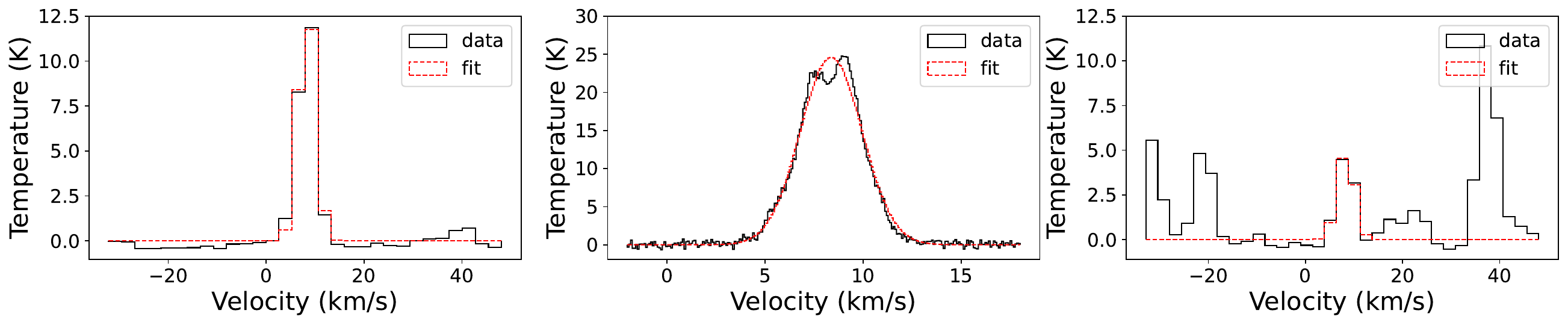}
        \caption{SVS13A}
        \label{fig:spec_SVS13A}
    \end{subfigure}
    \hspace{0.9cm}
    \begin{subfigure}{0.47\textwidth}
        \centering
        \includegraphics[width=\linewidth]{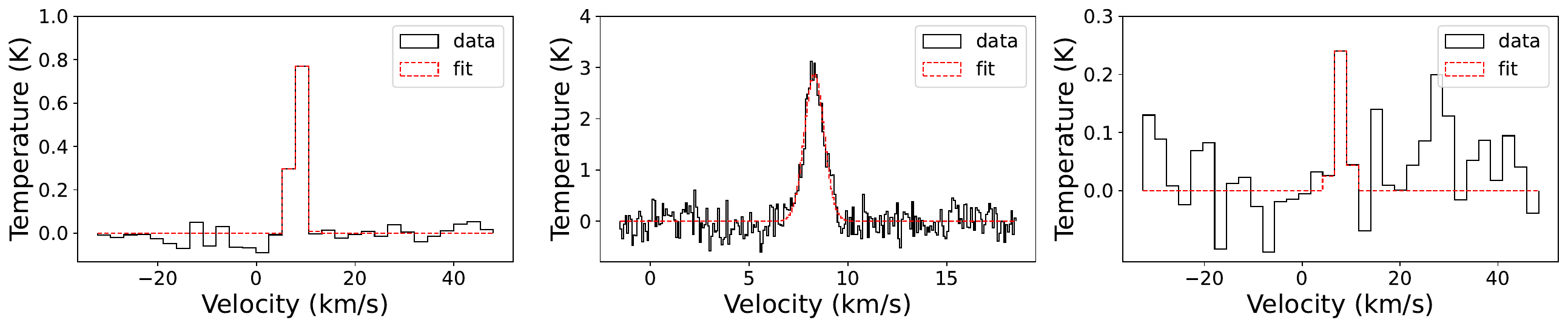}
        \caption{SVS13B}
        \label{fig:spec_SVS13B}
    \end{subfigure}

    \caption{(Continuation)}
\end{figure*}

\end{document}